\documentclass[prd,twocolumn,showpacs,superscriptaddress]{revtex4-1}
\usepackage{graphicx}
\usepackage{units}
\usepackage{multirow}
\usepackage{bigstrut}
\usepackage{overpic}
\usepackage{array}
\usepackage{color}
\usepackage{amsmath}
\usepackage{bm}
\usepackage{times}
\RequirePackage{lineno}
\newcommand{\gevc}{\,\unit{GeV}/c}

\begin{document}

\modulolinenumbers[2]

\setlength{\oddsidemargin}{-0.5cm} \addtolength{\topmargin}{15mm}

\title{\boldmath Model-independent determination of the relative strong-phase difference between $D^0$ and $\bar{D}^0\rightarrow K^0_{S,L}\pi^+\pi^-$ and its impact on the measurement of the CKM angle $\gamma/\phi_3$}

\author{
  \small
M.~Ablikim$^{1}$, M.~N.~Achasov$^{10,d}$, P.~Adlarson$^{59}$, S. ~Ahmed$^{15}$, M.~Albrecht$^{4}$, M.~Alekseev$^{58A,58C}$, D.~Ambrose$^{51}$, A.~Amoroso$^{58A,58C}$, F.~F.~An$^{1}$, Q.~An$^{55,43}$
% ...
      \\
      \vspace{0.2cm}
      (BESIII Collaboration)\\
      \vspace{0.2cm} {\it
$^{1}$ Institute of High Energy Physics, Beijing 100049, People's Republic of China\\
% ...
$^{4}$ Bochum Ruhr-University, D-44780 Bochum, Germany\\
% ...
$^{10}$ G.I. Budker Institute of Nuclear Physics SB RAS (BINP), Novosibirsk 630090, Russia\\
% ...
$^{15}$ Helmholtz Institute Mainz, Johann-Joachim-Becher-Weg 45, D-55099 Mainz, Germany\\
% ...
$^{43}$ State Key Laboratory of Particle Detection and Electronics, Beijing 100049, Hefei 230026, People's Republic of China\\
% ...
$^{51}$ University of Minnesota, Minneapolis, Minnesota 55455, USA\\
% ...
$^{55}$ University of Science and Technology of China, Hefei 230026, People's Republic of China\\
% ...
$^{58}$ (A)University of Turin, I-10125, Turin, Italy; (B)University of Eastern Piedmont, I-15121, Alessandria, Italy; (C)INFN, I-10125, Turin, Italy\\
$^{59}$ Uppsala University, Box 516, SE-75120 Uppsala, Sweden\\
% ...
$^{d}$ Also at the Novosibirsk State University, Novosibirsk, 630090, Russia\\
% ...
     \vspace{0.4cm}
}
}

\begin{abstract}
Crucial inputs for a variety of $CP$-violation studies can be determined through the analysis of pairs of quantum-entangled neutral $D$ mesons, which are produced in the decay of the $\psi(3770)$ resonance. The relative strong-phase parameters between $D^0$ and $\bar{D}^0$ in the decays $D^0\rightarrow K^0_{S,L}\pi^+\pi^-$ are studied using 2.93~${\rm fb}^{-1}$ of $e^+e^-$ annihilation data delivered by the BEPCII collider and collected by the BESIII detector at a center-of-mass energy of 3.773 GeV. Results are presented in regions of the phase space of the decay. These are the most precise measurements to date of the strong-phase parameters in $D \to K_{S,L}^0\pi^+\pi^-$ decays. Using these parameters, the associated uncertainty on the Cabibbo-Kobayashi-Maskawa angle $\gamma/\phi_3$ is expected to be between $0.7^\circ$ and $1.2^\circ$, for an analysis using the decay $B^{\pm}\rightarrow DK^{\pm}$, $D\rightarrow K^0_S\pi^+\pi^-$, where $D$ represents a superposition of $D^0$ and $\bar{D^0}$ states. This is a factor of three smaller than that achievable with previous measurements. Furthermore, these results provide valuable input for charm-mixing studies, other measurements of $CP$ violation, and the measurement of strong-phase parameters for other $D$-decay modes.

\end{abstract}


\maketitle

\section{\bf Introduction }

The study of quantum-correlated charm-meson pairs produced at
threshold allows unique access to hadronic decay properties that are
of great interest across a wide range of physics applications.  In
particular, determination of the strong-phase parameters provides
vital input to measurements of the Cabibbo-Kobayashi-Maskawa
(CKM)~\cite{CKM} angle $\gamma$ (also denoted $\phi_3$) and other
$CP$-violating observables.  The same parameters are required for
studies of $D^0\bar{D}^0$ mixing and $CP$ violation in charm at
experiments above threshold.  The angle $\gamma$ is a parameter of the
unitarity triangle (UT), which is a geometrical representation of the
CKM matrix in the complex plane. Within the standard model (SM) all
measurements of unitarity-triangle parameters should be
self-consistent.  The parameter $\gamma$ is of particular interest
since it is the only angle of the UT that can easily be extracted in
tree-level processes, in which the contribution of non-SM effects is
expected to be very small~\cite{jhep1401_051}. Therefore, a measurement
of $\gamma$ provides a benchmark of the SM with negligible theoretical
uncertainties.  A precise measurement of $\gamma$ is an essential
ingredient in testing the SM description of $CP$ violation. A comparison
between this, direct, measurement of gamma, and the indirect
determination coming from the other constraints of the UT is a
sensitive probe for new physics.

One of the most sensitive decay channels for measuring $\gamma$ is $B^{-}\rightarrow DK^{-}$, $D\rightarrow K^0_S\pi^+\pi^-$~\cite{prd68_054018} where $D$ represents a superposition of $D^0$ and $\bar{D}^0$ mesons. Throughout this paper, charge conjugation is assumed unless otherwise explicitly noted.
The amplitude of the $B^-$ decay can be written as
\begin{equation}
f_{B^-}(m^2_+,m^2_-)\propto f_D(m^2_+,m^2_-)+r_Be^{i(\delta_B-\gamma)}f_{\bar{D}}(m^2_+,m^2_-).
\label{eq:amp_bdecay}
\end{equation}
Here, $m^2_+$ and $m^2_-$ are the squared invariant masses of the $K^0_S\pi^+$ and $K^0_S\pi^-$ pairs from the $D^0\rightarrow K^0_S\pi^+\pi^-$ decay, $f_D(m^2_+,m^2_-)(f_{\bar{D}}(m^2_+,m^2_-))$ is the
amplitude of the $D^0(\bar{D}^0)$ decay to $K^0_S\pi^+\pi^-$ at $(m^2_+,m^2_-)$ in the Dalitz plot, $r_B$ is the ratio of the suppressed amplitude to the favored amplitude, and $\delta_B$ is the $CP$-conserving strong-phase difference between them. If the small second-order effects of charm mixing and $CP$ violation~\cite{prd68_054018,prd72_031501,prd82_034033,epjc47_347,epjc73_2476} are ignored Eq.~(\ref{eq:amp_bdecay}) can be written as
\begin{equation}
f_{B^-}(m^2_+,m^2_-)\propto f_D(m^2_+,m^2_-)+r_Be^{i(\delta_B-\gamma)}f_{D}(m^2_-,m^2_+)
\label{eq:amp_bdecay_2}
\end{equation}
through the use of the relation $f_{\bar{D}}(m^2_+,m^2_-)=f_{D}(m^2_-,m^2_+)$.
The square of the amplitude clearly depends on the
strong-phase difference $\Delta \delta_D\equiv\delta_D(m^2_+,m^2_-)-\delta_D(m^2_-,m^2_+)$, where $\delta_D(m^2_+,m^2_-)$ is the strong phase of $f_D(m^2_+,m^2_-)$.
While the strong-phase difference can be inferred from an amplitude model of the decay $D^0\rightarrow K^0_S\pi^+\pi^-$, such an approach introduces model-dependence in the measurement. This property is undesirable as the systematic uncertainty associated with the model is difficult to estimate reliably, since common approaches to amplitude-model building break the optical theorem~\cite{acta46_257}.
Instead, the strong-phase differences may be measured directly in the decays of quantum-correlated neutral $D$ meson pairs created in the decay of the $\psi(3770)$ resonance~\cite{prd68_054018,epjc47_347}. This approach ensures a model-independent ~\cite{plb718_43,jhep10_097,jhep06_131,jhep08_176,prd85_112014} measurement of $\gamma$ where the uncertainty in the strong-phase knowledge can be reliably propagated.

Knowledge of the strong-phase difference in $D\rightarrow
K^0_S\pi^+\pi^-$ has important applications beyond the measurement of
the angle $\gamma$ in $B^{\pm}\rightarrow DK^{\pm}$ decays. First,
this information can be used in $\gamma$ measurements based on other
$B$ decays ~\cite{jhep06_131,prd81_014025}. Second, it can be
exploited to provide a model-independent measurement of the CKM angle
$\beta$ through a time-dependent analysis of $\bar{B}^0\rightarrow
Dh^0$ where $h$ is a light meson~\cite{prd94_052004} and
$B^0\rightarrow D\pi^+\pi^-$~\cite{jhep03_195}. Finally, $D\rightarrow
K^0_S\pi^+\pi^-$ is also a powerful decay mode for performing
precision measurements of oscillation parameters and $CP$ violation in
$D^0\bar{D}^0$
mixing~\cite{jhep10_185,jhep04_033,prd99_012007,prl122_231802}. Again,
knowledge of the strong-phase differences allows these measurements to
be executed in a model-independent
manner~\cite{prd99_012007,prl122_231802}. The ability to have
model-independent results is critical as these measurements become
increasingly precise with the large data sets that will be analyzed at
LHCb and Belle~II, over the coming decade.

The strong-phase differences in $D\rightarrow K^0_S\pi^+\pi^-$ have
been studied by the CLEO collaboration using 0.82 fb$^{-1}$ of
data~\cite{prd80_032002,prd82_112006}. These measurements are limited
by their statistical precision and would contribute major
uncertainties to the measurements of $\gamma$, and mixing and
$CP$ violation in the charm sector, anticipated in the near future. The
BESIII detector at the BEPCII collider has the largest data sample
collected at the $\psi(3770)$ resonance, corresponding to an
integrated luminosity of 2.93 fb$^{-1}$. Therefore it is possible to
substantially improve the knowledge of the strong-phase differences,
which will reduce the associated uncertainty when used in other
$CP$ violation measurements.

The observables measured in this analysis are the amplitude-weighted average cosine and sine of the strong-phase difference for $D\rightarrow K^0_S\pi^+\pi^-$ and $D\rightarrow K^0_L\pi^+\pi^-$ in regions of phase space. The paper is organized as follows. In Sec.~\ref{sec:theory} the formalism of how the strong-phase information can be accessed is discussed along with the description of the phase space regions. The BESIII detector and the simulated data are described in Sec.~\ref{sec:detector}. The event selection is presented in Sec.~\ref{sec:evtsel}. Sections~\ref{sec:cisi} and \ref{sec:syst} describe the measurement of the strong-phase parameters and their systematic uncertainties. The impact of these results on measurements of $\gamma$ is assessed in Sec.~\ref{sec:gamma}.
This paper is accompanied by a letter submitted to Physical Review Letters~\cite{PRL}.

\section{Formalism}
\label{sec:theory}

\subsection{Division of phase space}

\begin{figure*}[htbp]
\begin{center}
\includegraphics[width=\linewidth]{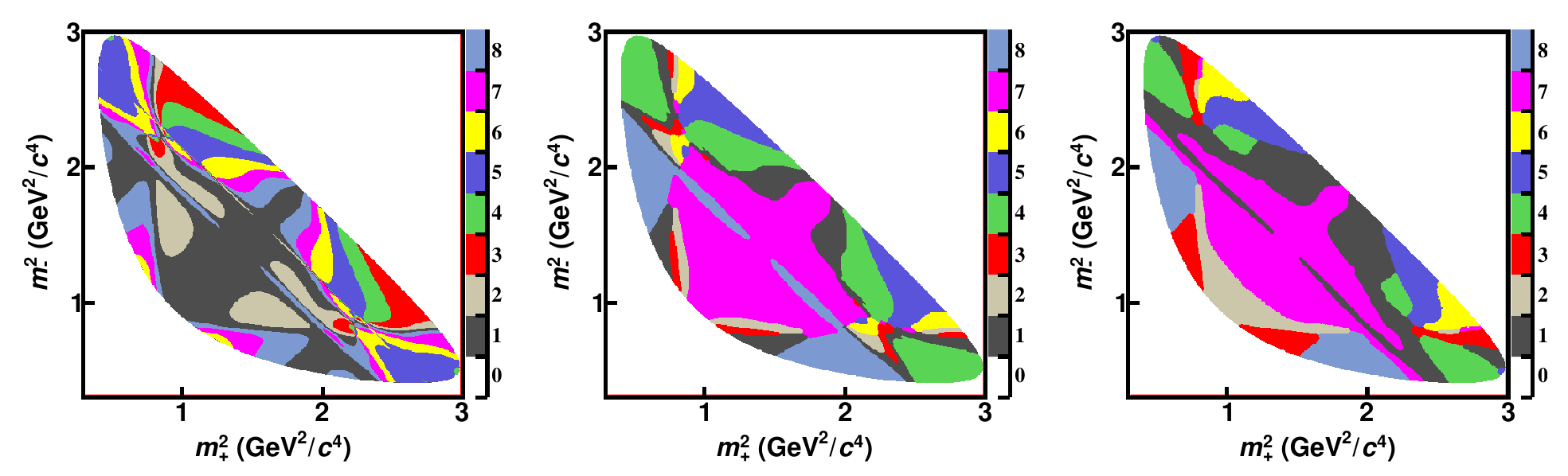}
\caption{ The (left) equal $\Delta \delta_D$, (middle) optimal and (right) modified optimal binnings of the $D\rightarrow K^0_{S,L}\pi^+\pi^-$ Dalitz plot from Ref.~\cite{prd82_112006}. The color scale represents the absolute value of the bin number $|i|$.}
\label{Fig:Bin}
\end{center}
\end{figure*}

The analysis of the data is performed in regions of phase space. Measurements are presented in three schemes which are identical to those used in Ref.~\cite{prd82_112006}. All schemes divide the phase space into eight pairs of bins, symmetrically along the $m^2_+=m^2_-$ line. The bins are indexed with $i$, running from $-8$ to $8$ excluding zero. The bins have a positive index if their position satisfies $m^2_+ < m^2_-$, and the exchange of coordinates $(m^2_+,m^2_-) \leftrightarrow (m^2_-,m^2_+)$, changes the sign of the bin. The choice of division of the phase space has an impact on the sensitivity of the $CP$ violation measurements that use this strong-phase information as input.
The schemes are irregular in shape and are shown in Fig.~\ref{Fig:Bin}. Detailed information on the choice of these regions is given in Ref.~\cite{prd82_112006}.
The scheme denoted ``equal binning'' defines regions such that the variation in $\Delta\delta_D$ over each bin is minimized, and is based on a model developed on flavor-tagged data~\cite{prd78_034023} to partition the phase space. In the half of the Dalitz plot $m^2_+<m^2_-$, the $i$th bin is defined by the condition
\begin{equation}
2\pi(i-3/2)/8<\Delta\delta_D(m^2_+,m^2_-)<2\pi(i-1/2)/8.
\label{eq:equalbin}
\end{equation}
A more sensitive scheme for the measurement of $\gamma$, denoted as
``optimal binning'', takes into account both the model of the
$D^0\rightarrow K^0_S\pi^+\pi^-$ decay and the expected distribution
of $D$ decays arising from the process $B^{-}\rightarrow DK^{-}$ when
determining the bins. This choice improves the sensitivity of $\gamma$
measurements compared to the equal binning by approximately
10$\%$. The third binning scheme, denoted the ``modified optimal
binning'' is useful in analyzing samples with low
yields~\cite{jhep06_131}. Although these three binning schemes are
based on the $D^0\rightarrow K^0_S\pi^+\pi^-$ model reported in
Ref.~\cite{prd78_034023}, this procedure does not introduce
model-dependence into the analyses that employ the resulting
strong-phase measurements. The determination of $CP$ violation
parameters will remain unbiased, but they may have a loss in
sensitivity with respect to expectation, due to the differences
between the model and the true strong-phase variation.

\subsection{Event yields in quantum-correlated data}

The interference between the amplitudes of the $D^0$ and $\bar{D}^0$ decays can be parameterized by two quantities $c_i$ and $s_i$, which are the amplitude-weighted averages of ${\rm cos}\Delta\delta_D$ and ${\rm sin}\Delta\delta_D$ over each Dalitz plot bin. They are defined as
\begin{eqnarray}
c_i=&&\frac{1}{\sqrt{F_iF_{-i}}}\int_i|f_D(m^2_+,m^2_-)||f_D(m^2_-,m^2_+)| \nonumber \\
      &&\times {\rm cos}[\Delta\delta_D(m^2_+,m^2_-)]dm^2_+dm^2_-,
\label{eq:ci}
\end{eqnarray}
and
\begin{eqnarray}
s_i=&&\frac{1}{\sqrt{F_iF_{-i}}}\int_i|f_D(m^2_+,m^2_-)||f_D(m^2_-,m^2_+)| \nonumber \\
      &&\times {\rm sin}[\Delta\delta_D(m^2_+,m^2_-)]dm^2_+dm^2_-,
\label{eq:si}
\end{eqnarray}
where $F_i$ is the fraction of events found in the $i$th bin of the flavor-specific decay $D^0\rightarrow K^0_S\pi^+\pi^-$.

The $\psi(3770)$ has a $C=-1$ quantum number and this is conserved in the strong decay in which two neutral $D$ mesons are produced. Hence the two neutral $D$ mesons have an anti-symmetric wave function. This also means that the two $D$ mesons do not decay independently of one another.

For example, if one $D$ meson decays to a $CP$-even eigenstate, {\it
  e.g.} $K^+K^-$, then the other $D$ meson is known to be a $CP$-odd
state. The analysis strategy is to use double-tagged events in which
both charm mesons are reconstructed.  The yield of events in which one
meson is flavor-tagged, {\it e.g.} through the decay $K^-e^+\nu_e$,
and the other decays to $D^0 \to K^0_S \pi^+\pi^-$ in bin $i$ can be
used to determine $K_i\propto\int_i |f_D(m^2_+,m^2_-)|^2
dm^2_+dm^2_-$~\cite{epjc47_347}. The details of determining $K_i$
through using flavor-specific decays are described in
Sec.~\ref{sec:kikip}.

Considering a pair of decays where one $D$ meson decays to $CP$
eigenstate, referred to as ``the tag'', and the other $D$ meson decays
to the $K^0_S\pi^+\pi^-$ final state, the decay amplitude of the
$D\rightarrow K^0_S\pi^+\pi^-$ decay is given by
\begin{equation}
f_{CP\pm}(m^2_+,m^2_-)=\frac{1}{\sqrt{2}}[f_D(m^2_+,m^2_-)\pm f_D(m^2_-,m^2_+)],
\label{eq:cpamp}
\end{equation}
where $f_{CP^{\pm}}$ refers to the $CP$ eigenvalue of the $D\rightarrow K^0_S\pi^+\pi^-$ decay.  It is possible to generalize this expression to include decays where the tag $D$ meson decays to a self-conjugate final state rather than a $CP$ eigenstate, assuming that the $CP$-even fraction, $F_{CP}$, is known. The number of events observed in the $i$th bin, $M_i$, where the tag $D$ meson decays to a self-conjugate final state is then given by
\begin{equation}
M_i=h_{CP}(K_i-(2F_{CP}-1)2c_i\sqrt{K_iK_{-i}}+K_{-i}),
\label{eq:mi}
\end{equation}
where $h_{CP}$ is a normalization factor. The value of $F_{CP}$ is 1 for $CP$-even tags and 0 for $CP$-odd tags.
This parameterization is valuable since it allows for final states with very high or very low $CP$-even fractions to be used to provide sensitivity to the $c_i$ parameters. A good example of such a decay is the mode $D\rightarrow\pi^+\pi^-\pi^0$ where the fractional $CP$-even content is measured to be $F^{\pi\pi\pi^0}_{CP}=0.973\pm0.017$~\cite{plb747_9}.

However, from Eq.~(\ref{eq:ci}), the sign of $\Delta\delta_D$ is undetermined if only the values of $c_i$ are known from the $CP$-tagged $D\rightarrow K^0_S\pi^+\pi^-$ decay.
Important additional information can be gained  to determine the $s_i$ parameters by studying the Dalitz plot distributions where both $D$ mesons decay to $K^0_S\pi^+\pi^-$.
The amplitude of the $\psi(3770)$ decay is in this case given by
\begin{eqnarray}
&&f(m^2_+,m^2_-,m^{2\dagger}_+,m^{2\dagger}_-) \nonumber \\
&&=\frac{f_D(m^2_+,m^2_-)f_D(m^{2\dagger}_-,m^{2\dagger}_+)-f_D(m^{2\dagger}_+,m^{2\dagger}_-)f_D(m^2_-,m^2_+)}{\sqrt{2}},  \nonumber \\
\end{eqnarray}
where the use of the $'\dagger'$ symbol differentiates the Dalitz plot coordinates of the two $D\rightarrow K^0_S\pi^+\pi^-$ decays. The variable $M_{ij}$ is defined as the event yield observed in the $i$th bin of the first and the $j$th bin of the second $D\rightarrow K^0_S\pi^+\pi^-$ Dalitz plot, and is given by
\begin{eqnarray}
M_{ij}=&&h_{\rm corr}[ K_iK_{-j}+K_{-i}K_j \nonumber \\
           &&-2\sqrt{K_iK_{-j}K_{-i}K_j}(c_ic_j+s_is_j) ]
\label{eq:mij},
\end{eqnarray}
where $h_{\rm corr}$ is a normalization factor.
Equation~(\ref{eq:mij}) is not sensitive to the sign of $s_i$, however, this ambiguity can be resolved using a weak model assumption.

In order to improve the precision of the $c_i$ and $s_i$ parameters it is useful to increase the possible tags to include $D\rightarrow K^0_L\pi^+\pi^-$ which is closely related to the $D\rightarrow K^0_S\pi^+\pi^-$ decay.
The convention $A(D^0\rightarrow K^0_S\pi^+\pi^-)=A(\bar{D}^0\rightarrow K^0_S\pi^-\pi^+)$ is used,
making the good approximation that the $K^0_S$ meson is $CP$-even. Similarly,
it follows that $A(D^0\rightarrow K^0_L\pi^+\pi^-)=-A(\bar{D}^0\rightarrow K^0_L\pi^-\pi^+)$. Hence, where the $D\rightarrow K^0_L\pi^+\pi^-$ is used as the signal decay, and the tag is a self-conjugate final state, the observed event yield $M_i^\prime$ is given by
\begin{equation}
M^{\prime}_i=h^{\prime}_{CP}(K^{\prime}_i+(2F_{CP}-1) 2c_i\sqrt{K^{\prime}_iK^{\prime}_{-i}}+K^{\prime}_{-i}),
\label{eq:mip}
\end{equation}
where $K^{\prime}_i$ and $c^{\prime}_i$ are associated to the $D\rightarrow K^0_L\pi^+\pi^-$ decay. The event yield $M^\prime_{ij}$, corresponding to the yield of events where the $D\rightarrow K^0_S\pi^+\pi^-$ decay is observed in the
$i$th bin and the $D\rightarrow K^0_L\pi^+\pi^-$ decay is observed in the $j$th bin, is given by
\begin{eqnarray}
M^{\prime}_{ij}=&&h^{\prime}_{\rm corr}[K_iK^{\prime}_{-j}+K_{-i}K^{\prime}_j \nonumber \\
                           &&+2\sqrt{K_iK^{\prime}_{-j}K_{-i}K^{\prime}_j}(c_ic^{\prime}_j+s_is^{\prime}_j)],
\label{eq:mijp}
\end{eqnarray}
where $s_i^{\prime}$ is the amplitude-weighted average sine of the strong-phase difference for the $D\rightarrow K^0_L\pi^+\pi^-$ decay.

In Eqs.~(\ref{eq:mi}), (\ref{eq:mij}), (\ref{eq:mip}) and (\ref{eq:mijp}),
the normalization factors $h^{(\prime)}_{CP}$ and $h^{(\prime)}_{\rm corr}$ can be related to the yields of reconstructed signal and tag final states, the reconstruction efficiencies, and the number of neutral $D$-meson pairs $N_{D\bar{D}}$ produced in the data set, with $h^{(\prime)}_{CP}=S_{CP}/2S_{\rm FT^{(\prime)}}\times \epsilon^{K^0_{S(L)}\pi^+\pi^-}$, $h_{\rm corr}=N_{D\bar{D}}/(2S_{\rm FT}^2)\times \epsilon^{K^0_{S}\pi^+\pi^- {\it vs. }K^0_{S}\pi^+\pi^-}$ and $h^{\prime}_{\rm corr}=N_{D\bar{D}}/(S_{\rm FT}S^{\prime}_{\rm FT})\times \epsilon^{K^0_{S}\pi^+\pi^- {\it vs. }K^0_{L}\pi^+\pi^-}$.
Here $S_{CP}$ is the yield of events in which one charm meson is reconstructed as the $CP$-tag where no requirement is placed on the decay of the other charm meson, and $S_{\rm FT^{(\prime)}}$ refers to the analogous quantity summed over flavor-tagged decays that are used in the determination of $K_i^{(\prime)}$. The effective efficiency for detecting the $D\rightarrow K^0_{S(L)}\pi^+\pi^-$ decay recoiling against the particular $CP$-tag under consideration, is defined as $\epsilon^{K^0_{S(L)}\pi^+\pi^-}=\epsilon_{\rm DT}/\epsilon_{\rm ST}$, where $\epsilon_{\rm ST}$ is the detection efficiency for finding the $CP$-tagged candidate,
while $\epsilon_{\rm DT}$ is the efficiency for simultaneously finding the $CP$-tagged candidate and the signal decay
$D\rightarrow K^0_{S(L)}\pi^+\pi^-$. Furthermore, $\epsilon^{K^0_{S}\pi^+\pi^- {\it vs. }K^0_{S}\pi^+\pi^-}$ and $\epsilon^{K^0_{S}\pi^+\pi^- {\it vs. } K^0_{L}\pi^+\pi^-}$ are efficiencies for detecting $D\rightarrow K^0_{S}\pi^+\pi^-$ {\it vs. }$D\rightarrow K^0_{S}\pi^+\pi^-$ and $D\rightarrow K^0_{L}\pi^+\pi^-$ {\it vs. }$D\rightarrow K^0_{S}\pi^+\pi^-$, respectively. Note that, as is discussed in Sec.~\ref{sec:kikip}, finite detector resolution results in the migration of reconstructed events between Dalitz plot bins.  In order to avoid biases arising from these migration effects it is necessary to modify  Eqs.~(\ref{eq:mi}), (\ref{eq:mij}), (\ref{eq:mip}) and (\ref{eq:mijp}) by substituting the efficiencies in the normalization factors $h_{CP}^{(\prime)}$ and $h_{\rm corr}^{(\prime)}$ by efficiency matrices, as described in
Sec.~\ref{sec:expdt}.

\section{ The BESIII detector}
\label{sec:detector}
BEPCII is a double-ring $e^+e^-$ collider with a center-of-mass energy
ranging from 2 to 5~GeV and a design luminosity of
$10^{33}$~cm$^{-2}$s$^{-1}$ at a beam energy of 1.89~GeV.  The BESIII
detector at BEPCII is a cylindrical detector with a solid-angle
coverage of 93\% of $4\pi$. The detector consists of a helium-gas
based main drift chamber (MDC), a plastic scintillator time-of-flight
(TOF) system, a CsI(Tl) electromagnetic calorimeter (EMC), a
superconducting solenoid providing a 1.0\,T magnetic field and a muon
counter. The charged-particle momentum resolution is 0.5\% at a
transverse momentum of 1\,$\gevc$, and the specific energy loss
($dE/dx$) resolution is 6\% for the electrons from Bhabha scattering.
The photon energy resolution in the EMC is 2.5\% in the barrel and
5.0\% in the end-caps at energies of 1\,GeV.  The time resolution of
the TOF barrel part is 68 ps, while that of the end-cap part is 110
ps.  More details about the design and performance of the detector are
given in Ref.~\cite{Ablikim:2009aa}.

A {\sc geant4}-based~\cite{geant4} simulation package, which includes
the geometric description of the detector and the detector response,
is used to determine signal detection efficiencies and to estimate
potential backgrounds. The production of the $\psi(3770)$,
initial-state radiation (ISR) production of the $\psi(2S)$ and
$J/\psi$, and the continuum processes $e^+e^-\rightarrow \tau^+\tau^-$
and $e^+e^-\rightarrow q\bar{q}$ ($q=u$,~$d$ and $s$) are simulated
with the event generator {\sc kkmc}~\cite{kkmc}, with the inclusion of
ISR effects up to second-order corrections~\cite{SJNP41_466}. The
final-state radiation effects are simulated via the {\sc photos}
package~\cite{plb303_163}. The known decay modes are generated by {\sc
  evtgen}~\cite{nima462_152} with the branching fractions (BFs) set to
the world average values from the Particle Data Group~\cite{pdg18},
while the remaining unknown decay modes are modeled by {\sc
  lundcharm}~\cite{lundcharm}. The generation of simulated signals
$D^0\rightarrow K_S^0\pi^+\pi^-$ and $D^0\rightarrow K_L^0\pi^+\pi^-$
is based on the knowledge of isobar resonance amplitudes from the
Dalitz plot analysis of $D^0\rightarrow K_S^0\pi^+\pi^-$.
The $D^0\rightarrow \pi^+\pi^-\pi^0\pi^0$ decay is simulated with a phase-space model since the relative contributions of intermediate resonances in the decay are poorly known.
For other multibody decay modes the simulated data are based on amplitude
models, where available, or through an estimate of the expected
intermediate resonances participating in the decay.

\section{\bf Event selection}
\label{sec:evtsel}

In order to measure $c_i$, $s_i$, $c_i^{\prime}$ and $s_i^{\prime}$, a range of single-tag (ST) and double-tag (DT) samples of $D$ decays are reconstructed.
The ST samples are those where the decay products of only one $D$ meson are reconstructed. The DT samples are those where one $D$ meson decays to the signal mode $K^0_S\pi^+\pi^-$ or $K^0_L\pi^+\pi^-$ and the other $D$ meson decays to one of the tag modes listed in Table~\ref{tab:tagmode}. Tag decay modes fall into the categories of flavor, $CP$ eigenstates or mixed-$CP$. Flavor tags identify the flavor of the decaying meson through a semi-leptonic decay or a Cabibbo-favored hadronic decay (contamination from doubly-Cabibbo-suppressed (DCS) decays is discussed later). $CP$ eigenstates and mixed-$CP$ tags identify a decay from an initial state which is a superposition of $D^0$ and $\bar{D}^0$. The $D\rightarrow \pi^+\pi^-\pi^0$ tag is used for the first time to measure the strong-phase parameters in $D \to K^0_{S,L}\pi^+\pi^-$ decays. It has a relatively high BF and selection efficiency resulting in a large increase to the $CP$-tagged yields.
The use of this tag is possible through the knowledge of $F_{CP}$ for this decay~\cite{plb747_9}. In this paper the $D\rightarrow \pi^+\pi^-\pi^0$ is referred to as a $CP$-even eigenstate, although its small $CP$-odd component is always taken into account, as in Eq.~(\ref{eq:mi}).

Due to the hermetic nature of the detector it is possible to use missing energy and momentum constraints to infer the presence of the neutrino in the $K^+ e^-\bar{\nu}_{e}$ final state that does not leave a response in the detector. Similarly, the $K^0_L$ meson, which does not decay within the detector, can be inferred by requiring the missing energy and momentum to be consistent with a $K^0_L$ particle. Tag decay modes such as $D\rightarrow K^0_L\omega$ are not included in the analysis as the systematic uncertainty due to the need to estimate their BFs would be larger than the impact on statistical precision brought from the increased $CP$-tag yields. The principles of missing energy and momentum can also be used to increase the selection efficiency in highly sensitive decay modes by only partially reconstructing the $D\rightarrow K_S^0\pi^+\pi^-$ candidate. The DT combinations that result in two missing particles are not pursued due to the inability to reliably allocate the missing energy and momentum between two missing particles. The ST yields are only measured in decay modes that are fully reconstructable.
\begin{table}
\caption{A list of tag decay modes used in the analysis.}
\begin{center}
\begin{tabular}
{lc} \hline Tag group &  \\
\hline
Flavor          & $K^{+}\pi^{-}$, $K^{+}\pi^{-}\pi^0$, $K^{+}\pi^{-}\pi^{-}\pi^{+}$, $K^{+}e^{-}\bar{\nu}_e$ \\
$CP$-even       & $K^{+}K^-$, $\pi^+\pi^-$, $K^0_S\pi^0\pi^0$, $K^0_L\pi^0$, $\pi^+\pi^-\pi^0$ \\
$CP$-odd        & $K^0_S\pi^0$, $K^0_S\eta$, $K^0_S\omega$, $K^0_S\eta^{\prime}$, $K^0_L\pi^0\pi^0$ \\
Mixed-$CP$     & $K^0_S\pi^+\pi^-$ \\
\hline
\end{tabular}
\label{tab:tagmode}
\end{center}
\end{table}

In this paper, we use the following selection criteria to reconstruct the ST and DT samples.
The charged tracks are required to be well reconstructed in the MDC detector with the polar angle $\theta$ satisfying $|\cos\theta|<0.93$.
Their distances of the closest approach to the interaction point (IP) are required to be less than 10 cm along the beam direction and less than 1 cm in the perpendicular plane.
For tracks originating from $K^0_S$, their distances of closest approach to the IP are required to be within 20 cm along the beam direction.

To discriminate pions from kaons, the $dE/dx$ and TOF information are
used to obtain particle identification (PID) likelihoods for the pion ($\mathcal{L}_{\pi}$) and
kaon ($\mathcal{L}_K$) hypotheses. Pion and kaon candidates are
selected using $\mathcal{L}_{\pi} > \mathcal{L}_{K}$ and
$\mathcal{L}_{K} > \mathcal{L}_{\pi}$, respectively.
To identify the electron, the information measured by the $dE/dx$, TOF, and EMC are used to construct likelihoods for electron, pion and kaon hypotheses ($\mathcal{L}^{\prime}_e$,
$\mathcal{L}^{\prime}_\pi$ and $\mathcal{L}^{\prime}_K$).  The electron candidate must satisfy $\mathcal{L}^{\prime}_{e} > 0.001$ and
$\mathcal{L}^{\prime}_e/(\mathcal{L}^{\prime}_e+\mathcal{L}^{\prime}_{\pi}+\mathcal{L}^{\prime}_K)>0.8$.
$K^0_S$ mesons are reconstructed from two oppositely charged tracks with an invariant mass within $(0.485,~0.510)$~GeV$/c^2$.
A fit is applied to constrain these two charged tracks to a common vertex, and the decay vertex is required
to be separated from the interaction point by more than twice the standard deviation ($\sigma$) of the measured flight distance ($L$), {\it i.e.}, $L/\sigma_L>2$, in order to suppress the background from pion pairs that do not originate from a $K^0_S$ meson.

Photon candidates are reconstructed from isolated clusters in the
EMC in the regions $|\cos\theta| \le 0.80$ (barrel) and $0.86 \le
|\cos\theta|\le 0.92$ (end cap). The deposited energy of a neutral cluster is required to be larger than 25 (50) MeV in
barrel (end cap) region.
To suppress electronic noise and energy deposits unrelated to the
event, the difference between the EMC time and the event start time
is required to be within (0, 700)~ns. To reconstruct $\pi^0(\eta)$
candidates, the invariant mass of the accepted photon pair is
required to be within $(0.110,~0.155)[(0.48,0.58)]$~GeV$/c^2$. To improve the momentum resolution, a kinematic fit is
applied to constrain the $\gamma\gamma$ invariant mass to the
nominal $\pi^0(\eta)$ mass~\cite{pdg18}, and the $\chi^2$ of the
kinematic fit is required to be less than 20. The fitted momenta of
the $\pi^0(\eta)$ are used in the further analysis.
When reconstructing $\eta$ candidates decaying through $\eta\rightarrow \pi^+\pi^-\pi^0$, it is required that their invariant masses be within $(0.530,~0.655)$~GeV$/c^2$.
Similarly, $\omega$ candidates are selected by requiring the invariant mass of $\pi^+\pi^-\pi^0$ to be within $(0.750,~0.820)$~GeV/$c^2$.
The decay modes $\eta^{\prime}\rightarrow \pi^+\pi^-\eta$ and $\eta^{\prime}\rightarrow \gamma\pi^+\pi^-$ are used to reconstruct
$\eta^{\prime}$ mesons, with the invariant masses of the $\pi^+\pi^-\eta$ and $\gamma\pi^+\pi^-$
required to be within (0.942,~0.973) and $(0.935,~0.973)$~GeV$/c^2$, respectively.

\subsection{Single-tag yields}
\label{sec:stag}

The ST $D$ signals are identified using the beam-constrained mass,
\begin{equation}
{\rm M}_{\rm BC}=\sqrt{(\sqrt{s}/2)^2-|\overrightarrow{p}_{D_{\rm tag}}|^2},
\end{equation}
where $\overrightarrow{p}_{D_{\rm tag}}$ is the momentum of the $D$
candidate. To improve the signal purity, the energy difference $\Delta
E=\sqrt{s}/2-E_{D_{\rm tag}}$ for each candidate is required to be
within approximately $\pm3\sigma_{\Delta E}$ around the $\Delta E$
peak, where $\sigma_{\Delta E}$ is the $\Delta E$ resolution and
$E_{D_{\rm tag}}$ is the reconstructed ST $D$ energy. The explicit
$\Delta E$ requirements for all reconstructed ST modes are listed in
the second column of Table~\ref{tab:numST}. If multiple combinations are selected, the
one with the minimum $|\Delta E|$ is retained. For the ST channels of
$K^+\pi^-$, $K^+K^-$ and $\pi^+\pi^-$, backgrounds of cosmic rays and
Bhabha events are removed with the following requirements.  First, the
two charged tracks must have a TOF time difference of less than 5~ns
and they must not be consistent with being a muon pair or an $e^+e^-$
pair.  Second, there must be at least one EMC shower with an energy
larger than 50 MeV or at least one additional charged track detected
in the MDC.

\begin{figure*}[tp!]
\begin{center}
\includegraphics[width=\linewidth]{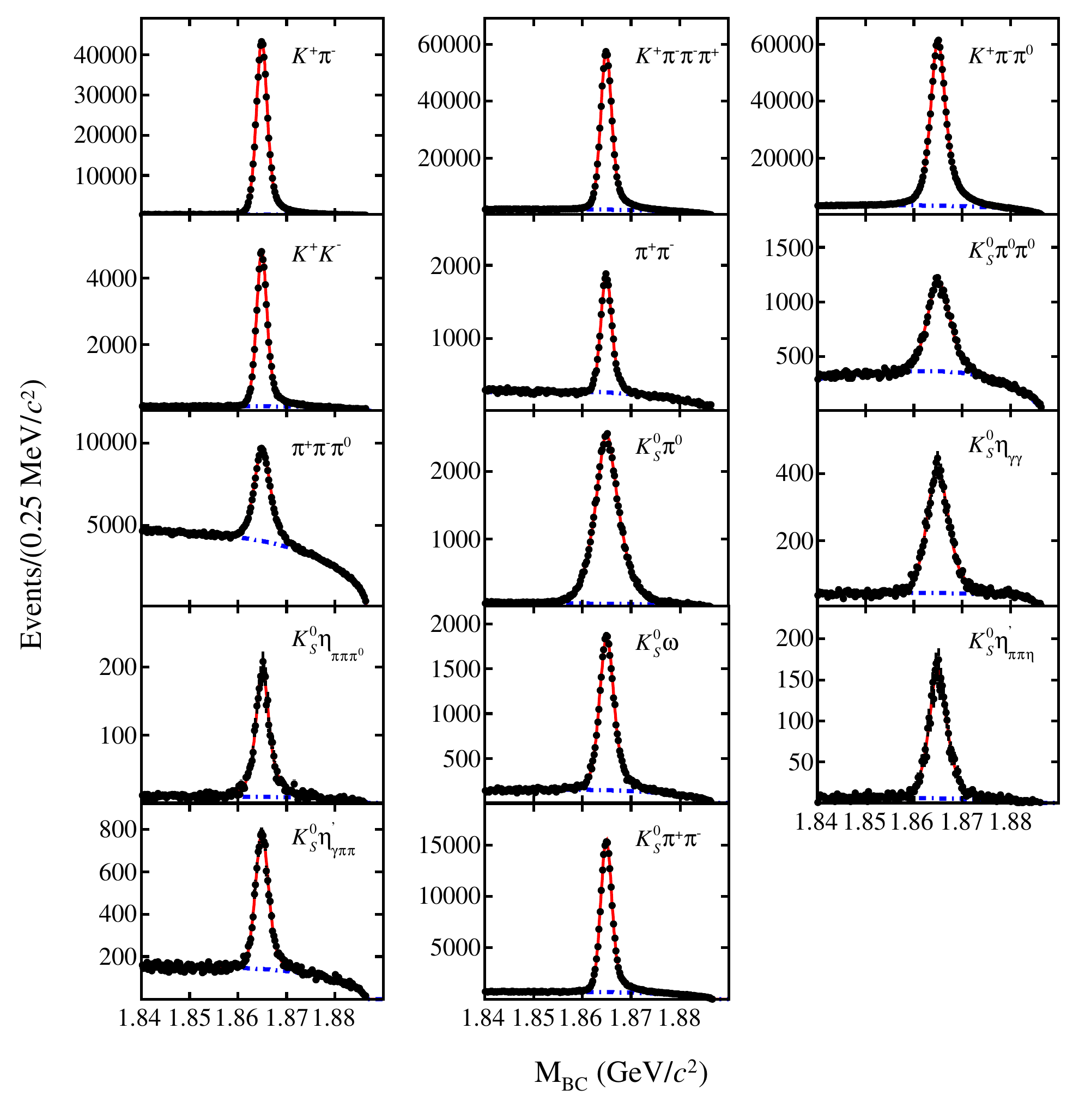}
\caption{Fits to ${\rm M}_{\rm BC}$ distributions for the candidates for the ST decay modes as denoted by the labels on each plot. The black points represent data. Overlaid is the fit to data which is indicated by the continuous red line. The blue dashed line indicates the combinatorial background component of the fit.}
\label{fig:stagmd0}
\end{center}
\end{figure*}

The ${\rm M}_{\rm BC}$ distributions for the ST modes are
shown in Fig.~\ref{fig:stagmd0}.  To obtain the ST yields
reconstructed by these modes, maximum likelihood fits are performed to
these spectra, where the signal peak is described by a Monte Carlo
(MC) simulated shape convolved with a double-Gaussian function, and
the combinatorial background is modeled with an ARGUS
function~\cite{plb241_278}.  In addition to the combinatorial
background, there are also some peaking backgrounds in the signal
region of ${\rm M}_{\rm BC}$.  These peaking backgrounds are included
in the yields obtained from fits to ${\rm M}_{\rm BC}$ spectra and
hence must be subtracted.  For example, for the ST modes of
$K^{+}\pi^{-}$, $K^{+}\pi^{-}\pi^0$ and $K^{+}\pi^{-}\pi^{-}\pi^{+}$,
there are small contributions of wrong-sign (WS) peaking backgrounds
in the ST $\bar{D}^0$ samples, which originate from the
DCS-dominated decays of $D^0\rightarrow K^+\pi^-$, $K^+\pi^-\pi^0$ and
$K^+\pi^-\pi^-\pi^+$.  In addition, the $D^0\rightarrow
K^0_SK^{+}\pi^{-}$ ($K^0_S\rightarrow \pi^+\pi^-$) decay is a source
of WS peaking background for the ST decay $\bar{D}^0\rightarrow
K^+\pi^-\pi^-\pi^+$.  Overall, the peaking background contamination
rates are less than 1\% for the ST modes of $K^{+}\pi^{-}$,
$K^{+}\pi^{-}\pi^0$ and $K^{+}\pi^{-}\pi^{-}\pi^{+}$.  For the
$CP$-eigenstate ST channels $K^0_S\pi^0(\pi^0)$ and $\pi^+\pi^-\pi^0$,
the peaking background rates are 0.8\%(3.9\%) and 3.9\%, dominated by
the $D$ meson decays to $\pi^+\pi^-\pi^0(\pi^0)$ and $K^0_S\pi^0$,
respectively.  The $D\rightarrow K^0_S\pi^+\pi^-\pi^0$ decay forms the
dominant peaking backgrounds and accounts for contamination rates of
13.7\%, 6.3\% and 3.8\% in the fitted ST yields for $K^0_S\omega$,
$K^0_S\eta_{\pi^+\pi^-\pi^0}$ and
$K^0_S\eta^{\prime}_{\gamma\pi^+\pi^-}$, respectively.  Additionally,
the sample of ST $K^0_S\pi^+\pi^-$ decays includes a 2\% contamination
from the peaking-background $D\rightarrow \pi^+\pi^-\pi^+\pi^-$.  The
sizes of these peaking backgrounds are all estimated from MC
simulation and then subtracted from the fitted ST yields.  The
background-subtracted yield and the efficiency for each of the ST
modes are summarized in the third and fourth columns of
Table~\ref{tab:numST}, respectively.  The ST efficiencies are
determined from the simulated data where one $D$ meson is forced to
decay to the reconstructed final states and the other $D$ meson is
allowed to decay to any final state.  The values of $\epsilon_{\rm
  ST}$ vary from $\sim$65$\%$ for decay modes with two charged
particles in the final state to $\sim$13$\%$ for final states with
multiple composite and neutral particles such as
$K^0_S\eta^{\prime}_{\pi^+\pi^-\eta}$.

The ST yields of the modes $K^+e^-\bar{\nu}_e$, $K^0_L\pi^0$ and $K^0_L\pi^0\pi^0$, which cannot be directly reconstructed, are estimated from knowledge of the number of neutral $D$ meson pairs $N_{D\bar{D}}$,
the estimated ST efficiencies $\epsilon^{\rm ST}_{\rm tag}$, and their BFs $\mathcal{B}_{\rm tag}$ reported in Ref.~\cite{pdg18}, where the $D\rightarrow K_S^0\pi^0\pi^0$ BF is used as a proxy for $D\rightarrow K_L^0\pi^0\pi^0$.
The yields are calculated from the relations
$$N^{\rm ST}_{\rm tag}=2N_{D\bar{D}}\times \mathcal{B}_{\rm tag}\times \epsilon^{\rm ST}_{\rm tag}$$
where $N_{D\bar{D}}=(10597\pm28\pm98)\times10^{3}$~\cite{cpc42_083001}.
The ST efficiencies, $\epsilon^{\rm ST}_{\rm tag}$, of detecting these three decays are estimated by evaluating the ratios between the corresponding DT (discussed later in Sec.~\ref{sec:DPs}) and ST efficiencies, which are determined to be
61.35\%, 48.97\% and 26.20\% for $D\rightarrow K^+e^-\bar{\nu}_e$, $D\rightarrow K^0_L\pi^0$ and $D\rightarrow K^0_L\pi^0\pi^0$, respectively.
The ST yields of $D\rightarrow K^-e^+\nu_e$, $D\rightarrow K^0_L\pi^0$ and $D\rightarrow K^0_L\pi^0\pi^0$ are also included in Table~\ref{tab:numST},
in which the uncertainties from the BFs, $N_{D\bar{D}}$ and the detection efficiencies are presented.

\begin{table*}[tp!]
\caption{ Summary of $\Delta E$ requirements, ST yields ($N_{\rm ST}$) and ST efficiencies ($\epsilon_{\rm ST}$) for various tags, as well as DT yields ($N_{\rm DT}$) and DT efficiencies ($\epsilon_{\rm DT}$) for $K^0_{S,L}\pi^+\pi^-$ {\it vs.} various tags; where the $K^0_S$ decay BF is not included in $\epsilon_{\rm DT}^{K^0_S\pi^+\pi^-}$. The listed uncertainties are statistical only. }
\begin{center}
\begin{tabular}
{l|ccr|cccr} \hline\hline Mode~~~  & \multicolumn{3}{c|}{ST }  & \multicolumn{4}{c}{DT  } \\
 & ~~$\Delta E$ (GeV) ~~ & ~~~$N_{\rm ST}$ & ~~~~~~$\epsilon_{\rm ST}$ (\%)~~~~  &  $N_{\rm DT}^{K^0_S\pi^+\pi^-}$  & ~~~~$\epsilon_{\rm DT}^{K^0_S\pi^+\pi^-}$ (\%) & ~~~~$N_{\rm DT}^{K^0_L\pi^+\pi^-}$ & ~~~~$\epsilon_{\rm DT}^{K^0_L\pi^+\pi^-}$ (\%)  \\
\hline
$K^+\pi^-$                                                 &  [$-$0.025, 0.028]  &$549373\pm756$         & $67.28\pm0.03$  &   $4740\pm71$        & $27.28\pm0.07$    & ~~$9511\pm115$     & $35.48\pm0.05$     \\
$K^+\pi^-\pi^0$                                         &  [$-$0.044, 0.066]  &$1076436\pm1406$     & $35.12\pm0.02$  &   $5695\pm78$        & $14.45\pm0.05$    & $11906\pm132$        &  $18.21\pm0.04$     \\
$K^+\pi^-\pi^-\pi^+$                                 &  [$-$0.020, 0.023]  &~~$712034\pm1705$  & $39.20\pm0.02$  &   $8899\pm95$        & $13.75\pm0.05$    & $19225\pm176$        &  $18.40\pm0.04$     \\
$K^+e^-\nu_e$                                           &                              &~~$458989\pm5724$  & $61.35\pm0.02$  &   $4123\pm75$        & $26.11\pm0.07$   &                                 &       \\
$CP$-even tags                                                &                              &                            &                                    &                            &                             &                                 &                             \\
$K^+K^-$                                                   &  [$-$0.020, 0.021]  &~~$57050\pm231$      & $63.90\pm0.05$  &~~$443\pm22$       & $25.97\pm0.07$     &$1289\pm41$            & $33.60\pm0.07$     \\
$\pi^+\pi^-$                                               &  [$-$0.027, 0.030]  &~~$20498\pm263$      & $68.44\pm0.08$  &~~$184\pm14$       & $27.27\pm0.07$     &~~$531\pm28$         & $35.60\pm0.08$      \\
$K^0_S\pi^0\pi^0$                                     &  [$-$0.044, 0.066]  &~~$22865\pm438$       & $15.81\pm0.04$  &~~$198\pm16$      &~~$6.47\pm0.03$   &~~$612\pm35$          & $ 8.57\pm0.03$     \\
$\pi^+\pi^-\pi^0$                                       &  [$-$0.051, 0.063]  &$107293\pm716$         & $37.26\pm0.04$   &~~$790\pm31$      & $14.28\pm0.06$     & $2571\pm74$           & $20.29\pm0.06$     \\
$K^0_L\pi^0$                                              &                             &~$103787\pm7337$    &  $48.97\pm0.11$  &~~$913\pm41$       & $20.84\pm0.04$     &                                &                    \\
$CP$-odd tags                                                  &                            &                            &                                    &                            &                               &                                 &                             \\
$K^0_S\pi^0$                                              &  [$-$0.040, 0.070]  &~~$66116\pm324$       & $35.98\pm0.04$  &~~$643\pm26$     & $14.84\pm0.05$      &~~$861\pm46$         & $18.76\pm0.06$     \\
$K^0_S\eta_{\gamma\gamma}$                    &  [$-$0.035, 0.038]  &~~~~$9260\pm119$    & $30.70\pm0.11$   &~~~$89\pm10$     & $12.86\pm0.05$      &~~$105\pm15$         & $16.78\pm0.06$     \\
$K^0_S\eta_{\pi^+\pi^-\pi^0}$                    &  [$-$0.027, 0.032]  &~~$2878\pm81$          & $16.61\pm0.13$   &~~$23\pm5$         & ~~$6.98\pm0.03$    &~~$40\pm 9$           & $ 8.88\pm0.03$     \\
$K^0_S\omega$                                            &  [$-$0.030, 0.039]  &~~$24978\pm448$      & $16.79\pm0.05$  &~~$245\pm17$       & ~~$6.30\pm0.03$    &~~$321\pm25$        & $ 8.14\pm0.03$     \\
$K^0_S\eta^{\prime}_{\pi^+\pi^-\eta}$        &  [$-$0.028, 0.031]  &~~$3208\pm88$          & $13.17\pm0.09$   &~~$24\pm6$          & ~~$5.06\pm0.02$    &~~$ 38\pm 8$         & $ 6.86\pm0.03$     \\
$K^0_S\eta^{\prime}_{\gamma\pi^+\pi^-}$  &  [$-$0.026, 0.034]  &~~~$9301\pm139$     & $23.80\pm0.10$   &~~~~$81\pm10$          & ~~$9.87\pm0.03$   &~~$120\pm14$        & $12.43\pm0.04$     \\
$K^0_L\pi^0\pi^{0}$                                   &                              &~~~$50531\pm6128$ &  $26.20\pm0.07$ &~~$620\pm32$       & $11.15\pm0.03$ &                               &                    \\
Mixed $CP$ tags                                                  &                            &                           &                                    &                           &                               &                                &                             \\
$K^0_S\pi^+\pi^-$                                      &  [$-$0.022, 0.024]  &$188912\pm756$        & $42.56\pm0.03$   &~~$899\pm31$      & $18.53\pm0.06$      & $3438\pm72$           & $21.61\pm0.05$     \\
$K^0_S\pi^+\pi^-_{\rm miss}$                    &                              &                           &                                     &~~$224\pm17$   & ~~$5.03\pm0.02$   &                  &                    \\
$K^0_S(\pi^0\pi^0_{\rm miss})\pi^+\pi^-$  &                              &                          &                                      &~~$710\pm34$   & $18.30\pm0.04$  &                  &                    \\
\hline\hline
\end{tabular}
\label{tab:numST}
\end{center}
\end{table*}

\subsection{\boldmath Double tags with $K^0_S\pi^+\pi^-$ }
\label{sec:DTKspipi}

In those cases where the decay products of the tag mode are fully
reconstructed and the signal mode is $D\rightarrow K^0_S\pi^+\pi^-$,
the signal decay is built by using the other tracks in the event
recoiling against the ST $D$ meson. The same selection on track
parameters and the $K^0_S$ candidate is imposed as described for the
$D\rightarrow K^0_S\pi^+\pi^-$ ST case. The energy difference, $\Delta
E^{\prime}=\sqrt{s}/2-E_{\rm sig}$, where $E_{\rm sig}$ is the energy
of the $D\rightarrow K^0_S\pi^+\pi^-$ candidate, is required to be
between $-30$ and $33$ MeV. If multiple combinations are selected, the
one with the minimum $|\Delta E^{\prime}|$ is retained. The
beam-constrained mass is defined as ${\rm M}^{\rm sig}_{\rm
  BC}=\sqrt{(\sqrt{s}/2)^2-|\vec{p}_{\rm sig}|^2}$, where
$\vec{p}_{\rm sig}$ is the momentum of the signal-decay candidate.

\begin{figure}[tp!]
\begin{center}
\includegraphics[width=0.9\linewidth]{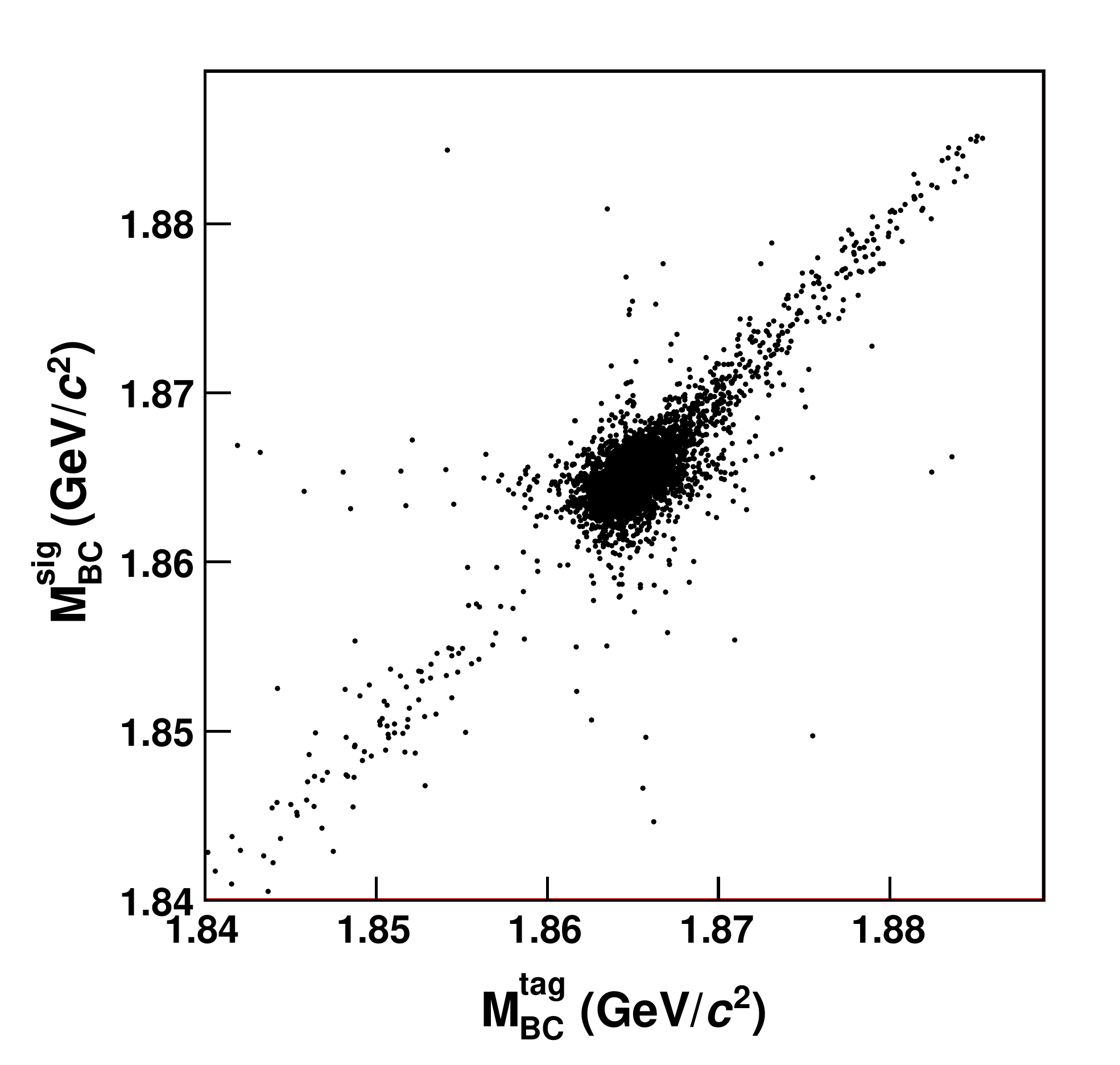}
\caption{The two-dimensional ${\rm M}_{\rm BC}$ distribution. The signal is visible at the center. The concentration of events along the diagonal is from  mis-reconstructed $D\overline{D}$ decays and from $q\overline{q}$ events.}
\label{fig:mbcmbc}
\end{center}
\end{figure}

\begin{figure*}[tp!]
\begin{center}
\includegraphics[width=\linewidth]{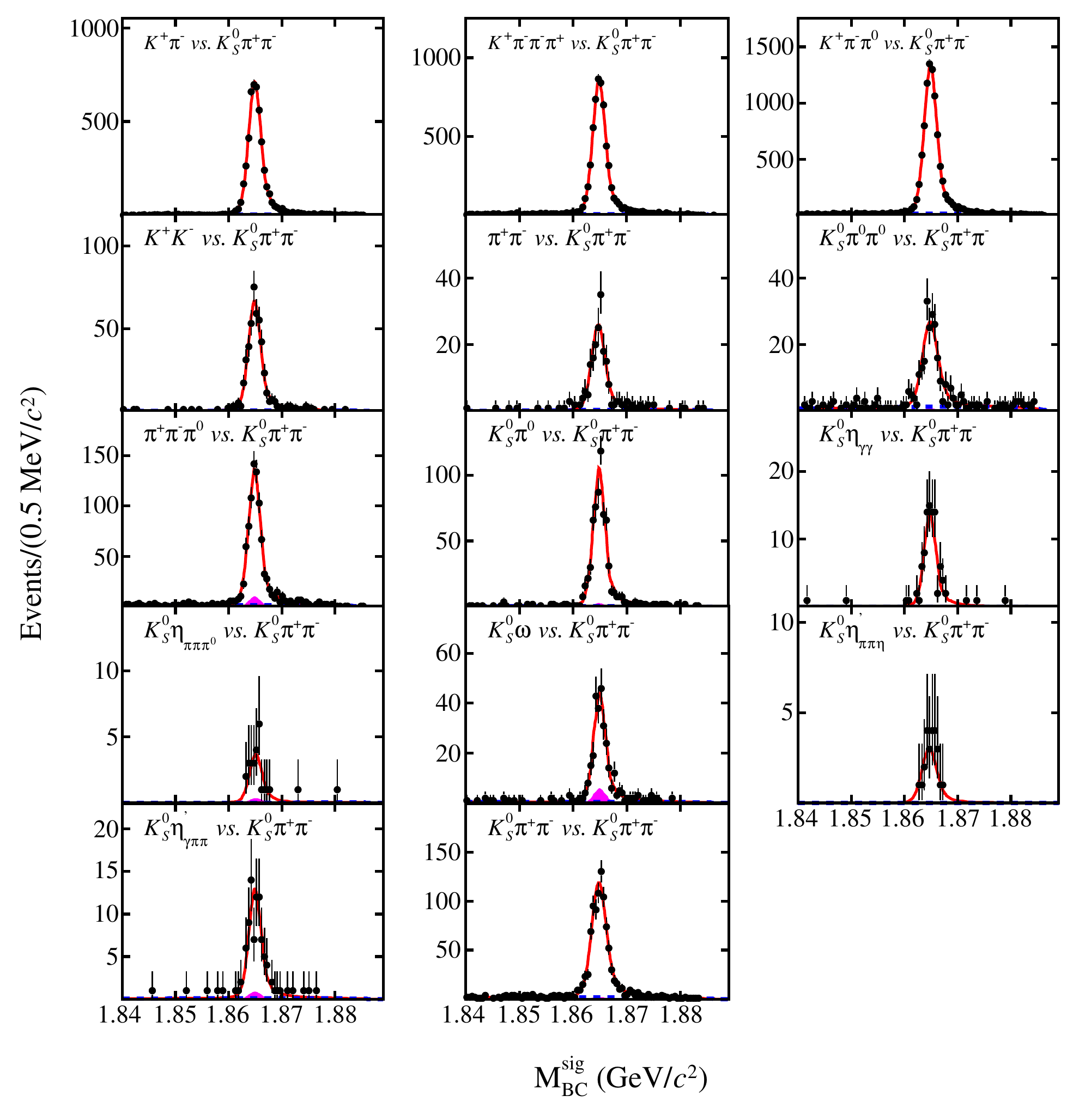}
\caption{The projections of the two-dimensional fits of
  $D^0\rightarrow K^0_S\pi^+\pi^-$ {\it vs. } various ST on the ${\rm
    M}_{\rm BC}^{\rm sig}$ distribution. The black points represent
  the data. Overlaid is the fit projection in the continuous red
  line. The blue dashed line indicates the combinatorial component and
  the peaking background contribution is shown by the shaded areas (pink)}.
\label{fig:dtksp}
\end{center}
\end{figure*}

The DT yield is determined by performing a two-dimensional unbinned maximum-likelihood fit to the ${\rm M}^{\rm sig}_{\rm BC}$ (signal) {\it vs.} ${\rm M}^{\rm tag}_{\rm BC}$ (tag) distribution. An example distribution for the tag mode $D \to K^+\pi^-$ is shown in Fig.~\ref{fig:mbcmbc}. The signal shape of the ${\rm M}^{\rm sig}_{\rm BC}$ {\it vs.} ${\rm M}^{\rm tag}_{\rm BC}$ distributions is modeled with a two-dimensional shape derived from simulated data convolved with two independent Gaussian functions representing the resolution differences between data and simulation. The parameters of the Gaussian functions are fixed at the values obtained from the one-dimensional fits of the ${\rm M}^{\rm sig}_{\rm BC}$ and ${\rm M}^{\rm tag}_{\rm BC}$ distributions in data, respectively. The combinatorial backgrounds in the ${\rm M}^{\rm sig}_{\rm BC}$ and ${\rm M}^{\rm tag}_{\rm BC}$ distributions are modeled by an ARGUS function in each dimension where the parameters are determined in the fit. The events that are observed along the diagonal arise from mis-reconstructed $D\overline{D}$ decays and from $q\overline{q}$ events. They are described with a product of a double-Gaussian function and an ARGUS function rotated by $45^{\circ}$~\cite{cpc42_083001}. The kinematic limit and exponent parameters of the rotated ARGUS function are fixed, while the slope parameter is determined by the fit. The peaking backgrounds in the ${\rm M}^{\rm sig}_{\rm BC}$ and ${\rm M}^{\rm tag}_{\rm BC}$ distributions are described by using a shape derived from simulation convolved with the same Gaussian function as used for the signal.
The decay $D\rightarrow \pi^+\pi^-\pi^+\pi^-$, which accounts for about 2\% peaking background to $D\rightarrow K^0_S\pi^+\pi^-$ signal, is predominantly $CP$-even~\cite{jhep01_144}, and hence the yields of this peaking background are adjusted from the expectation of simulation to account for the effects of quantum correlation. Figure~\ref{fig:dtksp} shows the projections of the two-dimensional fits on the ${\rm M}^{\rm sig}_{\rm BC}$ distribution for all the fully reconstructed ST decay modes.

The DT yield of $K^0_{S}\pi^+\pi^-$ {\it vs.} $K^0_{S}\pi^+\pi^-$ is
crucial for determining the $s_i$ values and thus it is desirable to
increase the reconstruction efficiency for these events.  Therefore
three independent selections are introduced in order to maximize the
yield of $D\rightarrow K^0_S\pi^+\pi^-$ {\it vs.} $D\rightarrow
K^0_S\pi^+\pi^-$ candidates.  The first selection requires that both
$K^0_S\pi^+\pi^-$ final states on the signal and tag side are fully
reconstructed.  However, in order to increase the efficiency, the PID
requirements on the pions originating from both the signal and tag $D$
mesons are removed and the $K^0_S$ candidate needs only satisfy
$L/\sigma_L>0$ (i.e., only candidates where $L$ is negative due to detector resolution are removed). This looser selection is applied to both $D$ mesons
and allows for an increase in yield of approximately 20$\%$ with only
a slight increase in background.

\begin{figure*}[tp!]
\begin{center}
\includegraphics[width=\linewidth]{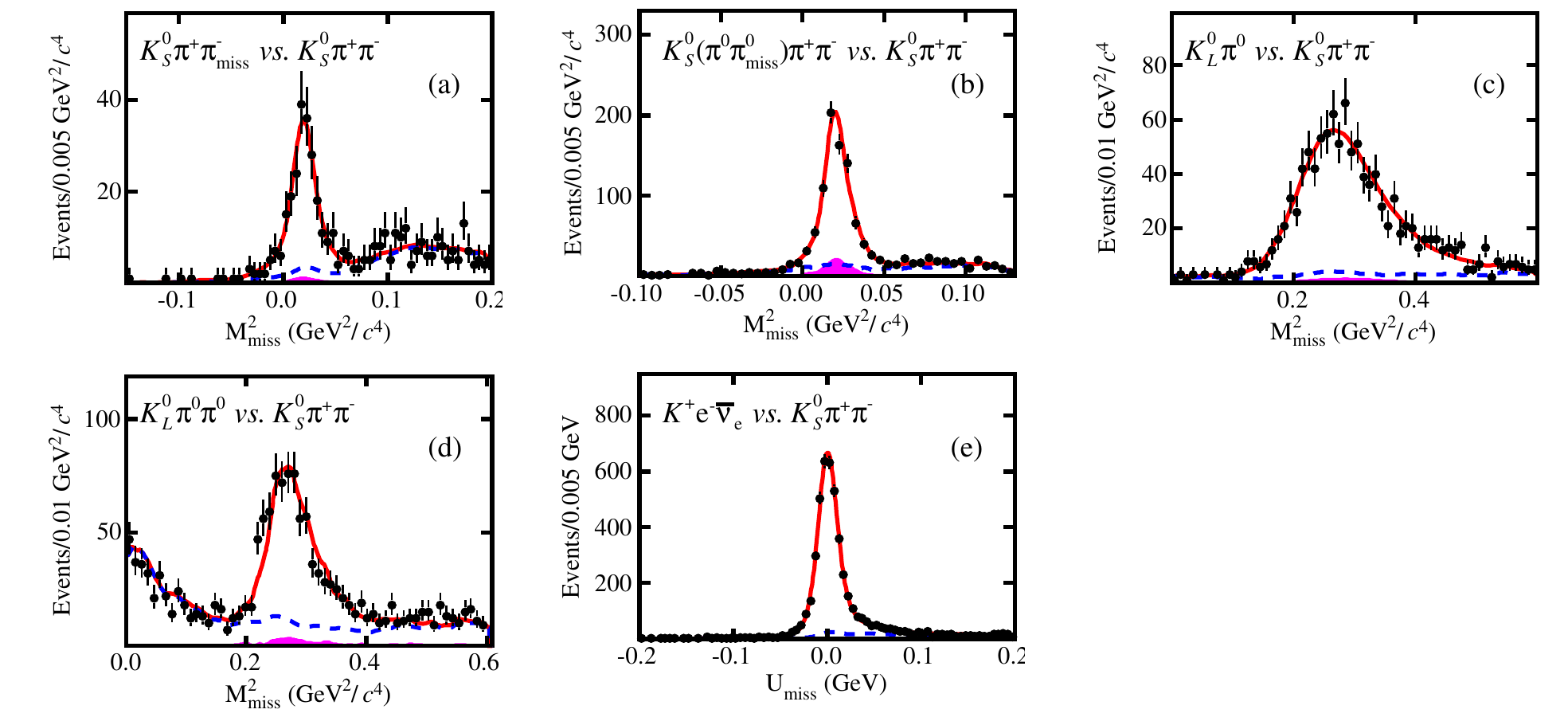}
\caption{Fits to ${\rm M}^2_{\rm miss}$ or ${\rm U}_{\rm miss}$
  distributions for the candidates of $D^0\rightarrow K^0_S\pi^+\pi^-$
  {\it vs.} various tags in data. Points with error bars represent
  data, the blue dashed curves are the fitted combinatorial
  backgrounds, the shaded areas (pink) show the MC-simulated peaking
  backgrounds, and the red solid curves show the total fits. }
\label{fig:dtmix}
\end{center}
\end{figure*}

The second selection class allows for one pion originating from the $D$ meson to be unreconstructed in the MDC, denoted as $K^0_S\pi^+\pi^-_{\rm miss}$.
Events with only three remaining charged tracks recoiling against the $D\rightarrow K^0_S\pi^+\pi^-$ ST are searched for. The $K^0_S$ and pion are identified with the same criteria used to select the ST candidates.
The missing pion is inferred by calculating the missing-mass squared (${\rm M}^2_{\rm miss}$) of the event, which is defined as
\begin{equation}
{\rm M}^2_{\rm miss}= (\sqrt{s}/2 - \sum_i E_i)^2 - |\vec{p}_{\rm sig}-\sum_i\vec{p_i}|^2,
\label{eq:missmass}
\end{equation}
where $\vec{p}_{\rm sig}$ is the momentum of the fully reconstructed $D\rightarrow K^0_S\pi^+\pi^-$ candidate and $\sum_i E_i$ and $\sum_i\vec{p_i}$ are the sum of the energy and momentum of the other reconstructed particles that form the partially reconstructed $D$ meson candidate.
Throughout this paper, in order to determine the signal yields of the DT containing a missing particle, an unbinned maximum-likelihood fit is performed to the defined kinematic distribution, {\it i.e.} ${\rm M}^2_{\rm miss}$ (or ${\rm U}_{\rm miss}$ discussed in Sec.~\ref{sec:DTKev}).
The signal and background components are described using shapes from simulated data where the signal shape is further convolved with a Gaussian function. The relative yields of the peaking backgrounds to the signals are fixed in the fits from information of the simulated data.
Figure~\ref{fig:dtmix}(a) shows the ${\rm M}^2_{\rm miss}$ distribution from the partially reconstructed $D\rightarrow K^0_S\pi^+\pi^-$ {\it vs.} $D\rightarrow K^0_S\pi^+\pi^-_{\rm miss}$ candidates.
The distribution peaks at ${\rm M}^2_{\rm miss}\sim 0.02~{\rm GeV^2}/c^4$, which is consistent with the missing particle being a $\pi^{\pm}$.
The peaking backgrounds are approximately 3$\%$ of the signal yield and are primarily from the $D\rightarrow \pi^+\pi^-\pi^+\pi^-$ decay.

The third $D\rightarrow K^0_S\pi^+\pi^-$ {\it vs.} $D\rightarrow K^0_S\pi^+\pi^-$ selection identifies those events where one $K^0_S$ meson decays to a $\pi^0\pi^0$ pair.
Events where there are only two remaining oppositely-charged tracks recoiling against the ST $D\rightarrow K^0_S\pi^+\pi^-$ are selected and these tracks are classified as the $\pi^+$ and $\pi^-$ from the $D$ meson. To avoid the reduced efficiency associated with reconstructing both $\pi^0$ mesons from the $K^0_S$, only one of the them is searched for. This type of tag is referred to as $K^0_{S}(\pi^0\pi^0_{\rm miss})\pi^+\pi^-$. The missing-mass squared of the event is defined in the same way as in Eq.~(\ref{eq:missmass}) and the summation is over the $\pi^+$, $\pi^-$, and $\pi^0$ mesons that are reconstructed on the tag side. A further variable, ${\rm M}^{\prime2}_{\rm miss}$, where the reconstructed $\pi^0$ is also not included in the summed energies and momenta of the tag-side particles is also computed. For true $D\rightarrow K^0_S\pi^+\pi^-$ decays this variable should be consistent with the square of the $K^0_S$ meson nominal mass. Therefore, candidates that do not satisfy $0.22<{\rm M}^{'2}_{\rm miss}<0.27$ GeV$^2/c^4$ are removed from the analysis in order to suppress background from $D\rightarrow \pi^+\pi^-\pi^0\pi^0$ decays. Figure~\ref{fig:dtmix}(b) shows the resultant $M^2_{\rm miss}$ distribution of the accepted candidates in data. There remains a contribution of peaking background dominated from $D \rightarrow \pi^+\pi^-\pi^0\pi^0$ decays, where the rate relative to signal is determined from simulated data to be around 15\%.

\begin{figure*}[tp!]
\begin{center}
\includegraphics[width=\linewidth]{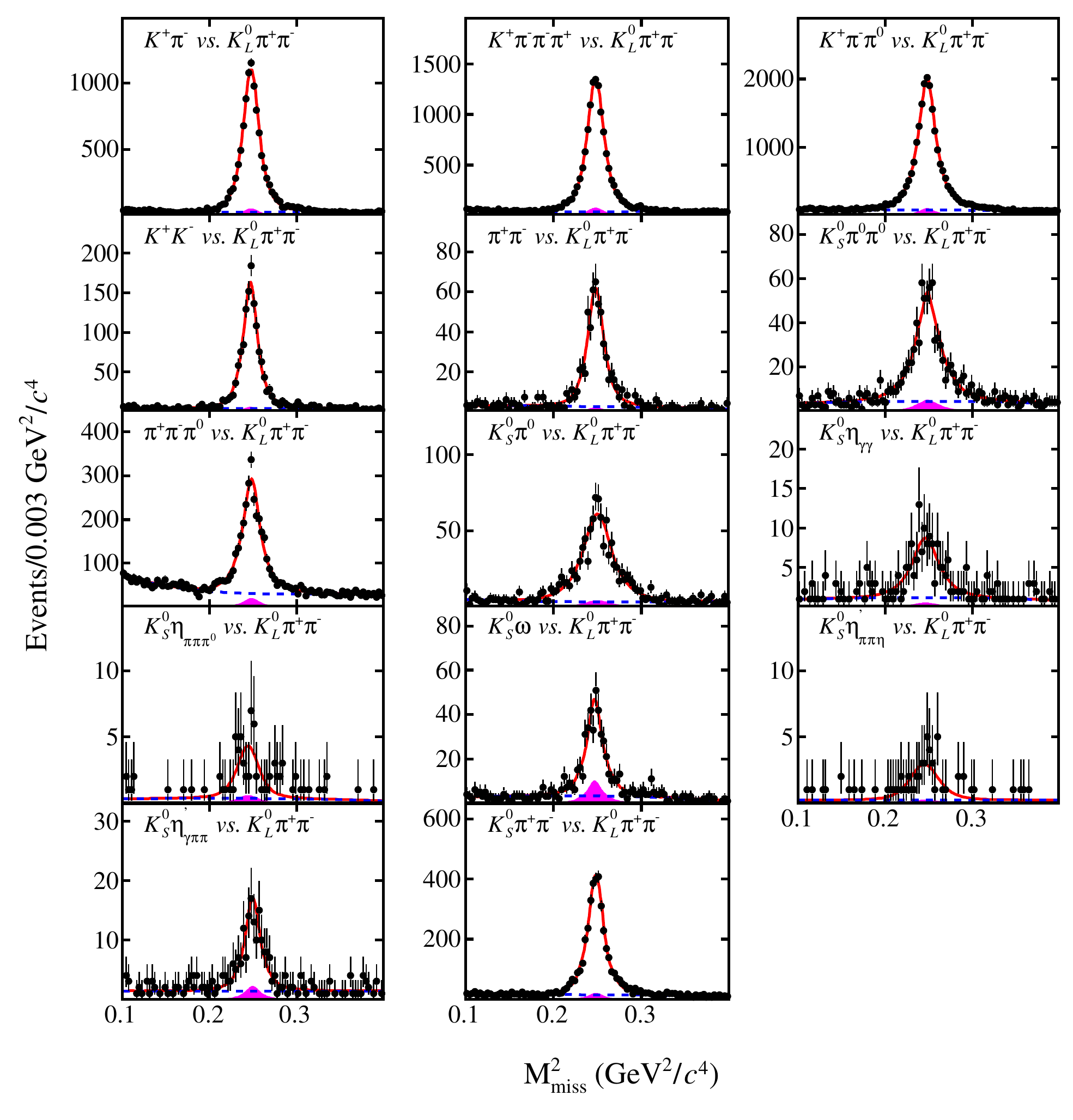}
\caption{Fits to ${\rm M}^2_{\rm miss}$ distributions for the
  candidates of $D^0\rightarrow K^0_L\pi^+\pi^-$ {\it vs.} various
  tags in data. Points with error bars are data, the blue dashed curves
  are the fitted combinatorial backgrounds, the shaded areas (pink) show
  the MC-simulated peaking backgrounds, and the red solid curves are the
  total fits.}
\label{fig:dtklp}
\end{center}
\end{figure*}

\begin{figure*}[tp!]
\begin{center}
\includegraphics[width=\linewidth]{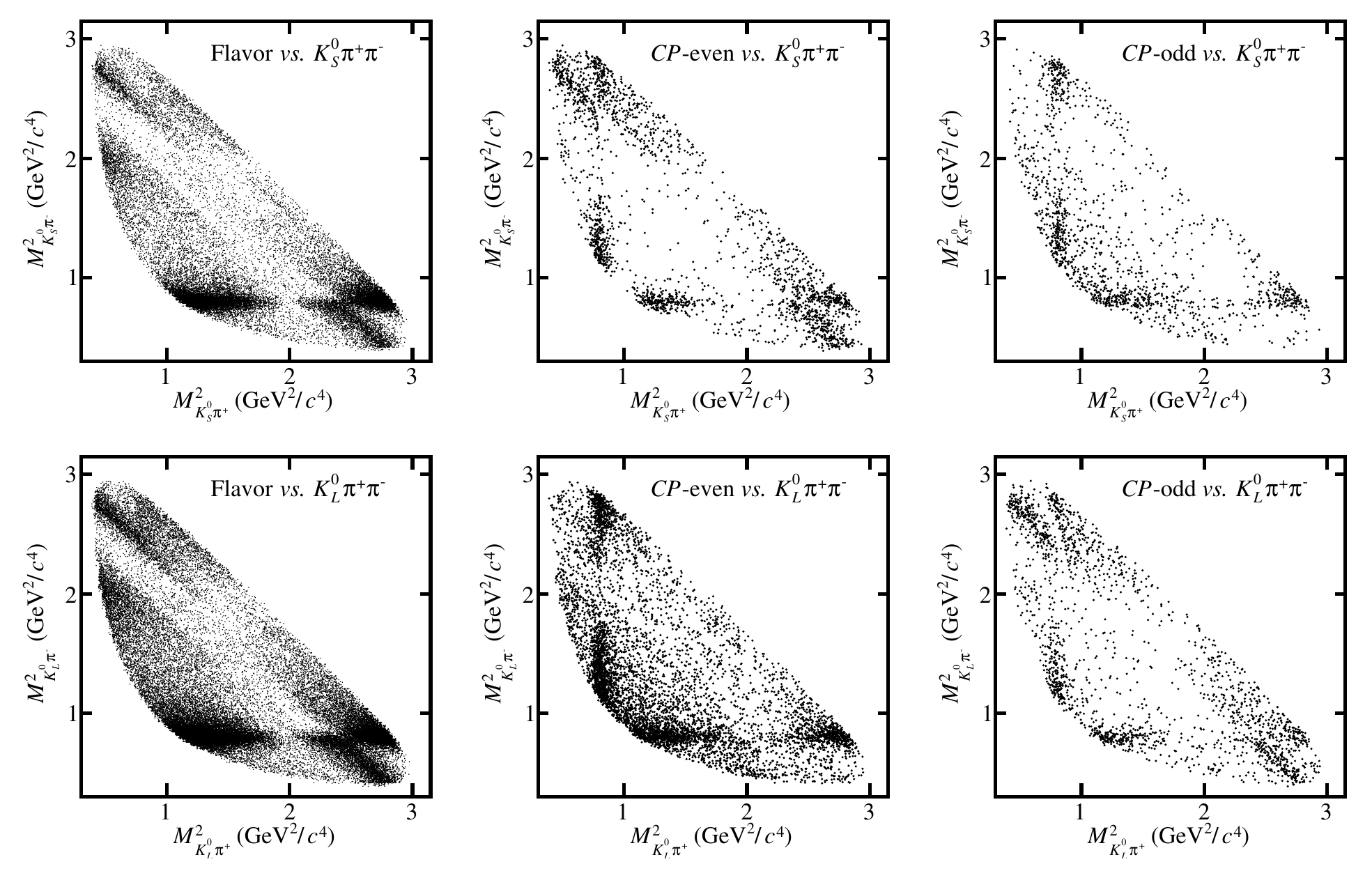}
\caption{Dalitz plots of $K^0_S\pi^+\pi^-$ and $K^0_L\pi^+\pi^-$ events in data. }
\label{fig:dalitz}
\end{center}
\end{figure*}

\subsection{\boldmath Double tags with $K^0_L\pi^0$ and $K^0_L\pi^0\pi^0$ }
\label{sec:DTKlpi0}

The $D\rightarrow K^0_S\pi^+\pi^-$ {\it vs.} $D\rightarrow
K^0_{L}\pi^0(\pi^0)$ DT candidates are also reconstructed with the
missing-mass squared technique as the $K^0_L$ particle is not directly
detectable in the BESIII detector.  In the rest of the event
containing a $D\rightarrow K^0_S\pi^+\pi^-$ ST, a further $\pi^0$ or
$\pi^0\pi^0$ pair is reconstructed. The event is removed if there are
any additional charged tracks in the event.  Figures~\ref{fig:dtmix}(c) and \ref{fig:dtmix}(d) show the resultant ${\rm M}^2_{\rm miss}$ distributions
for $D\rightarrow K^0_S\pi^+\pi^-$ {\it vs.} $D\rightarrow
K^0_{L}\pi^0$ and $D\rightarrow K^0_S\pi^+\pi^-$ {\it vs.}
$D\rightarrow K^0_{L}\pi^0\pi^0$ candidates, respectively. A peak at
the square of the mass of the $K^0_L$ meson is clearly visible.  In
this case the peaking backgrounds come from events where the decay
products of the $K^0_S$ have not been reconstructed and therefore the
$K^0_S$ meson has been identified as a $K^0_L$ meson.  The peaking
backgrounds from $D\rightarrow K^0_S\pi^0$ and $D\rightarrow
K^0_S\pi^0\pi^0$ comprise 5\% and 9\%, respectively, of the signal
sample.

\subsection{\boldmath Double tags with $K^-e^+\nu_e$ }
\label{sec:DTKev}

The $D^0\rightarrow K^-e^+\nu_e$ {\it vs.} $\bar{D}^0\rightarrow K^0_{S}\pi^+\pi^-$ DT candidates are reconstructed by combining an ST $K^0_S\pi^+\pi^-$ candidate with a $K^-$ and a positron candidate from the remaining tracks in the event.
Events with more than two additional charged tracks that have not been used in the ST selection are vetoed.
Information concerning the undetected neutrino is obtained through the kinematic variable
\begin{equation}
{\rm U}_{\rm miss} \equiv (\sqrt{s}/2-E_K-E_e)-|\vec{p}_{\rm miss}|,
\end{equation}
where $E_K$ and $E_e$ are the energy of the kaon and electron from the semi-leptonic $D$ decay candidate, and $\vec{p}_{\rm miss}$ is the missing momentum carried by the neutrino. The momentum $\vec{p}_{\rm miss}$ is defined as $\vec{p}_{\rm miss}=\vec{p}_{\rm sig} - \vec{p}_K -\vec{p}_e$.  Figure~\ref{fig:dtmix}(e) shows the $U_{\rm miss}$ distribution for $D^0\rightarrow K^-e^+\nu_e$ candidates in data, where a peak centered on ${\rm U}_{\rm miss}=0$ is observed due to the negligible mass of the neutrino.

\subsection{\boldmath Double tags with $K^0_L\pi^+\pi^-$ }
\label{sec:DTKlpipi}

To identify the signal candidates from $D\rightarrow K^0_L\pi^+\pi^-$ decays, only two additional and oppositely charged good tracks are required in an event where one of the ST has been selected. These two tracks are identified  as the $\pi^+$ and $\pi^-$ from the $D$ meson. Events that contain any additional charged tracks with the distance of closest approach to the IP less than 20 cm along the beam direction are vetoed. This requirement reduces background from $K^0_S\rightarrow \pi^+\pi^-$ decays. To reject the backgrounds containing $\pi^0$ and $\eta$ mesons, events are vetoed where the invariant mass of any further photon pairs are within the ranges $(0.098, 0.165)$~GeV/$c^2$ and $(0.48, 0.58)$~GeV/$c^2$. This requirement retains about $80\%$ of the signal while reducing more than 90\% of the peaking backgrounds from $D\rightarrow K^0_S\pi^+\pi^-$, where $K^0_S\rightarrow \pi^0\pi^0$. The residual peaking background rate in $D\rightarrow K^0_L\pi^+\pi^-$ selected candidates is $5\%$ of the signal yield and is primarily from the decay $D\to K^0_S(\pi^0\pi^0)\pi^+\pi^-$.
Figure~\ref{fig:dtklp} shows the ${\rm M}^2_{\rm miss}$ distributions of the accepted $D\rightarrow K^0_L\pi^+\pi^-$ candidates in data.

\subsection{Dalitz plot distributions }
\label{sec:DPs}

The DT yields of $K^0_S\pi^+\pi^-$ and $K^0_L\pi^+\pi^-$ tagged by different channels are shown in the fifth and seventh columns of Table~\ref{tab:numST}, respectively.
Their selection efficiencies ($\epsilon^{\rm DT}$) are also listed in the sixth and eighth columns of Table~\ref{tab:numST}.
The DT selection efficiencies are determined in simulation where the signal and tag $D$ meson are both forced to decay to the final states in which they are reconstructed.
The efficiency is determined as the number of DT candidates selected divided by the number of events generated.

The DT yields of $D \to K^0_{S(L)}\pi^+\pi^-$ involving a $CP$ eigenstate are a factor of 5.3(9.2) larger than those reported in Ref.~\cite{prd82_112006}. The yields of $K^0_S\pi^+\pi^-$ tagged with $D \to K^0_{S(L)}\pi^+\pi^-$ decays are a factor of 3.9(3.0) larger than those in Ref.~\cite{prd82_112006}. These increases come not only from the larger data set available at BESIII but also from the additional tag decay modes and partial reconstruction selection techniques.

The resolutions of $M_{K^0_{S}\pi^{\pm}}^2$ and $M_{K^0_{L}\pi^{\pm}}^2$ on the Dalitz plot are improved by requiring that the two neutral $D$ mesons
conserve energy and momentum in the center-of-mass frame, and the decay products from each $D$ meson are constrained to the nominal $D^0$ mass~\cite{pdg18}.
In addition the $K^0_S$ decay products are constrained to the $K^0_S$ nominal mass~\cite{pdg18}. Finally, the missing mass of $K^0_L$ candidates is constrained to the nominal value~\cite{pdg18}.
The study of simulated data indicates that the resulting resolutions of $M_{K^0_{S}\pi^{\pm}}^2$ and $M_{K^0_{L}\pi^{\pm}}^2$ are 0.0068 GeV$^2/c^4$ and 0.0105 GeV$^2/c^4$ for $D\rightarrow K^0_S\pi^+\pi^-$ and $D\rightarrow K^0_L\pi^+\pi^-$, respectively. It should be noted that the finite detector resolution can cause the selected events to migrate between Dalitz plot bins after reconstruction, which should be incorporated in evaluating the expected DT candidates observed in Dalitz plot bins.
More details are presented in Secs.~\ref{sec:kikip} and \ref{sec:expdt}.

The Dalitz plots for $D^0\rightarrow K^0_S\pi^+\pi^-$ and $D^0\rightarrow K^0_L\pi^+\pi^-$ {\it vs.}  the flavor tags  selected from the data are shown in Fig.~\ref{fig:dalitz}.
In order to merge the $D^0$ and $\bar{D}^0$ decays the exchange of coordinates $M_{K^0_{S,L}\pi^{\pm}}^2 \leftrightarrow M_{K^0_{S,L}\pi^{\mp}}^2  $ is performed for the $\bar{D}^0$ decays.
Figure~\ref{fig:dalitz} also shows the $CP$-even and $CP$-odd tagged signal channels selected in the data.
The effect of the quantum correlation in the data is immediately obvious by studying the differences in these plots. Most noticeably, the $CP$-odd component $D\rightarrow K^0_S \rho^0$ is visible in the $D\rightarrow K^0_S\pi^+\pi^-$ decay
when tagged by $CP$-even decays, but is absent when tagged by $CP$-odd decays.

\section{\boldmath Determination of $c_i^{(\prime)}$ and
  $s_i^{(\prime)}$}
\label{sec:cisi}

\subsection{Double-tag yields in Dalitz plot bins}
\label{sec:BinYields}

The fit used to determine the strong-phase parameters is based on the Poisson probability to observe $N$ events in a phase space region given the expectation value $\langle N \rangle$. To measure the observed yields, the data are divided into the phase space regions based on their Dalitz plot coordinates ($m^2_+$, $m^2_-$). A small fraction of candidates ($\sim$0.3\%) fall outside the defined bins. This is because the knowledge of the $D^0$ mass has improved since the model used to define the phase space regions was determined. This improvement leads to a slightly larger allowed phase space in the current analysis compared to the maps of the phase space regions. These outlying candidates are assigned to the bins to which they are closest.

In the $K^0_{S,L}\pi^+\pi^-$ Dalitz plots of the flavor-tagged samples,  the positive and negative bins are distinguishable, and hence yields are measured in 16 bins for each final state.
In contrast, the $CP$-tagged Dalitz plots are symmetric about the line $m_+^2=m_-^2$  (see Eqs.~(\ref{eq:mi}) and (\ref{eq:mip})) and so the entries are summed for bins $i$ and $-i$. Exploiting this symmetry reduces the statistical fluctuations for those $CP$ tags where the yields are low.

The $K^0_{S(L)}\pi^+\pi^-$ {\it vs.} $K^0_S\pi^+\pi^-$ samples are described by two Dalitz plots.
Therefore it is necessary to determine the yields for the $i$th bin ($\mathcal{D}_i$) of one plot and the $j$th bin ($\mathcal{D}_j$) of the other,
in order to obtain the quantities $M_{ij}$ ($M^\prime_{ij}$) that occur in Eq.~(\ref{eq:mij}) [Eq.~(\ref{eq:mijp})].
Considering each half of both plots gives the possibilities $M^{(\prime)}_{ij}$, $M^{(\prime)}_{i-j}$, $M^{(\prime)}_{-ij}$ and $M^{(\prime)}_{-i-j}$, which obey the following relations:
\begin{equation}
M^{(\prime)}_{ij}=M^{(\prime)}_{-i-j},~~M^{(\prime)}_{-ij}=M^{(\prime)}_{i-j},~~{\rm and}~~M^{(\prime)}_{ij}=M^{(\prime)}_{i-j}.
\label{eq:mij_sysmmetry}
\end{equation}
It follows that events can be classified into those where both decays occur in the Dalitz plots on the same side of the $m_+^2=m_-^2$ line, and those when they are on different sides.
For the case where both $D$ mesons are fully reconstructed as $D\rightarrow K^0_S\pi^+\pi^-$, it is not possible to distinguish between $\mathcal{D}_i$ and $\mathcal{D}_j$, and thus $M_{ij}$ is combined with $M_{ji}$.
The partially reconstructed $D\rightarrow K^0_S\pi^+\pi^-$ samples are treated in the same way, despite the distinguishability of the final states, in order to avoid low yields.
In the $K^0_L\pi^+\pi^-$ {\it vs.} $K^0_S\pi^+\pi^-$ sample $\mathcal{D}_i$ is chosen to specify the $K^0_S\pi^+\pi^-$ Dalitz plot bin, and $\mathcal{D}_j$ the $K^0_L\pi^+\pi^-$ bin.
In this case $M^{\prime}_{ij}$ and $M^{\prime}_{ji}$ are distinguishable and cannot be combined. Following these considerations, the samples with two Dalitz plots are divided into 72 and 128 bins.

In each bin of phase space there are candidates that are from signal, combinatorial background, and peaking backgrounds.
The yields for each DT mode are determined in the same way as in Sec.~\ref{sec:evtsel}, although in some regions where the yields are low it is necessary
to fix some parameters from the fit to data over the full phase space. The observed combinatorial background yield is determined in the fit and not considered further.
Although the expected peaking-background yield can be calculated with MC simulation, the fit cannot distinguish the observed peaking background yield from the signal yield.
Therefore the observed yield $N^{\rm obs}$ in each phase space region is the sum of signal and peaking background.

\subsection{\boldmath Determination of $K_i$ and $K_i^{\prime}$}
\label{sec:kikip}

The yields of $K_i$ and $K_i^\prime$ are necessary to determine the
expected yields in the decays sensitive to the strong-phase
parameters.  As discussed in Sec.~\ref{sec:DPs}, the finite detector
resolution can cause the individual decays to migrate between Dalitz
plot bins after reconstruction.  Furthermore, the migration effects
between $D^0\rightarrow K^0_S\pi^+\pi^-$ and $D^0\rightarrow
K^0_L\pi^+\pi^-$ are also different due to their resolution
differences.
Studies indicate that neglecting bin migration induces average biases of
0.7~(0.3) times the statistical uncertainty in the determination of
$c_i~(s_i)$, and hence it is important to correct for this effect in
the analysis.

\begin{table*}
\caption{ Efficiency matrix $\epsilon_{ij}$ (\%) for $K^0_{S,L}\pi^+\pi^-$ {\it vs.} $K^+\pi^-$ in the equal $\Delta\delta_D$ binning scheme. The column gives the true bins $j$, while the row gives the reconstructed bin $i$, in which the decay BF of $K^0_S\rightarrow \pi^+\pi^-$ is not included.  }
\begin{center}
\begin{tabular}{lrrrrrrrrrrrrrrrrrr}
\hline
Bins(True) & 1 & -1 & 2 & -2 & 3 & -3 & 4 & -4 & 5 & -5 & 6 & -6 & 7 & -7 & 8 & -8 \\
(Rec)        &    &     &    &     &   &     &    &     &    &     &    &     &    &     &    &    \\
\hline
    & \multicolumn{16}{c}{$\epsilon_{ij}$ for $K^0_S\pi^+\pi^-$ {\it vs.} $K^+\pi^-$} \\
\hline
~1  & 36.53 & 0.24 & 2.40 & 0.00 & 0.20 & 0.04 & 0.11 & 0.07 & 0.05 & 0.01 & 0.11 & 0.04 & 0.20 & 0.00 & 2.28 & 0.02  \\
-1  &  0.12 &  38.84 & 0.01 & 1.63 & 0.01 & 0.24 & 0.04 & 0.07 & 0.01 & 0.07 & 0.01 & 0.10 & 0.01 & 0.17 & 0.01 & 1.74  \\
~2  &  1.33 & 0.00 &  38.05 & 0.00 & 1.33 & 0.01 & 0.02 & 0.00 & 0.00 & 0.00 & 0.01 & 0.00 & 0.06 & 0.00 & 0.14 & 0.00  \\
-2  &  0.00 & 0.36 & 0.01 & 39.12 & 0.01 & 0.63 & 0.01 & 0.08 & 0.00 & 0.00 & 0.00 & 0.03 & 0.01 & 0.04 & 0.00 & 0.03  \\
~3  &  0.11 & 0.01 & 1.21 & 0.01 & 41.05 & 0.10 & 1.34 & 0.04 & 0.01 & 0.01 & 0.03 & 0.01 & 0.04 & 0.01 & 0.08 & 0.02  \\
-3  &  0.02 & 0.04 & 0.02 & 0.49 & 0.05 & 41.59 & 0.04 & 0.84 & 0.00 & 0.01 & 0.04 & 0.05 & 0.04 & 0.06 & 0.04 & 0.01  \\
~4  &  0.03 & 0.01 & 0.06 & 0.01 & 0.53 & 0.01 & 40.59 & 0.03 & 0.26 & 0.01 & 0.02 & 0.01 & 0.03 & 0.01 & 0.03 & 0.02  \\
-4  &  0.03 & 0.02 & 0.04 & 0.04 & 0.07 & 0.96 & 0.03 & 42.10 & 0.00 & 0.33 & 0.01 & 0.06 & 0.04 & 0.08 & 0.01 & 0.02  \\
~5  &  0.04 & 0.01 & 0.02 & 0.00 & 0.05 & 0.04 & 0.90 & 0.01 & 38.14 & 0.00 & 1.16 & 0.00 & 0.03 & 0.00 & 0.02 & 0.01  \\
-5  &  0.03 & 0.06 & 0.04 & 0.02 & 0.05 & 0.05 & 0.02 & 0.93 & 0.00 & 38.66 & 0.00 & 1.77 & 0.02 & 0.08 & 0.01 & 0.03  \\
~6  &  0.06 & 0.00 & 0.04 & 0.00 & 0.06 & 0.00 & 0.05 & 0.02 & 0.80 & 0.00 & 35.50 & 0.03 & 0.97 & 0.00 & 0.08 & 0.00  \\
-6  &  0.01 & 0.02 & 0.02 & 0.01 & 0.03 & 0.03 & 0.02 & 0.02 & 0.00 & 0.64 & 0.00 & 37.54 & 0.01 & 1.93 & 0.01 & 0.07  \\
~7  &  0.17 & 0.01 & 0.14 & 0.00 & 0.09 & 0.04 & 0.03 & 0.04 & 0.04 & 0.00 & 1.93 & 0.03 & 35.50 & 0.01 & 2.17 & 0.00  \\
-7  &  0.01 & 0.07 & 0.03 & 0.06 & 0.04 & 0.07 & 0.06 & 0.07 & 0.01 & 0.02 & 0.00 & 1.39 & 0.01 & 36.86 & 0.01 & 0.79  \\
~8  &  2.00 & 0.00 & 0.25 & 0.00 & 0.12 & 0.03 & 0.03 & 0.03 & 0.02 & 0.00 & 0.13 & 0.02 & 1.99 & 0.03 & 35.24 & 0.00  \\
-8  &  0.01 & 0.72 & 0.03 & 0.10 & 0.03 & 0.05 & 0.01 & 0.01 & 0.00 & 0.03 & 0.01 & 0.07 & 0.01 & 1.81 & 0.01 & 37.94  \\\hline
    & \multicolumn{16}{c}{$\epsilon_{ij}$ for $K^0_L\pi^+\pi^-$ {\it vs.} $K^+\pi^-$} \\
\hline
~1 & 45.66 & 0.61 & 4.14 & 0.00 & 0.20 & 0.00 & 0.07 & 0.00 & 0.06 & 0.00 & 0.37 & 0.00 & 0.58 & 0.00 & 4.62 & 0.01  \\
-1 & 0.36 & 51.88 & 0.00 & 2.96 & 0.00 & 0.20 & 0.00 & 0.13 & 0.00 & 0.08 & 0.01 & 0.13 & 0.02 & 0.49 & 0.02 & 3.32  \\
~2 & 2.25 & 0.00 & 46.88 & 0.00 & 1.83 & 0.00 & 0.11 & 0.00 & 0.00 & 0.00 & 0.03 & 0.00 & 0.10 & 0.00 & 0.30 & 0.00  \\
-2 & 0.00 & 0.68 & 0.00 & 50.04 & 0.00 & 1.04 & 0.00 & 0.06 & 0.00 & 0.01 & 0.00 & 0.06 & 0.00 & 0.08 & 0.00 & 0.14  \\
~3 & 0.12 & 0.00 & 1.66 & 0.00 & 50.44 & 0.00 & 2.10 & 0.00 & 0.03 & 0.00 & 0.08 & 0.00 & 0.02 & 0.00 & 0.05 & 0.00  \\
-3 & 0.00 & 0.06 & 0.00 & 0.62 & 0.00 & 52.50 & 0.00 & 1.30 & 0.00 & 0.01 & 0.00 & 0.05 & 0.00 & 0.06 & 0.00 & 0.03  \\
~4 & 0.02 & 0.00 & 0.03 & 0.00 & 0.74 & 0.00 & 51.02 & 0.00 & 0.47 & 0.00 & 0.04 & 0.00 & 0.01 & 0.00 & 0.01 & 0.00  \\
-4 & 0.00 & 0.01 & 0.00 & 0.04 & 0.00 & 1.35 & 0.00 & 51.33 & 0.00 & 0.50 & 0.00 & 0.08 & 0.00 & 0.01 & 0.00 & 0.00  \\
~5 & 0.04 & 0.00 & 0.02 & 0.00 & 0.04 & 0.00 & 1.41 & 0.00 & 51.23 & 0.00 & 1.98 & 0.00 & 0.06 & 0.00 & 0.04 & 0.00  \\
-5 & 0.00 & 0.09 & 0.00 & 0.01 & 0.00 & 0.07 & 0.00 & 1.50 & 0.00 & 50.45 & 0.00 & 3.01 & 0.00 & 0.18 & 0.00 & 0.07  \\
~6 & 0.11 & 0.00 & 0.04 & 0.00 & 0.03 & 0.00 & 0.11 & 0.00 & 1.24 & 0.00 & 45.69 & 0.00 & 1.70 & 0.00 & 0.12 & 0.00  \\
-6 & 0.00 & 0.07 & 0.00 & 0.01 & 0.00 & 0.05 & 0.00 & 0.08 & 0.00 & 0.99 & 0.00 & 47.92 & 0.00 & 3.10 & 0.00 & 0.12  \\
~7 & 0.31 & 0.00 & 0.19 & 0.00 & 0.10 & 0.00 & 0.08 & 0.00 & 0.07 & 0.00 & 3.52 & 0.00 & 44.30 & 0.00 & 3.92 & 0.00  \\
-7 & 0.00 & 0.09 & 0.00 & 0.10 & 0.00 & 0.09 & 0.00 & 0.04 & 0.00 & 0.04 & 0.00 & 2.23 & 0.00 & 45.24 & 0.00 & 1.47  \\
~8 & 3.53 & 0.01 & 0.44 & 0.00 & 0.12 & 0.00 & 0.04 & 0.00 & 0.03 & 0.00 & 0.36 & 0.00 & 4.01 & 0.00 & 42.92 & 0.00  \\
-8 & 0.01 & 1.35 & 0.00 & 0.32 & 0.00 & 0.09 & 0.00 & 0.02 & 0.00 & 0.02 & 0.00 & 0.19 & 0.00 & 3.04 & 0.00 & 49.95  \\
\hline
\end{tabular}
\label{tab:effmatrix_equal}
\end{center}
\end{table*}

To account for this effect, the number of observed signal events ($N_i^{\rm obs(\prime)}$) for flavor-tagged $K^0_{S(L)}\pi^+\pi^-$ decays in the $i$th bin of the Dalitz plot is written
\begin{equation}
N_i^{\rm obs(\prime)}=\sum\limits_{j=1}^{N_{\rm bins}}\epsilon_{ij}K^{(\prime)}_j,
\end{equation}
where $\epsilon_{ij}$ is the efficiency matrix which describes the reconstruction efficiency and migration effects across Dalitz plot bins associated with reconstruction of tag and signal decays.
The efficiency matrix $\epsilon_{ij}$ can be obtained
by analyzing a sample of signal MC events which are generated as $e^+e^-\rightarrow \psi(3770)\rightarrow D^0\bar{D}^0$, where the $\bar{D}^0$ meson decays to the ST modes and $D^0\rightarrow K^0_{S(L)}\pi^+\pi^-$. The efficiency matrix $\epsilon_{ij}$ for detecting $D\rightarrow K^0_{S(L)}\pi^+\pi^-$ decay is given by
\begin{equation}
\epsilon_{ij}=\frac{N^{\rm rec}_{ij}}{N_j^{\rm gen}}\times \frac{1}{\epsilon_{\rm ST}},
\label{eq:effmatrix}
\end{equation}
where $N^{\rm rec}_{ij}$ is the number of signal MC events generated in the $j$th Dalitz plot bin and reconstructed in the
$i$th Dalitz plot bin,
$N^{\rm gen}_j$ is the number of signal MC events which are generated in the $j$th Dalitz plot bin and $\epsilon_{\rm ST}$ is the ST efficiency.
An example of the efficiency matrix $\epsilon_{ij}$ for $K^0_{S(L)}\pi^+\pi^-$ {\it vs.} $K^+\pi^-$ in the equal $\Delta\delta_D$ binning scheme is shown in Table~\ref{tab:effmatrix_equal}.
Thus, the value of $K^{(\prime)}_i$ in the $i$th Dalitz plot bin for $D^0\rightarrow K^0_{S(L)}\pi^+\pi^-$ decay is obtained by
\begin{equation}
K^{(\prime)}_i=\sum\limits_{j=1}^{N_{\rm bins}}(\epsilon^{-1})_{ij}N_j^{\rm obs(\prime)}.
\label{eq:kikiprime_calcu}
\end{equation}
In addition, the migration effects in the $i$th Dalitz plot bin can be estimated by using $R_{i}=\epsilon_{ii}/\sum_j\epsilon_{ij}$,
which denotes the fraction of the reconstructed events falling outside the true Dalitz plot bins.
From the efficiency matrix $\epsilon_{ij}$ listed in Table~\ref{tab:effmatrix_equal},
it is estimated that the bin migration effects range within (3-12)\% and (3-18)\% for the $K^0_S\pi^+\pi^-$ and $K^0_L\pi^+\pi^-$ signals with the equal $\Delta\delta_D$ binning scheme, respectively.

Moreover, the event yields of $K^0_S\pi^+\pi^-$ and $K^0_L\pi^+\pi^-$ selected against hadronic flavored tags are also contaminated by DCS decays~\cite{prd80_032002,prd82_112006}.
To account for this effect, the flavor-tagged yield in each Dalitz plot bin is scaled by a correction factor $f^{(\prime)}_i$ ($f_i$ for $K^0_S\pi^+\pi^-$ and $f^{\prime}_i$ for $K^0_L\pi^+\pi^-$).
The correction factors for the hadronic tags $K^+\pi^-$, $K^+\pi^-\pi^0$ and $K^+\pi^-\pi^-\pi^+$ are calculated by
\begin{widetext}
\begin{eqnarray}
f_i&=&\frac{\int_i |f(m^2_+,m^2_-)|^2 dm^2_+dm^2_-}{\int_i (|f(m^2_+,m^2_-)|^2+(r_D^F)^2|f(m^2_-,m^2_+)|^2-2r^F_DR_F\mathcal{R}[e^{i\delta_D^F}f(m^2_+,m^2_-)f^*(m^2_-,m^2_+)])dm^2_+m^2_-}, \nonumber \\
f_i^\prime&=&\frac{\int_i |f^\prime(m^2_+,m^2_-)|^2 dm^2_+dm^2_-}{\int_i (|f^\prime(m^2_+,m^2_-)|^2+(r_D^F)^2|f^\prime(m^2_-,m^2_+)|^2+2r^F_DR_F\mathcal{R}[e^{i\delta_D^F}f^\prime(m^2_+,m^2_-)f^{*\prime}(m^2_-,m^2_+)])dm^2_+dm^2_-}.
\label{eq:fical}
\end{eqnarray}
\end{widetext}
Here $r_D^F$ is the ratio of the DCS amplitude to the Cabibbo-favored
decay amplitude and $\delta_D^F$ is the corresponding strong-phase
difference. For multibody final states these two quantities are
averaged over the decay phase space.  The coherence
factor~\cite{prd68_033303} $R_F$ equals unity for two-body decays, and
has been measured for $D\rightarrow K^+\pi^-\pi^-\pi^+$ and
$D\rightarrow K^+\pi^-\pi^0$~\cite{plb757_520,plb765_402}.  The values
of these parameters and the corresponding references are given in
Table~\ref{tab:kicorpara}.  Furthermore, $f(m^2_+,m^2_-)$ and
$f^{\prime}(m^2_+,m^2_-)$ are the amplitudes of $D^0\rightarrow
K^0_S\pi^+\pi^-$ and $D^0\rightarrow K^0_L\pi^+\pi^-$, respectively.
The amplitude of the decay $D^0\rightarrow K^0_S\pi^+\pi^-$ is taken
from the model given in Ref.~\cite{prd98_112012}.  The decay amplitude
of $D^0\rightarrow K^0_L\pi^+\pi^-$ has not been studied, however a
decay model can be estimated by adjusting the $D^0\rightarrow
K^0_S\pi^+\pi^-$ model in the same way as discussed in
Refs.~\cite{prd80_032002,prd82_112006}, and using that
$K^0_S$ and $K^0_L$ mesons are of opposite $CP$, to an excellent
approximation.  Starting with the $D^0\rightarrow K^0_S\pi^+\pi^-$
model in Ref.~\cite{prd98_112012}, the Cabibbo-favored amplitudes are
unchanged and the amplitudes of the DCS components gain a factor $-1$.
For the $CP$-eigenstate amplitudes, such as $K^0_S \rho(770)^0$, the
$D^0\rightarrow K^0_L\pi^+\pi^-$ amplitude can be related to the
$D^0\rightarrow K^0_S\pi^+\pi^-$ amplitude by multiplying the latter
by a factor $(1-2re^{i\delta})$~\cite{plb349_363}, where $r$ is of the
order of ${\rm tan}^2\theta_C$. Here, $\theta_C$ is the Cabibbo angle
and $\delta$ is an unknown phase. To determine central values for
$f_i^\prime$, the parameters $r$ and $\delta$ are varied a number of
times where $\delta$ is assumed to have an equal probability to lie
between $0^{\circ}$ and $360^{\circ}$ and $r$ is assumed to have a
Gaussian distribution with mean ${\rm tan}^2\theta_C$ and width
$0.5\times{\rm tan}^2\theta_C$. The mean value of the resulting
distribution of $f_i^\prime$ is taken as the nominal value of that
parameter.

\begin{figure*}[tp!]
\begin{center}
\includegraphics[width=\linewidth]{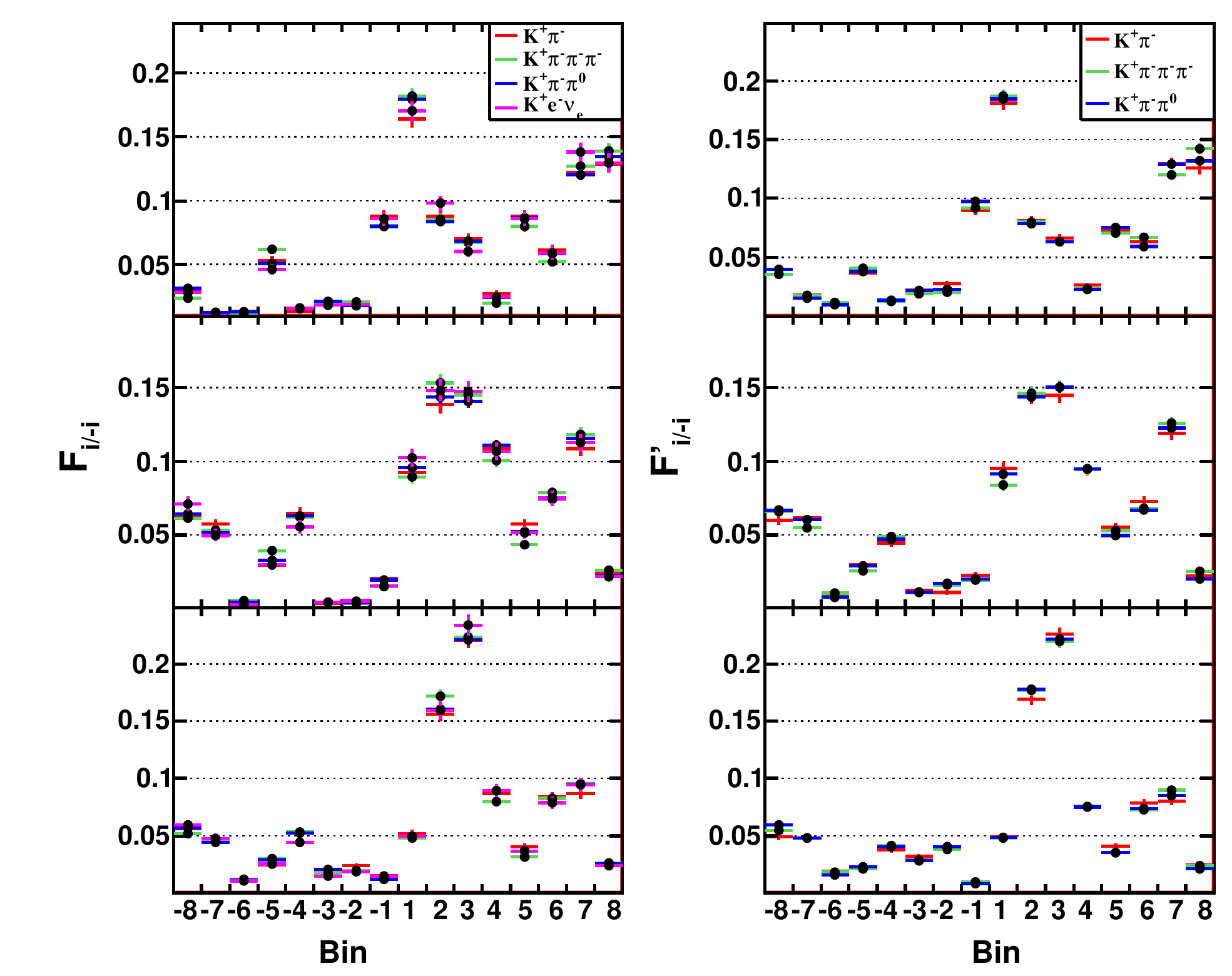}
\caption{Event fractions in Dalitz plot bins in data; (left) $F_{i/-i}$ for $K^0_S\pi^+\pi^-$ and (right) $F^{\prime}_{i/-i}$ for $K^0_L\pi^+\pi^-$ from (top) equal $\Delta \delta_D$, (middle) optimal and (bottom) modified optimal binning schemes, respectively. The horizontal and vertical error bars denote the bin intervals and the statistical uncertainties of the event fractions, respectively.}
\label{fig:evtfrac}
\end{center}
\end{figure*}

The event fraction $F^{(\prime)}_{i}$, defined as
$F^{(\prime)}_{i}=K^{(\prime)}_{i}/A_D$, where
$A_D=\sum\limits_{i=1}^{8}(K^{(\prime)}_i+K^{(\prime)}_{-i})$, is
computed for each flavor-tag.  Figure~\ref{fig:evtfrac} shows the
measured values of $F_i$ and $F^{\prime}_{i}$ in various Dalitz plot
bins for the flavor-tagged $D^0\rightarrow K^0_{S}\pi^+\pi^-$ and
$D^0\rightarrow K^0_{L}\pi^+\pi^-$ events, respectively, observed in
data.  The measured values of $F_i$ and $F^{\prime}_{i}$ are
consistent between the different categories of flavor tags, which
provides a good validation for the extracted $K_i$ and $K_i^{\prime}$
in data.  In order to recover the summed $K_i$ from all flavor-tagged
$K^0_S\pi^+\pi^-$ with $K^0_S\rightarrow \pi^+\pi^-$ used in this
analysis, the values of $F_i$ in Table~\ref{tab:fidata} should be
multiplied by 58607, 58647, and 58595 for the equal $\Delta\delta_D$,
optimal, and modified binning schemes, respectively. To obtain the
values of $K_i^\prime$ the corresponding factors are 80718, 80661, and
80706.

\begin{table}
\caption{ Values of the parameters used to make the corrections to the flavor-tagged yields. }
\begin{center}
\begin{tabular}{lccccccccc} \hline
\hline
$F$           & $r_D^F$ (\%)                              & $\delta_D^F$ ($^{\circ}$)                & $R_F$                                  \\ \hline
$K\pi\pi\pi$  & $5.49\pm0.06$~\cite{plb757_520,plb765_402}           & $128^{+28}_{-17}$~\cite{plb757_520,plb765_402}      & $0.43^{+0.17}_{-0.13}$~\cite{plb757_520,plb765_402}\\ \hline
$K\pi\pi^0$   & $4.47\pm0.12$~\cite{plb757_520,plb765_402}           & $198^{+14}_{-15}$~\cite{plb757_520,plb765_402}      & $0.81\pm0.06$~\cite{plb757_520,plb765_402}        \\ \hline
$K\pi$        & $5.86\pm0.02$~\cite{epjc77_895}           & $194.7^{+8.4}_{-17.6}$~\cite{epjc77_895} & 1                                      \\ \hline
\hline
\end{tabular}
\label{tab:kicorpara}
\end{center}
\end{table}

\begin{table*}
\caption{The values of $F_i$ and $F_i^\prime$ using all flavor-tagged DT samples, for each of the three binning schemes. The uncertainties are statistical only. }
\begin{center}
\begin{tabular}{lcccccc} \hline
      & \multicolumn{2}{c}{Equal $\Delta\delta_D$ binning }  & \multicolumn{2}{c}{Optimal binning}  & \multicolumn{2}{c}{Modified optimal binning }   \\
Bin  & $F_i$ &  $F^{\prime}_i$  & $F_i$ &  $F^{\prime}_i$  & $F_i$ &  $F^{\prime}_i$    \\
\hline
~1 &~  0.176 $\pm$  0.003  &~ 0.184 $\pm$   0.003  &~ 0.095 $\pm$     0.002  &~ 0.090 $\pm$     0.002  &~ 0.049 $\pm$     0.002  &~ 0.048 $\pm$     0.002 \\
-1&~   0.083 $\pm$   0.002 &~  0.094 $\pm$   0.002&~   0.018 $\pm$     0.001&~   0.021 $\pm$     0.001&~   0.013 $\pm$     0.001&~   0.009 $\pm$     0.001\\
~2 &~  0.087 $\pm$   0.002  &~ 0.079 $\pm$   0.002 &~  0.146 $\pm$     0.003 &~  0.145 $\pm$     0.003 &~  0.162 $\pm$     0.003 &~  0.176 $\pm$     0.003\\
-2 &~  0.019 $\pm$   0.001 &~  0.023 $\pm$   0.001&~   0.004 $\pm$     0.001&~   0.015 $\pm$     0.001&~   0.020 $\pm$     0.001&~   0.040 $\pm$     0.001\\
~3  &~ 0.067 $\pm$   0.002  &~ 0.064 $\pm$   0.002 &~  0.144 $\pm$     0.003 &~  0.149 $\pm$    0.003  &~ 0.224 $\pm$     0.003  &~ 0.222 $\pm$     0.003\\
-3 &~  0.021 $\pm$   0.001 &~  0.021 $\pm$   0.001&~   0.004 $\pm$     0.001&~   0.011 $\pm$     0.001&~   0.018 $\pm$     0.001&~   0.029 $\pm$     0.001\\
~4  &~ 0.024 $\pm$   0.001  &~ 0.023 $\pm$   0.001 &~  0.107 $\pm$     0.002 &~  0.095 $\pm$     0.002 &~  0.086 $\pm$     0.002 &~  0.075 $\pm$     0.002\\
-4 &~  0.016 $\pm$   0.001 &~  0.013 $\pm$   0.001&~   0.062 $\pm$     0.002&~   0.047 $\pm$     0.001&~   0.051 $\pm$    0.002 &~  0.040 $\pm$     0.001\\
~5  &~ 0.085 $\pm$   0.002  &~ 0.073 $\pm$   0.002 &~  0.051 $\pm$     0.002 &~  0.052 $\pm$     0.002 &~  0.036 $\pm$     0.001 &~  0.036 $\pm$     0.001\\
-5 &~  0.053 $\pm$   0.002 &~  0.038 $\pm$   0.001&~   0.033 $\pm$     0.001&~   0.028 $\pm$     0.001&~   0.028 $\pm$     0.001&~   0.022 $\pm$     0.001\\
~6  &~ 0.058 $\pm$   0.002  &~ 0.062 $\pm$   0.002 &~  0.076 $\pm$     0.002 &~  0.069 $\pm$     0.002 &~  0.081 $\pm$     0.002 &~  0.074 $\pm$     0.002\\
-6 &~  0.012 $\pm$   0.001 &~  0.010 $\pm$   0.001&~   0.004 $\pm$     0.001&~   0.009 $\pm$     0.001&~   0.011 $\pm$     0.001&~   0.017 $\pm$     0.001\\
~7  &~ 0.125 $\pm$   0.003  &~ 0.126 $\pm$  0.002  &~ 0.114 $\pm$     0.002  &~ 0.123 $\pm$     0.002 &~  0.093 $\pm$     0.002  &~ 0.085 $\pm$     0.002\\
-7 &~  0.012 $\pm$   0.001 &~  0.016 $\pm$  0.001 &~  0.053 $\pm$     0.002 &~   0.059 $\pm$     0.002&~   0.045 $\pm$     0.002&~   0.048 $\pm$     0.002\\
~8  &~ 0.134 $\pm$   0.003  &~ 0.134 $\pm$     0.003&~   0.024 $\pm$     0.001&~   0.022 $\pm$     0.001&~   0.025 $\pm$     0.001&~   0.022 $\pm$     0.001\\
-8 &~  0.028 $\pm$   0.001 &~  0.037 $\pm$     0.001&~   0.065 $\pm$     0.002&~   0.065 $\pm$     0.002&~   0.056 $\pm$    0.002 &~  0.055  $\pm$    0.001\\
\hline
\end{tabular}
\label{tab:fidata}
\end{center}
\end{table*}

\subsection {Expected DT yields in Dalitz plot bins}
\label{sec:expdt}

The expected yields, $\langle  N \rangle$, are a sum of expected signal and peaking background contributions. The expected signal yields are calculated from the equations given in Sec.~\ref{sec:theory}, with adjustments made to account for bin migration and selection and reconstruction efficiencies, so that they can be compared to the yields from data.
For $CP$-tagged $D\rightarrow K^0_S\pi^+\pi^-$ decays, the expected signal yield in the $i$th bin is given by
\begin{eqnarray}
M_i=&&h_{CP} \sum\limits_j^8 \epsilon^{K^0_S\pi^+\pi^-}_{ij}\times \nonumber \\
&& [K_j-(2F_{CP}-1)2c_j\sqrt{K_jK_{-j}}+K_{-j}],
\label{eq:miexp}
\end{eqnarray}
where $\epsilon^{K^0_S\pi^+\pi^-}_{ij}$ is the efficiency matrix for detecting $D\rightarrow K^0_S\pi^+\pi^-$ {\it vs.} the particular $CP$ tag under consideration defined similarly as in Eq.~(\ref{eq:effmatrix}).
The efficiency matrix is defined to take into account the merging of the $i$th and $-i$th regions in data ({\it i.e.} it has size 8 $\times$ 8). The normalization factor $h_{CP}$ is defined as $S_{CP}/S_{\rm FT}$.
For $CP$-tagged $D\rightarrow K^0_L\pi^+\pi^-$, the expected signal yield in the $i$th bin is given by
\begin{eqnarray}
M_i^\prime= &&h^\prime_{CP} \sum\limits_j^8 \epsilon^{K^0_L\pi^+\pi^-}_{ij} \times \nonumber \\
&& [K^\prime_j-(2F_{CP}-1)2c_j^\prime\sqrt{K^\prime_jK^\prime_{-j}}+K^\prime_{-j}],
\label{eq:mipexp}
\end{eqnarray}
where $h^\prime_{CP}$ is given by $S_{CP}/S_{\rm FT^\prime}$, and $S_{\rm FT^\prime}$ is the sum of ST yields for the three hadronic flavor tags used to determine the values of $K_i^\prime$.

For the DT $D\rightarrow K^0_S\pi^+\pi^-$ {\it vs.} $D\rightarrow K^0_S\pi^+\pi^-$ the efficiency matrix, $\epsilon_{nm}^{K^0_S\pi^+\pi^- {\it vs.} K^0_S\pi^+\pi^-}$, is a 72 $\times$ 72 matrix where each value of the indices $n$ and $m$ corresponds to one of the 72 distinct bin-pairs. The expected signal yields are expressed as
\begin{eqnarray}
M_{n}=&&h_{\rm corr}\sum\limits_{m=1}^{m=72} \epsilon_{nm}^{K^0_S\pi^+\pi^- {\it vs.} K^0_S\pi^+\pi^-} [ K_{i_m}K_{-j_m}+K_{-i_m}K_{j_m} \nonumber \\
&&-2\sqrt{K_{i_m}K_{-j_m}K_{-i_m}K_{j_m}}(c_{i_m}c_{j_m}+s_{i_m}s_{j_m}) ],
\label{eq:mijexp}
\end{eqnarray}
where $i_m$ and $j_m$ correspond to the $i$th and $j^{\rm th}$ bins of the $m$th bin-pair and $h_{\rm corr} = (N_{DD}/S_{\rm FT}^2)\times \alpha\beta$. The constant $\alpha$ results from the symmetry relations used to combine bin pairs. The value of $\alpha$ is 1 when the $i,j$ values of the index $n$ satisfy $|i|=|j|$ and 2 otherwise. The constant $\beta$ arises from the symmetry of the signal and tag decays and has value 1 for the
selections where both $K^0_S$ mesons decay via $K^0_S\rightarrow \pi^+\pi^-$, and value $2\times\mathcal{B}(K^0_S\rightarrow \pi^0\pi^0)/\mathcal{B}(K^0_S\rightarrow \pi^+\pi^-)$ when one $K^0_S$ meson decays to the $\pi^0\pi^0$ final state. Here, $\mathcal{B}(K^0_S\rightarrow \pi^0\pi^0)$ and $\mathcal{B}(K^0_S\rightarrow \pi^+\pi^-)$ are BFs for $K^0_S\rightarrow \pi^0\pi^0$ and $K^0_S\rightarrow \pi^+\pi^-$, respectively.
For the DT $D\rightarrow K^0_L\pi^+\pi^-$ {\it vs.} $D\rightarrow K^0_S\pi^+\pi^-$ the expected signal yields are expressed as
\begin{eqnarray}
M_{n}^\prime=&&h_{\rm corr}^\prime\sum\limits_{m=1}^{m=128} \epsilon_{nm}^{\prime K^0_L\pi^+\pi^- {\it vs.} K^0_S\pi^+\pi^-} [ K_{i_m}K^\prime_{-j_m}+K_{-i_m}K^\prime_{j_m}\nonumber \\
 &&-2\sqrt{K_{i_m}K^\prime_{-j_m}K_{-i_m}K^\prime_{j_m}}(c_{i_m}c^\prime_{j_m}+s_{i_m}s^\prime_{j_m}) ],
\label{eq:mijpexp}
\end{eqnarray}
where $h^\prime_{\rm corr} = 2N_{DD}/(S_{\rm FT^\prime}S_{\rm FT})$. The DT efficiency matrix of detecting $D\rightarrow K^0_L\pi^+\pi^-$ {\it vs.} $D\rightarrow K^0_S\pi^+\pi^-$, $\epsilon_{nm}^{\prime K^0_L\pi^+\pi^- {\it vs.} K^0_S\pi^+\pi^-}$, is a 128 $\times$ 128 matrix where each value of the indices $n$ and $m$ corresponds to one of the 128 distinct bin-pairs.

The peaking-background yields integrated over the Dalitz plot have been estimated for each DT in Sec.~\ref{sec:evtsel}. The majority of the peaking backgrounds are $CP$ eigenstates, for example the
decay $D\rightarrow K^0_S\pi^0$ forms a peaking background to the tag $D\rightarrow K^0_L\pi^0$.
The simulated data cannot accurately describe the distribution of the peaking background over the Dalitz plot since it does not account for quantum correlations. For the peaking backgrounds which are $CP$ eigenstates
the expected yields are distributed according to Eq.~(\ref{eq:mi}) with an appropriate normalization factor to take into account the expected yield integrated over the Dalitz plot.
The values of $c_i$ and $s_i$ used to make the initial estimate of the expected peaking background yields are computed from the $D\rightarrow K^0_S\pi^+\pi^-$ amplitude model~\cite{prd98_112012}.
As the yields of peaking backgrounds are small compared to the signal yields the effects of migration or small variations in efficiency over the Dalitz plot are ignored.
A similar estimate for the peaking background of $D\rightarrow K^0_S\pi^+\pi^-$ {\it vs.} $D\rightarrow K^0_S\pi^+\pi^-$ in the $D\rightarrow K^0_L\pi^+\pi^-$ {\it vs.} $D\rightarrow K^0_S\pi^+\pi^-$ DT
can be estimated through Eq.~(\ref{eq:mij}). The peaking background $D\rightarrow K^0_S\pi^+\pi^-\pi^0$ in the $D\to K^0_S\omega$ tag is treated as $CP$-odd,
as indicated by the results in Ref.~\cite{jhep01_082}.
The strong-phase parameters of $D\rightarrow \pi^+\pi^-\pi^0\pi^0$ are not known and in this case the peaking background is distributed as observed in simulated data.
The expected distribution of the remaining peaking backgrounds that occur at low rates, such as $D\rightarrow \pi^+\pi^-\pi^+\pi^-$, are also taken from simulation.

\subsection{\boldmath Fit to determine $c_i$, $c^{\prime}_i$, $s_i$ and $s^{\prime}_i$}

To determine the values of $c^{(\prime)}_i$ and $s^{(\prime)}_i$, a log-likelihood fit is performed where the likelihood is given by
\begin{eqnarray}
-2{\rm log}\mathcal{L}&=&-2\sum\limits_{i=1}^{8} {\rm ln}P(N^{\rm obs}_i,\langle N_i^{\rm exp}\rangle)_{CP,K^0_S\pi^+\pi^-} \nonumber \\
                       &&-2\sum\limits_{i=1}^8 {\rm ln}P(N^{\rm obs}_i,\langle N_i^{\rm exp}\rangle)_{CP,K^0_L\pi^+\pi^-} \nonumber \\
                       &&-2\sum\limits_{n=1}^{72} {\rm ln}P(N^{\rm obs}_{n},\langle N^{\rm exp}_{n}\rangle)_{K^0_S\pi^+\pi^-,K^0_S\pi^+\pi^-} \nonumber \\
                       &&-2\sum\limits_{n=1}^{128} {\rm ln}P(N^{\rm obs\prime}_{n},\langle N^{\rm exp}_{n}\rangle)_{K^0_L\pi^+\pi^-,K^0_S\pi^+\pi^-} \nonumber \\
                       &&+\chi^2,
\label{eq:likelihood}
\end{eqnarray}
where $P(N^{\rm obs},\langle N^{\rm exp} \rangle)$ is the Poisson probability to observe $N^{\rm obs}$ events given the expected number $\langle N^{\rm exp} \rangle$.
The observed yield of signal and peaking background in the $i$th bin or $n$th bin-pair is denoted $N^{\rm obs}_{i,n}$, and $N^{\rm exp}_{i,n}$ is defined to account for both expected signal and peaking background from the same region.
Biases can occur in the case where $N^{\rm obs}_{i,n}$ is close to zero. To mitigate this effect the observed and expected yields of the three selections of $D\rightarrow K^0_S\pi^+\pi^-$ {\it vs.} $D\rightarrow K^0_S\pi^+\pi^-$ DT candidates are summed together. The observed and expected yields of the two final states of the $D\rightarrow K^0_S\eta$ tag are also added together and the same is done for both final states of the $D\rightarrow K^0_S\eta^{\prime}$ tag. The $\chi^2$ term in Eq.~(\ref{eq:likelihood}) is
\begin{equation}
\chi^2=\sum\limits_{i}(\frac{c_i^{\prime}-c_i-\Delta c_i}{\delta\Delta c_i})^2+\sum\limits_{i}(\frac{s_i^{\prime}-s_i-\Delta s_i}{\delta\Delta s_i})^2,
\end{equation}
which constrains the measured differences $c_i^{\prime}-c_i$ $(s_i^{\prime}-s_i)$ to the predicted differences, $\Delta c_i$ ($\Delta s_i$), where $\delta\Delta c_i$ ($\delta\Delta s_i$) are the uncertainties in the predictions. The presence of the constraint is necessary in order to improve the precision of $s_i$ and $s_i^{\prime}$, and introduces very weak model assumptions in the fit.
The expected values of $c_i$ and $s_i$ are determined from the $D\rightarrow K^0_S\pi^+\pi^-$ amplitude model in Ref.~\cite{prd98_112012}. The expected values of $c_i^{\prime}$ and $s_i^{\prime}$ are determined from
the assumed $D\rightarrow K^0_L\pi^+\pi^-$ amplitude model described in Sec.~\ref{sec:kikip}, where the central values come from the mean of the strong-phase distributions generated using different values of $r$ and $\delta$. In order to determine $\delta\Delta c_i$ and $\delta\Delta s_i$, the values of  $\Delta c_i$ and $\Delta s_i$ are also estimated using the models of $D\rightarrow K^0_S\pi^+\pi^-$ reported in Refs.~\cite{prd81_112002,prl95_121802} with the same transformation to estimate the $D\rightarrow K^0_L\pi^+\pi^-$ decay model. In order to assign $\delta\Delta c_i$ and $\delta\Delta s_i$, the larger deviation of the central values of $\Delta c_i$ and $\Delta s_i$ using these two alternative models is taken as part of the uncertainty and added in quadrature to the uncertainty from the choice of $r$ and $\delta$.  The $CP$-tagged data can also be used to fit only $c_i$ and $c_i^{\prime}$ where the likelihood does not contain a constraint on the difference between these parameters. The measured differences from this fit are consistent with the predicted values of $\Delta c_i$, which gives further confidence in the transformations used to define the $D\rightarrow K^0_L\pi^+\pi^-$ amplitude model. Table~\ref{tab:cisimodel} summarizes the expected $c_i$, $s_i$, $\Delta c_i$ and $\Delta s_i$ and uncertainties for the three binning schemes.
\begin{table*}
\caption{The predicted values of $c_i$, $s_i$, $\Delta c_i$ and $\Delta s_i$ for three binning schemes using the model reported in Ref.~\cite{prd98_112012}. } \begin{center}
\resizebox{!}{1.8cm}{
\begin{tabular}{l|rrrr|rrrr|rrrr} \hline\hline
  &  \multicolumn{4}{c|}{Equal $\Delta\delta_D$ binning}        & \multicolumn{4}{c|}{Optimal binning}                                    & \multicolumn{4}{c}{Modified optimal binning}  \\ \hline
bin & $c_i$~~    & $s_i$~~    & $\Delta c_i$~~~~~   & $\Delta s_i$~~~~~   & $c_i$~~    & $s_i$~~    & $\Delta c_i$~~~~~   & $\Delta s_i$~~~~~ & $c_i$~~    & $s_i$~~    & $\Delta c_i$~~~~~   & $\Delta s_i$~~~~~  \\ \hline
1 &$~~0.662$ &$~~0.003$ & $~~0.11\pm0.03$ &$~~0.01\pm0.04$ &$ -0.018$ &$ -0.811$ &$~~0.34\pm0.10$ &$~~0.08\pm0.07$ &$ -0.356$ &$ -0.282$ &$~~0.16\pm0.12$ &$ -0.05\pm0.10$ \\
2 &$~~0.622$ &$~~0.423$ & $~~0.17\pm0.02$ &$ -0.06\pm0.07$ &$~~0.844$ &$ -0.133$ &$~~0.12\pm0.05$ &$~~0.09\pm0.12$ &$~~0.805$ &$ -0.005$ &$~~0.12\pm0.01$ &$~~0.01\pm0.04$ \\
3 &$~~0.094$ &$~~0.828$ & $~~0.28\pm0.08$ &$ -0.05\pm0.06$ &$~~0.187$ &$ -0.865$ &$~~0.63\pm0.04$ &$~~0.38\pm0.20$ &$~~0.068$ &$ -0.727$ &$~~0.50\pm0.07$ &$~~0.17\pm0.14$ \\
4 &$ -0.505$ &$~~0.751$ & $~~0.12\pm0.09$ &$~~0.06\pm0.05$ &$ -0.913$ &$ -0.080$ &$~~0.03\pm0.04$ &$ -0.02\pm0.06$ &$ -0.943$ &$ -0.112$ &$~~0.03\pm0.02$ &$ -0.03\pm0.04$ \\
5 &$ -0.948$ &$ -0.035$ & $~~0.02\pm0.02$ &$ -0.02\pm0.05$ &$ -0.155$ &$~~0.857$ &$~~0.18\pm0.12$ &$~~0.01\pm0.04$ &$ -0.354$ &$~~0.807$ &$~~0.13\pm0.10$ &$~~0.04\pm0.04$ \\
6 &$ -0.574$ &$ -0.562$ & $~~0.24\pm0.13$ &$ -0.06\pm0.06$ &$~~0.362$ &$~~0.794$ &$~~0.39\pm0.16$ &$ -0.26\pm0.11$ &$~~0.257$ &$~~0.782$ &$~~0.34\pm0.08$ &$ -0.13\pm0.09$ \\
7 &$~~0.027$ &$ -0.794$ & $~~0.49\pm0.09$ &$~~0.15\pm0.12$ &$~~0.864$ &$~~0.206$ &$~~0.04\pm0.01$ &$ -0.03\pm0.05$ &$~~0.713$ &$~~0.231$ &$~~0.09\pm0.03$ &$ -0.02\pm0.06$ \\
8 &$~~0.442$ &$ -0.403$ & $~~0.25\pm0.04$ &$~~0.11\pm0.08$ &$~~0.857$ &$ -0.333$ &$~~0.01\pm0.04$ &$~~0.06\pm0.13$ &$~~0.784$ &$ -0.378$ &$~~0.03\pm0.04$ &$~~0.08\pm0.11$ \\
\hline\hline
\end{tabular}
\label{tab:cisimodel}
}
\end{center}
\end{table*}

To resolve the ambiguity in the sign of $s_i$ present in Eqs.~(\ref{eq:mijexp}) and~(\ref{eq:mijpexp}), the starting values of the parameters of the fit are set to be consistent with the model prediction. An iterative fit is performed to the data. After each iteration the expectation values of the peaking backgrounds that use $c_i$ and $s_i$ as input are recalculated using the $c_i$ and $s_i$ values determined by the fit. Three iterations are required to provide a stable result. The fitted strong-phase parameters $c_i$, $s_i$, $c_i^{\prime}$, and $s_i^{\prime}$ are summarized in Table~\ref{tab:cisifit_final}, in which both statistical and systematic uncertainties are included. Pseudo-experiments are used to validate the fit procedure. For each pseudo-experiment the simulated data yields in each bin are generated according a Poisson distribution based on the expectation using the values of $c_i^{(\prime)}$ and $s_i^{(\prime)}$ found in data. The resultant pull distributions for all strong-phase parameters are found to be consistent with normal distributions and hence the fit procedure is unbiased and returns Gaussian uncertainties.

Furthermore, several checks are performed to assess the stability of the fit results.
The fits are repeated on different subsets of the data, for example, separating partially and fully-reconstructed $K^0_S\pi^+\pi^-$ events. Further tests involve removing specific tags, such as $\pi^+\pi^-\pi^0$, $K^0_L\pi^0$ and $K^0_L\pi^0\pi^0$. The results from these checks are consistent with the default values of $c_i^{(\prime)}$ and $s_i^{(\prime)}$.
Furthermore, the $c_i$ and $s_i$ results are found to be robust when $K^0_L\pi^+\pi^-$ tags are removed from the fit.

\begin{table*}
\caption{ The measured strong-phase difference parameters $c_i$, $s_i$,  $c^{\prime}_i$  and $s^{\prime}_i$, where the first uncertainties are statistical, including that related to the $\Delta c_i$ and $\Delta s_i$ constraints, and the second are systematic. }
\begin{center}
\begin{tabular}{crrrr} \hline
\hline
    & \multicolumn{4}{c}{Equal $\Delta\delta_D$ binning} \\
    & $c_i$~~~~~~~~~~~~~~~~~  & $s_i$~~~~~~~~~~~~~~~~~  & $c^{\prime}_i$~~~~~~~~~~~~~~~~~  &  $s^{\prime}_i$~~~~~~~~~~~~~~~~~  \\
\hline
1 & $~~0.708\pm0.020\pm0.009$ & $~~0.128\pm0.076\pm0.017$ & $~~0.801\pm0.020\pm0.013$ & $~~0.137\pm0.078\pm0.017$\\
2 & $~~0.671\pm0.035\pm0.016$ & $~~0.341\pm0.134\pm0.015$ & $~~0.848\pm0.036\pm0.016$ & $~~0.279\pm0.137\pm0.016$\\
3 & $~~0.001\pm0.047\pm0.019$ & $~~0.893\pm0.112\pm0.020$ & $~~0.174\pm0.047\pm0.016$ & $~~0.840\pm0.118\pm0.021$\\
4 & $ -0.602\pm0.053\pm0.017$ & $~~0.723\pm0.143\pm0.022$ & $ -0.504\pm0.055\pm0.019$ & $~~0.784\pm0.147\pm0.022$\\
5 & $ -0.965\pm0.019\pm0.013$ & $~~0.020\pm0.081\pm0.009$ & $ -0.972\pm0.021\pm0.017$ & $ -0.008\pm0.089\pm0.009$\\
6 & $ -0.554\pm0.062\pm0.024$ & $ -0.589\pm0.147\pm0.031$ & $ -0.387\pm0.069\pm0.025$ & $ -0.642\pm0.152\pm0.034$\\
7 & $~~0.046\pm0.057\pm0.023$ & $ -0.686\pm0.143\pm0.028$ & $~~0.462\pm0.056\pm0.019$ & $ -0.550\pm0.159\pm0.030$\\
8 & $~~0.403\pm0.036\pm0.017$ & $ -0.474\pm0.091\pm0.027$ & $~~0.640\pm0.036\pm0.015$ & $ -0.399\pm0.099\pm0.026$\\
\hline
    & \multicolumn{4}{c}{Optimal binning} \\
    & $c_i$~~~~~~~~~~~~~~~~~  & $s_i$~~~~~~~~~~~~~~~~~  & $c^{\prime}_i$~~~~~~~~~~~~~~~~~  &  $s^{\prime}_i$~~~~~~~~~~~~~~~~~  \\
\hline
1 & $ -0.034\pm0.052\pm0.017$ & $ -0.899\pm0.094\pm0.030$ & $~~0.240\pm0.054\pm0.014$ & $ -0.854\pm0.106\pm0.032$\\
2 & $~~0.839\pm0.062\pm0.037$ & $ -0.272\pm0.166\pm0.031$ & $~~0.927\pm0.054\pm0.036$ & $ -0.298\pm0.162\pm0.029$\\
3 & $~~0.140\pm0.064\pm0.028$ & $ -0.674\pm0.172\pm0.038$ & $~~0.742\pm0.060\pm0.030$ & $ -0.350\pm0.180\pm0.039$\\
4 & $ -0.904\pm0.021\pm0.009$ & $ -0.065\pm0.062\pm0.006$ & $ -0.930\pm0.023\pm0.019$ & $ -0.075\pm0.075\pm0.007$\\
5 & $ -0.300\pm0.042\pm0.013$ & $~~1.047\pm0.055\pm0.019$ & $ -0.173\pm0.043\pm0.010$ & $~~1.053\pm0.062\pm0.018$\\
6 & $~~0.303\pm0.088\pm0.027$ & $~~0.884\pm0.191\pm0.043$ & $~~0.554\pm0.073\pm0.032$ & $~~0.605\pm0.184\pm0.043$\\
7 & $~~0.927\pm0.016\pm0.008$ & $~~0.228\pm0.066\pm0.015$ & $~~0.975\pm0.017\pm0.008$ & $~~0.198\pm0.071\pm0.014$\\
8 & $~~0.771\pm0.032\pm0.015$ & $ -0.316\pm0.123\pm0.021$ & $~~0.798\pm0.035\pm0.017$ & $ -0.253\pm0.141\pm0.019$\\
\hline
    & \multicolumn{4}{c}{Modified optimal binning} \\
    & $c_i$~~~~~~~~~~~~~~~~~  & $s_i$~~~~~~~~~~~~~~~~~  & $c^{\prime}_i$~~~~~~~~~~~~~~~~~  &  $s^{\prime}_i$~~~~~~~~~~~~~~~~~  \\
\hline
1 & $ -0.270\pm0.061\pm0.019$ & $ -0.140\pm0.168\pm0.028$ & $ -0.198\pm0.067\pm0.025$ & $ -0.209\pm0.181\pm0.028$\\
2 & $~~0.829\pm0.027\pm0.018$ & $ -0.014\pm0.100\pm0.018$ & $~~0.945\pm0.026\pm0.018$ & $ -0.019\pm0.100\pm0.017$\\
3 & $~~0.038\pm0.044\pm0.021$ & $ -0.796\pm0.095\pm0.020$ & $~~0.477\pm0.040\pm0.019$ & $ -0.709\pm0.119\pm0.028$\\
4 & $ -0.963\pm0.020\pm0.009$ & $ -0.202\pm0.080\pm0.014$ & $ -0.948\pm0.021\pm0.013$ & $ -0.235\pm0.086\pm0.014$\\
5 & $ -0.460\pm0.044\pm0.012$ & $~~0.899\pm0.078\pm0.021$ & $ -0.359\pm0.046\pm0.011$ & $~~0.943\pm0.084\pm0.022$\\
6 & $~~0.130\pm0.055\pm0.017$ & $~~0.832\pm0.131\pm0.031$ & $~~0.333\pm0.051\pm0.019$ & $~~0.701\pm0.137\pm0.029$\\
7 & $~~0.762\pm0.025\pm0.012$ & $~~0.178\pm0.094\pm0.016$ & $~~0.878\pm0.026\pm0.015$ & $~~0.188\pm0.098\pm0.016$\\
8 & $~~0.699\pm0.035\pm0.012$ & $ -0.085\pm0.141\pm0.018$ & $~~0.740\pm0.037\pm0.014$ & $ -0.025\pm0.149\pm0.019$\\
\hline\hline
\end{tabular}
\label{tab:cisifit_final}
\end{center}
\end{table*}

\section{Systematic uncertainties}
\label{sec:syst}

Uncertainties associated with the selection, tracking and PID efficiencies do not bias the measurement as the expected DT yields are calculated using the ST yields and determined values of $K_i^{(\prime)}$. This use of data-driven quantities to provide the normalization means that detector effects on the common selection affect the observed and expected DT yields in the same way, and hence these systematic uncertainties are not considered further.

Uncertainties on the ST yields, the $K_i^{(\prime)}$ parameters, and the efficiency matrices have an impact on the expected yields. Systematic uncertainties on the ST yields are determined by alternative fits to the ${\rm M}_{\rm BC}$ distribution, in which the endpoint of the ARGUS function and the number of bins in the ${\rm M}_{\rm BC}$ distribution are varied. An alternative data-driven method is used to determine the dominant peaking backgrounds. For example, the rate of the background from $D\rightarrow K^0_S\pi^+\pi^-\pi^0$ decays in $D\rightarrow K^0_S\omega$ candidates is determined by analyzing the ${\rm M}_{\rm BC}$ distribution of candidates whose $\pi^+\pi^-\pi^0$ invariant mass falls in the side band of the reconstructed $\omega$ candidate mass distribution. The difference between this estimate and the nominal one from simulation is assigned as a systematic uncertainty. The total uncertainty on the measured ST yields comes from these sources added in quadrature to the statistical uncertainty from the fit. For the calculated yields, the uncertainty comes from the propagated uncertainties on $N_{D\bar{D}}$, the BFs, and efficiencies. The impact of the uncertainties on the ST yields is investigated by performing multiple fits to data, where in each fit the ST yields used to calculate the expectation are varied according to their uncertainty. The resulting width of the distribution of the values of the strong-phase parameters is assigned as the systematic uncertainty associated with the ST yields.

The statistical uncertainties on the measured $K_i^{(\prime)}$ are propagated in a similar way, where the correlations between the measurements are taken into account. Systematic uncertainties also arise from the DCS correction factors, $f_i^{(\prime)}$, used to determine the $K^{(\prime)}$ parameters. The uncertainties on $K_i^{(\prime)}$ are assigned by varying the input parameters in Table~\ref{tab:kicorpara} and by assessing the impact of using the alternative $D\rightarrow K^0_S\pi^+\pi^-$ models reported in Refs.~\cite{prd81_112002} and ~\cite{prl95_121802} to calculate the $f_i^{(\prime)}$ factors. These systematic uncertainties on the $K_i^{(\prime)}$ are propagated to the strong-phase parameters. These different uncertainties on $K_i^{(\prime)}$ are combined in Tables~\ref{tab:systot_equal} $-$ ~\ref{tab:systot_mdoptimal}, where the statistical contribution is dominant.

A difference in the resolution between simulation and data introduces an uncertainty in the efficiency matrices. The difference in resolution is quantified by studying the mass spectrum of the $K^{*}(892)$ resonance found in $D\rightarrow K^0_S\pi^+\pi^-$ decays. The mass spectrum is fitted with a shape determined by simulation convolved with a Gaussian function, which defines the difference in resolution between data and simulation. The Gaussian has a mean of 0.23~MeV/$c^2$ and width 0.21~MeV/$c^2$. The variables $M_{K^0_{S}\pi^-}$ and $M_{K^0_{S}\pi^+}$ of all simulated events used to determine the efficiency matrices are smeared by a Gaussian with these parameters and new efficiency matrices are calculated. The same procedure is performed on the mass spectrum of $K^{*}(892)$ from $D\rightarrow K^0_L\pi^+\pi^-$ decays and the differences here are described with a Gaussian with mean 4.0~MeV/$c^2$ and width 2.0~MeV/$c^2$. The fit to determine the strong-phase parameters is repeated with the new efficiency matrices and the differences between these fit results and the nominal values are assigned as the systematic uncertainty due to residual differences between the momentum resolution in data and simulation.
The impact of finite samples of simulated data to determine the efficiency matrices on the strong phases is assessed by varying the matrix elements by their statistical uncertainties.
This is repeated multiple times, and the data are refitted using these new matrices to determine the expected yields. The resulting width of the distribution of the values of the strong-phase parameters is assigned as the systematic uncertainty due to the size of the simulated samples.

The expectation values of the peaking background have systematic uncertainties due to the inputs used to calculate their integrated yields and the assumptions concerning the distribution over the Dalitz plot. For the uncertainty from the integrated yields, the expected yield of peaking background in each phase space region is varied according to a Gaussian distribution. This distribution has the nominal value of the peaking-background yield as the mean, and a width which combines the uncertainties from the BFs of peaking-background decays, and the uncertainties arising from tracking~\cite{prd99_011103}, PID~\cite{prd99_011103}, and $\pi^0$ reconstruction efficiencies~\cite{1903.04118}. The distributions over the Dalitz plots for peaking backgrounds that
are $CP$ eigenstates, or $D\rightarrow K^0_S\pi^+\pi^-$ for $D\rightarrow K^0_L\pi^+\pi^-$ signals, are dependent on the values of $c_i$ and $s_i$.
As the iterative fit procedure recalculates the peaking background with updated values of $c_i$ and $s_i$, no further systematic uncertainty is assigned for these backgrounds. The $D\rightarrow \pi^+\pi^-\pi^0\pi^0$ peaking background constitutes a significant contribution to the observed yields in $D\rightarrow K^0_S\pi^+\pi^-$ {\it vs.} $D\rightarrow K^0_S\pi^+\pi^-$ where one $K^0_S$ meson
decays to the $\pi^0\pi^0$ final state. To find an alternative distribution of this background a DT sample of $D\rightarrow K^0_S\pi^+\pi^-$ {\it vs. } $D\rightarrow \pi^+\pi^-\pi^0\pi^0$ events is fully reconstructed in data. The distribution over the Dalitz plot is found by assigning the $K^0_S$ mass to the $\pi^0$ pair. This distribution is used instead of the nominal one (from simulation) in the fit and small shifts are observed in the strong-phase parameters that are assigned as an additional contribution to the systematic uncertainties arising from the DT peaking backgrounds. Additionally, in Figs.~\ref{fig:dtmix}(a) and \ref{fig:dtmix}(d), a few peaking backgrounds of the fitted combinatorial curves are not included in the nominal fit to extract the strong-phase parameters. To estimate their effects, a new fit is performed by including these peaking backgrounds and the difference between the resulting fitted results and nominal values are taken as the other sources of systematic uncertainties. 

The effects from $D^0\bar{D}^0$ mixing are not considered in the nominal fit. The required correction factor for $CP_{\pm}$ eigenstate ST yields is $1/(1-\eta_{\pm}y_D)$, where $\eta_{\pm}=\pm1$ and the mixing parameter $y_D=(0.62\pm0.08)\%$~\cite{pdg18}. The data are fitted using the corrected ST yields and the difference with respect to the nominal results is assigned as the systematic uncertainty due to charm mixing.

Systematic uncertainties in the observed DT yields arise from the fit procedure and the description of combinatorial background. Due to the low candidate yields in multiple of the phase space regions, small biases in the fitted yields can be present. The sizes of these biases are determined in pseudo-experiments. An alternative combinatorial background shape is employed and the difference in $N^{\rm obs}$ between this fit and the nominal is added in quadrature to the bias estimate to determine the systematic uncertainty on the observed yields. All the observed yields are smeared within these uncertainties and the fit is repeated. The resulting width of the distribution of values of the strong-phase parameters is assigned as the systematic uncertainty due to the DT yields.

The systematic uncertainties of the measured strong-phase parameters $c_i$, $s_i$, $c_i^{\prime}$ and $s_i^{\prime}$ for the equal $\Delta\delta_D$, optimal, and modified optimal binning schemes are summarized in Tables~\ref{tab:systot_equal}, ~\ref{tab:systot_optimal} and ~\ref{tab:systot_mdoptimal}, respectively. There is no source of systematic uncertainty that is dominant for all strong-phase parameters. The statistical uncertainty obtained from the fit includes the contribution related to the associated uncertainties on $\Delta c_i$ and $\Delta s_i$ through the $\chi^2$ term of Eq.~(\ref{eq:likelihood}). In order to estimate this contribution, the fit is repeated in a configuration where $\Delta c_i$ and $\Delta s_i$ are fixed. The difference in quadrature between the uncertainties from this fit and the nominal approach provides an estimate of the contribution to the uncertainty from the constraint. This estimate is also given in Tables~\ref{tab:systot_equal}, ~\ref{tab:systot_optimal} and ~\ref{tab:systot_mdoptimal}, and it is seen that this contribution to the overall uncertainty is small. The measurements of the strong-phase parameters are limited by their statistical uncertainties. The correlation matrices for the statistical and systematic uncertainties associated with different binning schemes are given in Tables~\ref{tab:sta_equal_corela}$-$\ref{tab:sys_mdoptimal_corela}.

\begin{figure*}[tp!]
\begin{center}
\includegraphics[width=0.8\linewidth]{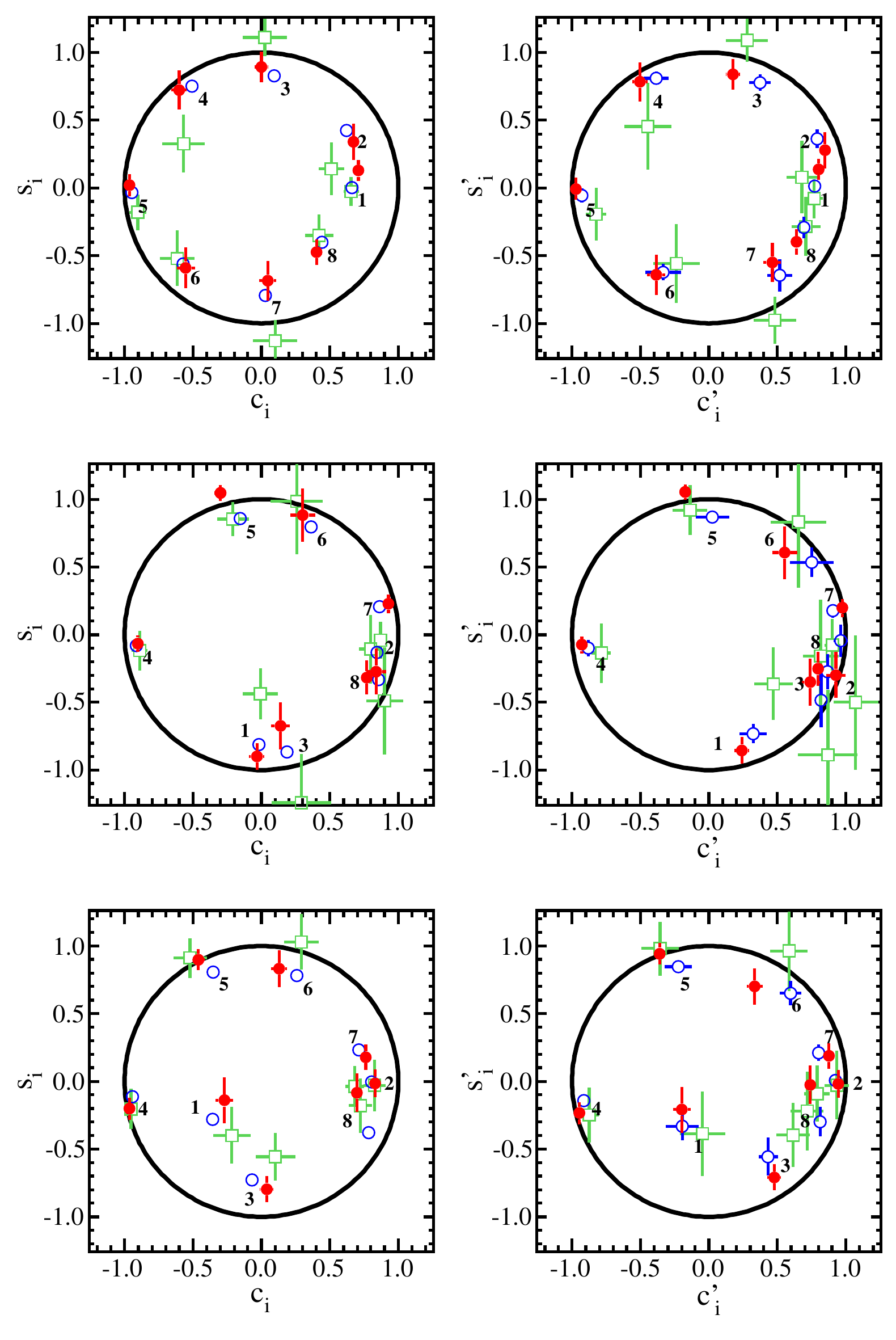}
\caption{
   The $c^{(\prime)}_i$ and $s^{(\prime)}_i$ measured in this work (red dots with error bars), the expected values from Ref.~\cite{prd98_112012} (blue open circles) as well as CLEO results (green open squares with error bars) in Ref.~\cite{prd82_112006}.
   The top plots are from the equal $\Delta\delta_D$ binning, the middle plots from the optimal binning and plots from the modified optimal binning scheme are on the bottom. The circle indicates the boundary of the physical region $c^{(\prime)2}_i+s^{(\prime)2}_i=1$.
}
\label{fig:cisi}
\end{center}
\end{figure*}

The measurements are displayed in Fig.~\ref{fig:cisi}, together with the model predictions from Ref.~\cite{prd98_112012}, which are seen to be in reasonable agreement.
Given the compatibility between the current measurements and those reported by the CLEO collaboration~\cite{prd82_112006} an additional set of fits is performed, where the CLEO results are imposed as a Gaussian constraint in Eq.~(\ref{eq:likelihood}). These results are presented in Appendix A.

\begin{table*}
\caption{ The uncertainties for $c_i$, $s_i$, $c^{\prime}_i$ and $s^{\prime}_i$ for the equal $\Delta\delta_D$ binning scheme.}
\begin{center}
\begin{tabular}{lccccccccc} \hline\hline
Uncertainty  & ~~~~~~$c_1$~~~~~~ & ~~~~~~$c_2$~~~~~~ & ~~~~~~$c_3$~~~~~~ & ~~~~~~$c_4$~~~~~~ & ~~~~~~$c_5$~~~~~~ & ~~~~~~$c_6$~~~~~~ & ~~~~~~$c_7$~~~~~~ & ~~~~~~$c_8$~~~~~~ \\
\hline
$K_i$ and $K_i^{\prime}$         & 0.004 & 0.013 & 0.005 & 0.007 & 0.005 & 0.014 & 0.006 & 0.007 \\
ST yields                         & 0.007 & 0.007 & 0.013 & 0.008 & 0.004 & 0.014 & 0.019 & 0.011 \\
MC statistics                     & 0.001 & 0.003 & 0.003 & 0.003 & 0.001 & 0.004 & 0.004 & 0.003 \\
DT peaking-background subtraction & 0.002 & 0.003 & 0.002 & 0.007 & 0.005 & 0.007 & 0.003 & 0.002 \\
DT yields                         & 0.001 & 0.002 & 0.002 & 0.001 & 0.001 & 0.002 & 0.003 & 0.002 \\
Momentum resolution               & 0.002 & 0.003 & 0.012 & 0.011 & 0.010 & 0.010 & 0.011 & 0.009 \\
$D^0\bar{D}^0$ mixing           & 0.001 & 0.000 & 0.002 & 0.001 & 0.000 & 0.002 & 0.002 & 0.001 \\
\hline
Total systematic                  & 0.009 & 0.016 & 0.019 & 0.017 & 0.013 & 0.024 & 0.023 & 0.017 \\
Statistical plus $K^0_L\pi^+\pi^-$ model &  0.020 & 0.035 & 0.047 & 0.053 & 0.019 & 0.062 & 0.057 & 0.036 \\
\hline
$K^0_L\pi^+\pi^-$ model alone            &0.011 & 0.009 & 0.027 & 0.030 & 0.007 & 0.034 & 0.033 & 0.017 \\
\hline
Total                                      &0.022 & 0.039 & 0.051 & 0.055 & 0.023 & 0.066 & 0.061 & 0.039 \\
\hline
\\
Uncertainty  &~~~~~~$s_1$~~~~~~ & ~~~~~~$s_2$~~~~~~ & ~~~~~~$s_3$~~~~~~ & ~~~~~~$s_4$~~~~~~ & ~~~~~~$s_5$~~~~~~ & ~~~~~~$s_6$~~~~~~ & ~~~~~~$s_7$~~~~~~ & ~~~~~~$s_8$~~~~~~ \\
\hline
$K_i$ and $K_i^{\prime}$         & 0.004 & 0.006 & 0.012 & 0.005 & 0.003 & 0.018 & 0.022 & 0.008 \\
ST yields                         & 0.001 & 0.001 & 0.001 & 0.001 & 0.001 & 0.002 & 0.001 & 0.001 \\
MC statistics                     & 0.007 & 0.011 & 0.009 & 0.010 & 0.005 & 0.009 & 0.011 & 0.006 \\
DT peaking-background subtraction & 0.007 & 0.005 & 0.007 & 0.018 & 0.005 & 0.009 & 0.011 & 0.004 \\
DT yields                         & 0.005 & 0.005 & 0.003 & 0.004 & 0.003 & 0.004 & 0.005 & 0.003 \\
Momentum resolution               & 0.012 & 0.005 & 0.011 & 0.001 & 0.003 & 0.022 & 0.006 & 0.025 \\
$D^0\bar{D}^0$ mixing           & 0.000 & 0.000 & 0.000 & 0.001 & 0.000 & 0.000 & 0.000 & 0.000 \\
\hline
Total systematic                  & 0.017 & 0.015 & 0.020 & 0.022 & 0.009 & 0.031 & 0.028 & 0.027 \\
Statistical plus $K^0_L\pi^+\pi^-$ model &  0.076 & 0.134 & 0.112 & 0.143 & 0.081 & 0.147 & 0.143 & 0.091 \\
\hline
$K^0_L\pi^+\pi^-$ model alone            &0.017 & 0.029 & 0.022 & 0.018 & 0.012 & 0.017 & 0.036 & 0.028 \\
\hline
Total                                      &0.078 & 0.135 & 0.114 & 0.144 & 0.081 & 0.150 & 0.146 & 0.095 \\
\hline
\\
Uncertainty  & ~~~~~~$c^{\prime}_1$~~~~~~ & ~~~~~~$c^{\prime}_2$~~~~~~ & ~~~~~~$c^{\prime}_3$~~~~~~ & ~~~~~~$c^{\prime}_4$~~~~~~ & ~~~~~~$c^{\prime}_5$~~~~~~ & ~~~~~~$c^{\prime}_6$~~~~~~ & ~~~~~~$c^{\prime}_7$~~~~~~ & ~~~~~~$c^{\prime}_8$~~~~~~ \\
\hline
$K_i$ and $K_i^{\prime}$         & 0.006 & 0.014 & 0.009 & 0.006 & 0.005 & 0.012 & 0.014 & 0.009 \\
ST yields                         & 0.003 & 0.006 & 0.005 & 0.004 & 0.003 & 0.005 & 0.007 & 0.006 \\
MC statistics                     & 0.002 & 0.003 & 0.004 & 0.004 & 0.001 & 0.006 & 0.005 & 0.003 \\
DT peaking-background subtraction & 0.003 & 0.004 & 0.003 & 0.008 & 0.010 & 0.009 & 0.005 & 0.002 \\
DT yields                         & 0.002 & 0.002 & 0.003 & 0.002 & 0.001 & 0.004 & 0.004 & 0.003 \\
Momentum resolution               & 0.010 & 0.003 & 0.009 & 0.015 & 0.012 & 0.017 & 0.001 & 0.009 \\
$D^0\bar{D}^0$ mixing           & 0.002 & 0.001 & 0.004 & 0.002 & 0.000 & 0.005 & 0.006 & 0.003 \\
\hline
Total systematic                  & 0.013 & 0.016 & 0.016 & 0.019 & 0.017 & 0.025 & 0.019 & 0.015 \\
Statistical plus $K^0_L\pi^+\pi^-$ model &  0.020 & 0.036 & 0.047 & 0.055 & 0.021 & 0.069 & 0.056 & 0.036 \\
\hline
$K^0_L\pi^+\pi^-$ model alone            &0.012 & 0.010 & 0.028 & 0.033 & 0.011 & 0.046 & 0.032 & 0.017 \\
\hline
Total                                      &0.024 & 0.039 & 0.050 & 0.058 & 0.027 & 0.073 & 0.059 & 0.039 \\
\hline
\\
Uncertainty  & ~~~~~~$s^{\prime}_1$~~~~~~ & ~~~~~~$s^{\prime}_2$~~~~~~ & ~~~~~~$s^{\prime}_3$~~~~~~ & ~~~~~~$s^{\prime}_4$~~~~~~ & ~~~~~~$s^{\prime}_5$~~~~~~ & ~~~~~~$s^{\prime}_6$~~~~~~ & ~~~~~~$s^{\prime}_7$~~~~~~ & ~~~~~~$s^{\prime}_8$~~~~~~ \\
\hline
$K_i$ and $K_i^{\prime}$         & 0.005 & 0.006 & 0.012 & 0.005 & 0.003 & 0.019 & 0.024 & 0.010 \\
ST yields                         & 0.002 & 0.001 & 0.001 & 0.001 & 0.000 & 0.002 & 0.001 & 0.001 \\
MC statistics                     & 0.007 & 0.011 & 0.009 & 0.010 & 0.005 & 0.009 & 0.013 & 0.007 \\
DT peaking-background subtraction & 0.007 & 0.004 & 0.008 & 0.019 & 0.005 & 0.010 & 0.009 & 0.004 \\
DT yields                         & 0.005 & 0.005 & 0.003 & 0.004 & 0.003 & 0.004 & 0.006 & 0.004 \\
Momentum resolution               & 0.011 & 0.006 & 0.012 & 0.000 & 0.004 & 0.024 & 0.007 & 0.022 \\
$D^0\bar{D}^0$ mixing           & 0.000 & 0.000 & 0.000 & 0.001 & 0.000 & 0.000 & 0.000 & 0.000 \\
\hline
Total systematic                  & 0.017 & 0.016 & 0.021 & 0.022 & 0.009 & 0.034 & 0.030 & 0.026 \\
Statistical plus $K^0_L\pi^+\pi^-$ model &  0.078 & 0.137 & 0.118 & 0.147 & 0.089 & 0.152 & 0.159 & 0.099 \\
\hline
$K^0_L\pi^+\pi^-$ model alone            &0.024 & 0.040 & 0.041 & 0.039 & 0.040 & 0.045 & 0.078 & 0.048 \\
\hline
Total                                      &0.080 & 0.137 & 0.119 & 0.147 & 0.090 & 0.156 & 0.162 & 0.103 \\
\hline\hline
\end{tabular}
\label{tab:systot_equal}
\end{center}
\end{table*}

\begin{table*}
\caption{ The uncertainties of $c_i$, $s_i$, $c^{\prime}_i$ and $s^{\prime}_i$ for the optimal binning scheme.}
\begin{center}
\begin{tabular}{lccccccccc} \hline\hline
Uncertainty  & ~~~~~~$c_1$~~~~~~ & ~~~~~~$c_2$~~~~~~ & ~~~~~~$c_3$~~~~~~ & ~~~~~~$c_4$~~~~~~ & ~~~~~~$c_5$~~~~~~ & ~~~~~~$c_6$~~~~~~ & ~~~~~~$c_7$~~~~~~ & ~~~~~~$c_8$~~~~~~ \\
\hline
$K_i$ and $K_i^{\prime}$         & 0.003 & 0.032 & 0.023 & 0.005 & 0.005 & 0.015 & 0.007 & 0.012 \\
ST yields                         & 0.016 & 0.013 & 0.013 & 0.005 & 0.011 & 0.021 & 0.003 & 0.006 \\
MC statistics                     & 0.003 & 0.005 & 0.005 & 0.001 & 0.003 & 0.006 & 0.001 & 0.002 \\
DT peaking-background subtraction & 0.003 & 0.008 & 0.003 & 0.004 & 0.004 & 0.003 & 0.002 & 0.007 \\
DT yields                         & 0.002 & 0.004 & 0.004 & 0.001 & 0.002 & 0.003 & 0.001 & 0.001 \\
Momentum resolution               & 0.004 & 0.007 & 0.002 & 0.005 & 0.001 & 0.001 & 0.002 & 0.001 \\
$D^0\bar{D}^0$ mixing           & 0.003 & 0.002 & 0.003 & 0.001 & 0.003 & 0.005 & 0.000 & 0.000 \\
\hline
Total systematic                  & 0.017 & 0.037 & 0.028 & 0.009 & 0.013 & 0.027 & 0.008 & 0.015 \\
Statistical plus $K^0_L\pi^+\pi^-$ model &  0.052 & 0.062 & 0.064 & 0.021 & 0.042 & 0.088 & 0.016 & 0.032 \\
\hline
$K^0_L\pi^+\pi^-$ model alone            &0.031 & 0.034 & 0.025 & 0.012 & 0.027 & 0.062 & 0.003 & 0.013 \\
\hline
Total                                      &0.055 & 0.073 & 0.070 & 0.023 & 0.044 & 0.092 & 0.018 & 0.035 \\
\hline
\\
Uncertainty  &~~~~~~$s_1$~~~~~~ & ~~~~~~$s_2$~~~~~~ & ~~~~~~$s_3$~~~~~~ & ~~~~~~$s_4$~~~~~~ & ~~~~~~$s_5$~~~~~~ & ~~~~~~$s_6$~~~~~~ & ~~~~~~$s_7$~~~~~~ & ~~~~~~$s_8$~~~~~~ \\
\hline
$K_i$ and $K_i^{\prime}$         & 0.018 & 0.026 & 0.033 & 0.002 & 0.006 & 0.028 & 0.004 & 0.005 \\
ST yields                         & 0.001 & 0.001 & 0.001 & 0.001 & 0.001 & 0.002 & 0.001 & 0.001 \\
MC statistics                     & 0.006 & 0.013 & 0.012 & 0.003 & 0.003 & 0.017 & 0.004 & 0.011 \\
DT peaking-background subtraction & 0.012 & 0.009 & 0.011 & 0.004 & 0.015 & 0.013 & 0.011 & 0.015 \\
DT yields                         & 0.003 & 0.006 & 0.005 & 0.002 & 0.001 & 0.005 & 0.003 & 0.004 \\
Momentum resolution               & 0.021 & 0.001 & 0.008 & 0.000 & 0.009 & 0.024 & 0.008 & 0.005 \\
$D^0\bar{D}^0$ mixing           & 0.000 & 0.000 & 0.000 & 0.000 & 0.001 & 0.000 & 0.000 & 0.000 \\
\hline
Total systematic                  & 0.030 & 0.031 & 0.038 & 0.006 & 0.019 & 0.043 & 0.015 & 0.021 \\
Statistical plus $K^0_L\pi^+\pi^-$ model &  0.094 & 0.166 & 0.172 & 0.062 & 0.055 & 0.191 & 0.066 & 0.123 \\
\hline
$K^0_L\pi^+\pi^-$ model alone            &0.018 & 0.064 & 0.081 & 0.013 & 0.010 & 0.069 & 0.018 & 0.033 \\
\hline
Total                                      &0.099 & 0.169 & 0.176 & 0.062 & 0.058 & 0.196 & 0.068 & 0.125 \\
\hline
\\
Uncertainty  & ~~~~~~$c^{\prime}_1$~~~~~~ & ~~~~~~$c^{\prime}_2$~~~~~~ & ~~~~~~$c^{\prime}_3$~~~~~~ & ~~~~~~$c^{\prime}_4$~~~~~~ & ~~~~~~$c^{\prime}_5$~~~~~~ & ~~~~~~$c^{\prime}_6$~~~~~~ & ~~~~~~$c^{\prime}_7$~~~~~~ & ~~~~~~$c^{\prime}_8$~~~~~~ \\
\hline
$K_i$ and $K_i^{\prime}$         & 0.009 & 0.032 & 0.027 & 0.005 & 0.006 & 0.028 & 0.007 & 0.015 \\
ST yields                         & 0.005 & 0.008 & 0.010 & 0.002 & 0.004 & 0.006 & 0.002 & 0.004 \\
MC statistics                     & 0.005 & 0.005 & 0.005 & 0.001 & 0.004 & 0.007 & 0.001 & 0.003 \\
DT peaking-background subtraction & 0.003 & 0.005 & 0.004 & 0.014 & 0.005 & 0.008 & 0.002 & 0.005 \\
DT yields                         & 0.003 & 0.004 & 0.005 & 0.001 & 0.002 & 0.004 & 0.001 & 0.002 \\
Momentum resolution               & 0.004 & 0.008 & 0.002 & 0.011 & 0.001 & 0.005 & 0.003 & 0.001 \\
$D^0\bar{D}^0$ mixing           & 0.006 & 0.005 & 0.004 & 0.001 & 0.004 & 0.006 & 0.001 & 0.002 \\
\hline
Total systematic                  & 0.014 & 0.036 & 0.030 & 0.019 & 0.010 & 0.032 & 0.008 & 0.017 \\
Statistical plus $K^0_L\pi^+\pi^-$ model &  0.054 & 0.054 & 0.060 & 0.023 & 0.043 & 0.073 & 0.017 & 0.035 \\
\hline
$K^0_L\pi^+\pi^-$ model alone            &0.033 & 0.013 & 0.013 & 0.014 & 0.029 & 0.038 & 0.006 & 0.020 \\
\hline
Total                                      &0.056 & 0.065 & 0.068 & 0.030 & 0.045 & 0.080 & 0.019 & 0.039 \\
\hline
\\
Uncertainty  & ~~~~~~$s^{\prime}_1$~~~~~~ & ~~~~~~$s^{\prime}_2$~~~~~~ & ~~~~~~$s^{\prime}_3$~~~~~~ & ~~~~~~$s^{\prime}_4$~~~~~~ & ~~~~~~$s^{\prime}_5$~~~~~~ & ~~~~~~$s^{\prime}_6$~~~~~~ & ~~~~~~$s^{\prime}_7$~~~~~~ & ~~~~~~$s^{\prime}_8$~~~~~~ \\
\hline
$K_i$ and $K_i^{\prime}$         & 0.019 & 0.023 & 0.031 & 0.003 & 0.006 & 0.025 & 0.004 & 0.008 \\
ST yields                         & 0.001 & 0.001 & 0.001 & 0.001 & 0.001 & 0.002 & 0.001 & 0.001 \\
MC statistics                     & 0.007 & 0.015 & 0.016 & 0.004 & 0.003 & 0.016 & 0.005 & 0.013 \\
DT peaking-background subtraction & 0.013 & 0.007 & 0.005 & 0.004 & 0.012 & 0.012 & 0.010 & 0.009 \\
DT yields                         & 0.003 & 0.006 & 0.005 & 0.002 & 0.001 & 0.005 & 0.003 & 0.005 \\
Momentum resolution               & 0.021 & 0.004 & 0.014 & 0.002 & 0.012 & 0.027 & 0.008 & 0.006 \\
$D^0\bar{D}^0$ mixing           & 0.000 & 0.000 & 0.000 & 0.000 & 0.000 & 0.000 & 0.000 & 0.000 \\
\hline
Total systematic                  & 0.032 & 0.029 & 0.039 & 0.007 & 0.018 & 0.043 & 0.014 & 0.019 \\
Statistical plus $K^0_L\pi^+\pi^-$ model &  0.106 & 0.162 & 0.180 & 0.075 & 0.062 & 0.184 & 0.071 & 0.141 \\
\hline
$K^0_L\pi^+\pi^-$ model alone            &0.051 & 0.054 & 0.097 & 0.045 & 0.030 & 0.044 & 0.031 & 0.076 \\
\hline
Total                                      &0.111 & 0.165 & 0.184 & 0.076 & 0.064 & 0.189 & 0.073 & 0.143 \\
\hline\hline
\end{tabular}
\label{tab:systot_optimal}
\end{center}
\end{table*}

\begin{table*}
\caption{ The uncertainties of $c_i$, $s_i$, $c^{\prime}_i$ and $s^{\prime}_i$ for the modified optimal binning scheme.}
\begin{center}
\begin{tabular}{lccccccccc} \hline\hline
Uncertainty  & ~~~~~~$c_1$~~~~~~ & ~~~~~~$c_2$~~~~~~ & ~~~~~~$c_3$~~~~~~ & ~~~~~~$c_4$~~~~~~ & ~~~~~~$c_5$~~~~~~ & ~~~~~~$c_6$~~~~~~ & ~~~~~~$c_7$~~~~~~ & ~~~~~~$c_8$~~~~~~ \\
\hline
$K_i$ and $K_i^{\prime}$         & 0.007 & 0.014 & 0.005 & 0.005 & 0.005 & 0.008 & 0.006 & 0.009 \\
ST yields                         & 0.013 & 0.006 & 0.018 & 0.004 & 0.008 & 0.014 & 0.005 & 0.007 \\
MC statistics                     & 0.004 & 0.002 & 0.003 & 0.001 & 0.003 & 0.004 & 0.002 & 0.002 \\
DT peaking-background subtraction & 0.005 & 0.004 & 0.005 & 0.004 & 0.006 & 0.002 & 0.002 & 0.003 \\
DT yields                         & 0.002 & 0.002 & 0.003 & 0.001 & 0.001 & 0.002 & 0.001 & 0.001 \\
Momentum resolution               & 0.010 & 0.008 & 0.007 & 0.006 & 0.000 & 0.005 & 0.008 & 0.000 \\
$D^0\bar{D}^0$ mixing           & 0.002 & 0.001 & 0.002 & 0.000 & 0.002 & 0.002 & 0.000 & 0.000 \\
\hline
Total systematic                  & 0.019 & 0.018 & 0.021 & 0.009 & 0.012 & 0.017 & 0.012 & 0.012 \\
Statistical plus $K^0_L\pi^+\pi^-$ model &  0.061 & 0.027 & 0.044 & 0.020 & 0.044 & 0.055 & 0.025 & 0.035 \\
\hline
$K^0_L\pi^+\pi^-$ model alone            &0.034 & 0.006 & 0.028 & 0.008 & 0.027 & 0.033 & 0.011 & 0.015 \\
\hline
Total                                      &0.064 & 0.032 & 0.048 & 0.022 & 0.046 & 0.058 & 0.027 & 0.037 \\
\hline
\\
Uncertainty  &~~~~~~$s_1$~~~~~~ & ~~~~~~$s_2$~~~~~~ & ~~~~~~$s_3$~~~~~~ & ~~~~~~$s_4$~~~~~~ & ~~~~~~$s_5$~~~~~~ & ~~~~~~$s_6$~~~~~~ & ~~~~~~$s_7$~~~~~~ & ~~~~~~$s_8$~~~~~~ \\
\hline
$K_i$ and $K_i^{\prime}$         & 0.010 & 0.011 & 0.014 & 0.004 & 0.005 & 0.013 & 0.005 & 0.008 \\
ST yields                         & 0.001 & 0.002 & 0.001 & 0.001 & 0.002 & 0.002 & 0.001 & 0.001 \\
MC statistics                     & 0.011 & 0.009 & 0.007 & 0.005 & 0.005 & 0.011 & 0.008 & 0.013 \\
DT peaking-background subtraction & 0.023 & 0.009 & 0.005 & 0.013 & 0.019 & 0.021 & 0.008 & 0.007 \\
DT yields                         & 0.006 & 0.005 & 0.004 & 0.003 & 0.002 & 0.004 & 0.005 & 0.006 \\
Momentum resolution               & 0.000 & 0.002 & 0.011 & 0.000 & 0.001 & 0.015 & 0.010 & 0.002 \\
$D^0\bar{D}^0$ mixing           & 0.000 & 0.001 & 0.000 & 0.000 & 0.001 & 0.000 & 0.000 & 0.000 \\
\hline
Total systematic                  & 0.028 & 0.018 & 0.020 & 0.014 & 0.021 & 0.031 & 0.016 & 0.018 \\
Statistical plus $K^0_L\pi^+\pi^-$ model &  0.168 & 0.100 & 0.095 & 0.080 & 0.078 & 0.131 & 0.094 & 0.141 \\
\hline
$K^0_L\pi^+\pi^-$ model alone            &0.029 & 0.021 & 0.037 & 0.010 & 0.013 & 0.035 & 0.026 & 0.041 \\
\hline
Total                                      &0.170 & 0.102 & 0.097 & 0.081 & 0.081 & 0.134 & 0.096 & 0.143 \\
\hline
\\
Uncertainty  & ~~~~~~$c^{\prime}_1$~~~~~~ & ~~~~~~$c^{\prime}_2$~~~~~~ & ~~~~~~$c^{\prime}_3$~~~~~~ & ~~~~~~$c^{\prime}_4$~~~~~~ & ~~~~~~$c^{\prime}_5$~~~~~~ & ~~~~~~$c^{\prime}_6$~~~~~~ & ~~~~~~$c^{\prime}_7$~~~~~~ & ~~~~~~$c^{\prime}_8$~~~~~~ \\
\hline
$K_i$ and $K_i^{\prime}$         & 0.008 & 0.014 & 0.012 & 0.005 & 0.004 & 0.014 & 0.007 & 0.012 \\
ST yields                         & 0.005 & 0.005 & 0.007 & 0.003 & 0.003 & 0.006 & 0.003 & 0.004 \\
MC statistics                     & 0.005 & 0.002 & 0.004 & 0.001 & 0.004 & 0.005 & 0.002 & 0.003 \\
DT peaking-background subtraction & 0.005 & 0.003 & 0.004 & 0.009 & 0.003 & 0.005 & 0.003 & 0.003 \\
DT yields                         & 0.004 & 0.002 & 0.004 & 0.001 & 0.002 & 0.003 & 0.002 & 0.002 \\
Momentum resolution               & 0.021 & 0.008 & 0.011 & 0.008 & 0.007 & 0.007 & 0.013 & 0.000 \\
$D^0\bar{D}^0$ mixing           & 0.004 & 0.002 & 0.006 & 0.001 & 0.003 & 0.004 & 0.002 & 0.002 \\
\hline
Total systematic                  & 0.025 & 0.018 & 0.019 & 0.013 & 0.011 & 0.019 & 0.015 & 0.014 \\
Statistical plus $K^0_L\pi^+\pi^-$ model &  0.067 & 0.026 & 0.040 & 0.021 & 0.046 & 0.051 & 0.026 & 0.037 \\
\hline
$K^0_L\pi^+\pi^-$ model alone            &0.043 & 0.004 & 0.021 & 0.010 & 0.031 & 0.027 & 0.014 & 0.019 \\
\hline
Total                                      &0.071 & 0.032 & 0.044 & 0.025 & 0.048 & 0.055 & 0.030 & 0.039 \\
\hline
\\
Uncertainty  & ~~~~~~$s^{\prime}_1$~~~~~~ & ~~~~~~$s^{\prime}_2$~~~~~~ & ~~~~~~$s^{\prime}_3$~~~~~~ & ~~~~~~$s^{\prime}_4$~~~~~~ & ~~~~~~$s^{\prime}_5$~~~~~~ & ~~~~~~$s^{\prime}_6$~~~~~~ & ~~~~~~$s^{\prime}_7$~~~~~~ & ~~~~~~$s^{\prime}_8$~~~~~~ \\
\hline
$K_i$ and $K_i^{\prime}$         & 0.011 & 0.011 & 0.020 & 0.005 & 0.005 & 0.012 & 0.006 & 0.009 \\
ST yields                         & 0.002 & 0.002 & 0.001 & 0.001 & 0.002 & 0.002 & 0.001 & 0.001 \\
MC statistics                     & 0.011 & 0.009 & 0.010 & 0.005 & 0.005 & 0.011 & 0.008 & 0.013 \\
DT peaking-background subtraction & 0.022 & 0.008 & 0.005 & 0.012 & 0.020 & 0.016 & 0.008 & 0.008 \\
DT yields                         & 0.007 & 0.005 & 0.005 & 0.003 & 0.002 & 0.004 & 0.005 & 0.007 \\
Momentum resolution               & 0.002 & 0.002 & 0.015 & 0.000 & 0.002 & 0.018 & 0.009 & 0.001 \\
$D^0\bar{D}^0$ mixing           & 0.000 & 0.001 & 0.000 & 0.000 & 0.001 & 0.000 & 0.000 & 0.000 \\
\hline
Total systematic                  & 0.028 & 0.017 & 0.028 & 0.014 & 0.022 & 0.029 & 0.016 & 0.019 \\
Statistical plus $K^0_L\pi^+\pi^-$ model &  0.181 & 0.100 & 0.119 & 0.086 & 0.084 & 0.137 & 0.098 & 0.149 \\
\hline
$K^0_L\pi^+\pi^-$ model alone            &0.073 & 0.022 & 0.081 & 0.033 & 0.034 & 0.054 & 0.037 & 0.061 \\
\hline
Total                                      &0.183 & 0.102 & 0.122 & 0.087 & 0.087 & 0.140 & 0.099 & 0.150 \\
\hline\hline
\end{tabular}
\label{tab:systot_mdoptimal}
\end{center}
\end{table*}

\hspace{2cm}
\section{\boldmath  Impact on $\gamma/\phi_3$ measurement}
\label{sec:gamma}

The model-independent measurement of $\gamma$ described in Ref.~\cite{prd68_054018} is performed by comparing the number of $B^-\rightarrow DK^-$, $D\rightarrow K^0_S\pi^+\pi^-$ events in a given Dalitz plot bin with the integral of the square of the amplitude given in Eq.~(\ref{eq:amp_bdecay}) over the same region. An analogous expression for the $B^+$ events is also used.
Therefore the expected yield of $B^{\pm}$ events in a Dalitz plot region is a function of $K_i$, $c_i$, $s_i$, and $\gamma$, $\delta_B$ and $r_B$, the underlying parameters of interest, and is given by
\begin{widetext}
\begin{eqnarray}
N_{\pm i}^{\rm exp}(B^-\rightarrow K^-D_{K^0_S\pi^-\pi^+})=h_B^{-}\left[K_{\pm i}+r^2_BK_{\mp i}+2r_B\sqrt{K_iK_{-i}}\times[c_i{\rm cos}(\delta_B-\gamma)\pm s_i{\rm sin}(\delta_B-\gamma)]\right], \nonumber \\
N_{\pm i}^{\rm exp}(B^+\rightarrow K^+D_{K^0_S\pi^-\pi^+}) =h_B^{+}\left[K_{\mp i}+r^2_BK_{\pm i}+2r_B\sqrt{K_iK_{-i}}\times[c_i{\rm cos}(\delta_B+\gamma)\mp s_i{\rm sin}(\delta_B+\gamma)]\right].
\end{eqnarray}
\end{widetext}

\begin{figure*}[tp!]
\begin{center}
\includegraphics[width=\linewidth]{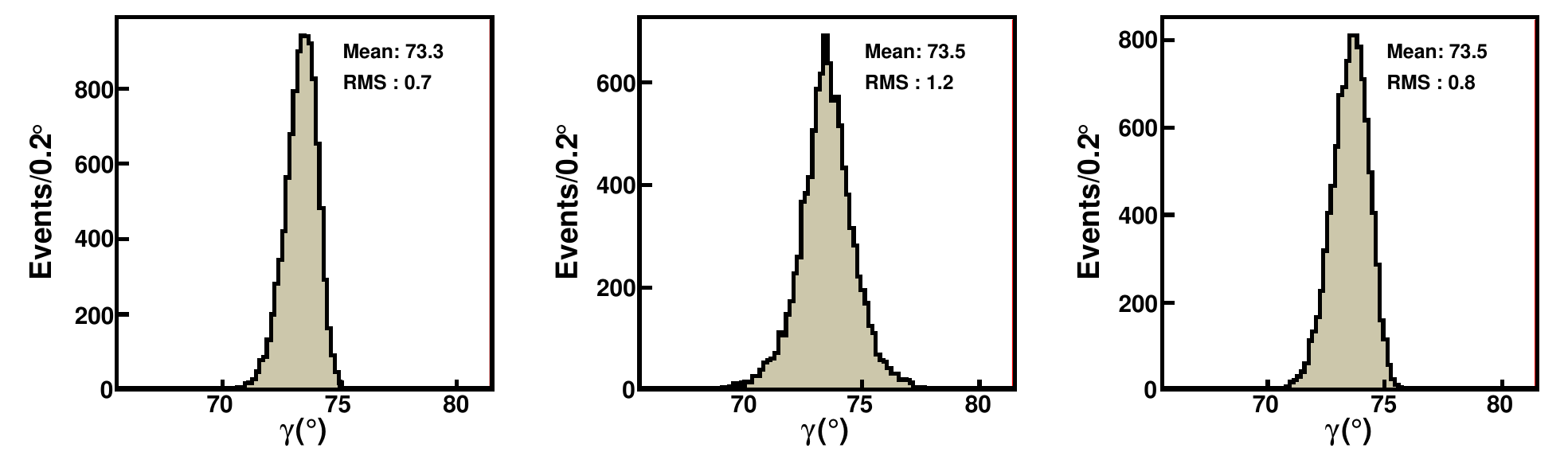}
\caption{The distribution of the fitted value of $\gamma$ in the (left) equal $\Delta\delta_D$, (middle) optimal and (right) modified optimal binning schemes, respectively.}
\label{fig:gamma}
\end{center}
\end{figure*}

In order to assess the impact of the uncertainty in the strong-phase parameters on a measurement of $\gamma$, a large simulated data set of $B^{\pm}$ events is generated according to the expected distribution given
the measured central values of $K_i$, $c_i$, and $s_i$ and the input values $\gamma=73.5^{\circ}$, $r_B=0.103$, and $\delta_B=136.9^{\circ}$, which are close to the current central values of these parameters from existing measurements~\cite{epjc77_895}. The simulated data are fit many times to determine $\gamma$, $\delta_B$ and $r_B$. The values of $c_i$ and $s_i$ used in each fit are sampled from the measured values smeared by their uncertainties, where the correlations between the measurements are taken into account.  The uncertainty on the measured $K_i$ is not considered since experiments are expected to use their own data to provide this input~\cite{jhep08_176}. The overall yield of the generated $B^{\pm}$ sample is sufficiently large to ensure that the statistical uncertainty from the fit is negligible. Therefore the width of the distribution of the fitted value of $\gamma$ is an estimate of the uncertainty on $\gamma$ due to the precision of the strong-phase parameters. The distribution of the fitted value of $\gamma$ in the three binning schemes is shown in Fig.~\ref{fig:gamma}.

Based on this study, the uncertainty on $\gamma$ due to the measured uncertainty on $c_i$ and $s_i$ is found to be 0.7$^{\circ}$, 1.2$^{\circ}$ and 0.8$^{\circ}$ for the equal $\Delta\delta_D$, optimal and modified optimal binning schemes, respectively. The very small phase space regions in the optimal binning scheme are the cause for the larger propagated uncertainty in this case. Very small biases of less than 0.2$^\circ$ are observed due to some values in the fit being unphysical, {\it i.e.} $c_i^2+s_i^2 > 1$. The size of the uncertainty on $\gamma$ is approximately a factor of three smaller than from the CLEO measurements~\cite{prd82_112006}. The predicted statistical uncertainties on $\gamma$ from LHCb prior to the start of High-Luminosity LHC operation in the mid 2020s, and from Belle II is expected to be 1.5$^\circ$~\cite{1808.08865,1808.10567}.
Therefore the uncertainty associated to the strong-phase measurements presented here will not be dominant in the determination of $\gamma$ for Belle II or for LHCb until then. The measurements of $c_i$ and $s_i$ can also be used for determination of strong-phase parameters in other multi-body decay modes of $D$ mesons, where the $D^0\rightarrow K^0_S\pi^+\pi^-$ decay is used as a tag~\cite{plb747_9,plb757_520,jhep01_082,jhep01_144,prd85_092016}.
Here, the improved precision leads to smaller systematic uncertainties on the strong-phase parameters in other $D$ decay modes, which subsequently reduces associated systematic uncertainties on $\gamma$ when these $D$ decay modes are use to measure $\gamma$ in $B^{\pm}\rightarrow DK^{\pm}$ decays.

\section{Summary}
\label{sec:sum}

Measurements of the relative strong-phase differences between $D^0$ and $\bar{D}^0\rightarrow K^0_{S,L}\pi^+\pi^-$ in bins of phase space have been performed using 2.93 fb$^{-1}$ of data collected at $\sqrt{s}$=3.773 GeV collected with the BESIII detector.
These results are on average a factor of 2.5 (1.9) more precise for $c_i$ ($s_i$) and a factor of 2.8 (2.2) more precise for $c^{\prime}_i$ ($s^{\prime}_i$) than the previous measurements of these parameters~\cite{prd82_112006}.
This improvement arises from the combination of a larger data sample, an increased variety of $CP$ tags, and broader use of the partial reconstruction technique to improve efficiency. The strong-phase parameters provide an important input in a wide range of $CP$ violation measurements in the beauty and charm sectors. The propagated uncertainty from these measurements on the CKM parameter $\gamma$ determined through the analysis of $B^{\pm}\rightarrow D_{K^0_S\pi^+\pi^-}K^{\pm}$ events is expected to be 0.7$^{\circ}$, 1.2$^{\circ}$ and 0.8$^{\circ}$ for the equal $\Delta\delta_D$, optimal and modified optimal binning schemes, respectively. This improved precision will ensure that measurements of $\gamma$ from LHCb and Belle II over the next decade are not limited by the knowledge of these strong-phase parameters, and also be invaluable
in studies of charm mixing and $CP$ violation.

\section{Acknowledgements}
The BESIII collaboration thanks the staff of BEPCII and the IHEP computing center for their strong support. This work is supported in part by National Key Basic Research Program of China under Contract No. 2015CB856700; National Natural Science Foundation of China (NSFC) under Contracts Nos. 11625523, 11635010, 11735014, 11775027, 11822506, 11835012; the Chinese Academy of Sciences (CAS) Large-Scale Scientific Facility Program; Joint Large-Scale Scientific Facility Funds of the NSFC and CAS under Contracts Nos. U1532257, U1532258, U1732263, U1832207; CAS Key Research Program of Frontier Sciences under Contracts Nos. QYZDJ-SSW-SLH003, QYZDJ-SSW-SLH040; 100 Talents Program of CAS; INPAC and Shanghai Key Laboratory for Particle Physics and Cosmology; ERC under Contract No. 758462; German Research Foundation DFG under Contracts Nos. Collaborative Research Center CRC 1044, FOR 2359; Istituto Nazionale di Fisica Nucleare, Italy; Koninklijke Nederlandse Akademie van Wetenschappen (KNAW) under Contract No. 530-4CDP03; Ministry of Development of Turkey under Contract No. DPT2006K-120470; National Science and Technology fund; STFC (United Kingdom); The Knut and Alice Wallenberg Foundation (Sweden) under Contract No. 2016.0157; The Royal Society, UK under Contracts Nos. DH140054, DH160214; The Swedish Research Council; U. S. Department of Energy under Contracts Nos. DE-FG02-05ER41374, DE-SC-0010118, DE-SC-0012069; University of Groningen (RuG) and the Helmholtzzentrum fuer Schwerionenforschung GmbH (GSI), Darmstadt; This paper is also supported by Beijing municipal government under Contract No CIT\&TCD201704047, and by the Royal Society under Contracts No. NF170002.


\newpage
\section{Appendix A}
\label{sec:appendixa}

As the results presented here and those from the CLEO collaboration~\cite{prd82_112006} are compatible it is legitimate to combine them in order to provide a single set of results that benefits from both measurements. The combination is performed by performing the fit described in Sec.~\ref{sec:cisi} to the double tags with an additional multi-dimensional Gaussian constraint present on the strong-phase parameters. This constraint comes from the central values and the covariance matrices in Ref.~\cite{prd82_112006}.
A small, additional contribution to these covariance matrices, determined through pseudo-experiments, accounts for the effects reported in Ref.~\cite{plb765_402}.

The systematic uncertainties reported in Tables~\ref{tab:systot_equal} $-$~\ref{tab:systot_mdoptimal} are added in quadrature to those from the fit, which include contributions from the BESIII statistical and CLEO statistical and systematic uncertainties. The central values and their uncertainties for the three binning schemes are reported in Table~\ref{tab:cisifit_cmbcleo} and Tables~\ref{tab:systot_equal_cmbcleo}, \ref{tab:systot_optimal_cmbcleo}, and \ref{tab:systot_mdoptimal_cmbcleo} show the associated covariance matrices.

\begin{table*}
\caption{ The statistical correlation coefficients (\%) between the $c_i$, $s_i$, $c^{\prime}_i$ and $s^{\prime}_i$ parameters for the equal $\Delta\delta_D$ binning scheme.}
\begin{center}
\resizebox{!}{4.5cm}{
\begin{tabular}{l|cccccccccccccccccccccccccccccccc}
\hline\hline
      & $c_2$          & $c_3$          & $c_4$          & $c_5$          & $c_6$          & $c_7$          & $c_8$          & $s_1$          & $s_2$          & $s_3$          & $s_4$          & $s_5$          & $s_6$          & $s_7$          & $s_8$          & $c^{\prime}_1$ & $c^{\prime}_2$ & $c^{\prime}_3$ & $c^{\prime}_4$ & $c^{\prime}_5$ & $c^{\prime}_6$ & $c^{\prime}_7$ & $c^{\prime}_8$ & $s^{\prime}_1$ & $s^{\prime}_2$ & $s^{\prime}_3$ & $s^{\prime}_4$ & $s^{\prime}_5$ & $s^{\prime}_6$ & $s^{\prime}_7$ & $s^{\prime}_8$ \\
\hline
$c_1$          & -10 & -1 & 1 & 4 & 0 & 1 & -12 & 0 & 0 & 0 & 0 & 1 & 0 & 0 & 0 & 38 & -10 & 0 & 1 & 4 & 0 & 0 & -11 & 0 & 0 & 0 & 0 & 1 & 0 & 0 & 0 \\
$c_2$          &   & -6 & 1 & 4 & 1 & -1 & 0 & 0 & 1 & 0 & 0 & 3 & 0 & 0 & -1 & -10 & 85 & -6 & 0 & 2 & -1 & -2 & 0 & 0 & 1 & 0 & 0 & 3 & 0 & 0 & -1 \\
$c_3$          &   &   & -5 & 0 & -3 & 0 & 0 & -5 & -4 & 1 & 4 & -4 & 3 & -2 & -2 & 0 & -5 & 31 & -4 & 0 & -2 & 0 & -1 & -6 & -4 & 1 & 4 & -4 & 3 & -2 & -2 \\
$c_4$          &   &   &   & -5 & -1 & -2 & 2 & 2 & -1 & 0 & 4 & 6 & 0 & 1 & -1 & 2 & 1 & -3 & 32 & -4 & 0 & -2 & 2 & 3 & -1 & 0 & 4 & 5 & 0 & 1 & -2 \\
$c_5$          &   &   &   &   & -6 & 0 & 1 & 0 & 0 & 0 & 0 & -1 & 0 & 0 & 0 & 7 & 4 & 1 & -3 & 62 & -5 & 1 & 2 & -1 & 0 & 0 & 0 & -1 & 0 & 0 & 0 \\
$c_6$          &   &   &   &   &   & -6 & 0 & 4 & 2 & 0 & -1 & 0 & -5 & 1 & 2 & 2 & 1 & -1 & 0 & -5 & 25 & -4 & 1 & 4 & 2 & 0 & -1 & 0 & -4 & 1 & 2 \\
$c_7$          &   &   &   &   &   &   & -9 & 4 & 3 & 1 & -1 & 0 & -1 & 0 & 2 & 1 & -1 & 1 & -1 & 0 & -3 & 35 & -7 & 4 & 2 & 1 & -1 & -1 & -1 & 0 & 3 \\
\hline\hline
\end{tabular}
}
\label{tab:sta_equal_corela}
\end{center}
\end{table*}
\begin{table*}
\caption{ The correlation coefficients (\%) of systematic uncertainties between $c_i$, $s_i$, $c^{\prime}_i$ and $s^{\prime}_i$ parameters for the equal $\Delta\delta_D$ binning scheme.}
\begin{center}
\resizebox{!}{4.8cm}{
\begin{tabular}{l|cccccccccccccccccccccccccccccccc}
\hline\hline
       & $c_2$          & $c_3$          & $c_4$          & $c_5$          & $c_6$          & $c_7$          & $c_8$          & $s_1$          & $s_2$          & $s_3$          & $s_4$          & $s_5$          & $s_6$          & $s_7$          & $s_8$          & $c^{\prime}_1$ & $c^{\prime}_2$ & $c^{\prime}_3$ & $c^{\prime}_4$ & $c^{\prime}_5$ & $c^{\prime}_6$ & $c^{\prime}_7$ & $c^{\prime}_8$ & $s^{\prime}_1$ & $s^{\prime}_2$ & $s^{\prime}_3$ & $s^{\prime}_4$ & $s^{\prime}_5$ & $s^{\prime}_6$ & $s^{\prime}_7$ & $s^{\prime}_8$ \\
\hline
$c_1$          & 35 & 65 & 50 & 32 & 56 & 72 & 60 & 14 & 11 & 14 & 12 & 14 & 6 & 10 & 17 & 48 & 30 & 31 & 26 & 24 & 20 & 21 & 34 & 13 & 14 & 15 & 11 & 13 & 7 & 10 & 17 \\
$c_2$          &   & 37 & 30 & 22 & 34 & 42 & 36 & 14 & 8 & 8 & 13 & 12 & 9 & 6 & 10 & 14 & 96 & 19 & 16 & 17 & 16 & 15 & 24 & 15 & 8 & 9 & 12 & 12 & 9 & 7 & 11 \\
$c_3$          &   &   & 57 & 42 & 56 & 72 & 66 & 29 & 14 & 20 & 7 & 13 & 19 & 7 & 31 & 38 & 34 & 45 & 35 & 32 & 29 & 20 & 45 & 29 & 17 & 22 & 5 & 14 & 20 & 7 & 29 \\
$c_4$          &   &   &   & 40 & 48 & 58 & 53 & 26 & 4 & 23 & 1 & 15 & 23 & 10 & 28 & 37 & 28 & 38 & 48 & 33 & 31 & 16 & 39 & 25 & 7 & 25 & 0 & 16 & 24 & 9 & 26 \\
$c_5$          &   &   &   &   & 32 & 41 & 39 & 35 & 15 & 25 & 11 & 20 & 27 & 12 & 38 & 40 & 23 & 34 & 41 & 58 & 35 & 15 & 35 & 34 & 19 & 26 & 9 & 22 & 28 & 11 & 36 \\
$c_6$          &   &   &   &   &   & 61 & 53 & 27 & 17 & 18 & 9 & 18 & 47 & 9 & 23 & 30 & 31 & 31 & 30 & 26 & 50 & 21 & 35 & 28 & 21 & 20 & 8 & 18 & 47 & 11 & 24 \\
$c_7$          &   &   &   &   &   &   & 69 & 25 & 13 & 16 & 8 & 13 & 13 & -6 & 24 & 34 & 37 & 38 & 33 & 31 & 26 & 42 & 45 & 25 & 16 & 17 & 6 & 12 & 14 & -6 & 22 \\
\hline\hline
\end{tabular}
}
\label{tab:sys_equal_corela}
\end{center}
\end{table*}

\begin{table*}
\caption{ The correlation coefficients (\%) of statistical uncertainties between the $c_i$, $s_i$, $c^{\prime}_i$ and $s^{\prime}_i$ parameters for the optimal binning scheme.}
\begin{center}
\resizebox{!}{4.5cm}{
\begin{tabular}{l|cccccccccccccccccccccccccccccccc}
\hline\hline
       & $c_2$          & $c_3$          & $c_4$          & $c_5$          & $c_6$          & $c_7$          & $c_8$          & $s_1$          & $s_2$          & $s_3$          & $s_4$          & $s_5$          & $s_6$          & $s_7$          & $s_8$          & $c^{\prime}_1$ & $c^{\prime}_2$ & $c^{\prime}_3$ & $c^{\prime}_4$ & $c^{\prime}_5$ & $c^{\prime}_6$ & $c^{\prime}_7$ & $c^{\prime}_8$ & $s^{\prime}_1$ & $s^{\prime}_2$ & $s^{\prime}_3$ & $s^{\prime}_4$ & $s^{\prime}_5$ & $s^{\prime}_6$ & $s^{\prime}_7$ & $s^{\prime}_8$ \\
\hline
$c_1$          & -2 & -4 & -3 & -1 & -2 & -2 & -1 & -2 & 5 & 3 & -2 & -1 & 2 & 4 & -4 & 27 & -1 & -4 & -2 & 0 & -1 & -2 & -1 & -1 & 5 & 3 & -1 & -1 & 2 & 4 & -6 \\
$c_2$          &   & -12 & 5 & 0 & -6 & -5 & 0 & 0 & -1 & 0 & -2 & 0 & 0 & 0 & 0 & -2 & 69 & -13 & 2 & 1 & -5 & -6 & 0 & 0 & -1 & 1 & -2 & 0 & 0 & 0 & 0 \\
$c_3$          &   &   & 0 & -2 & -2 & 0 & 0 & 0 & 3 & -1 & 0 & -1 & 2 & 1 & -1 & -8 & -14 & 81 & 0 & 0 & 0 & 0 & 0 & 0 & 4 & -1 & 0 & 0 & 2 & 2 & -2 \\
$c_4$          &   &   &   & -4 & 0 & 6 & 3 & 0 & 0 & 0 & -4 & 0 & 0 & 0 & 0 & 0 & 7 & 0 & 27 & -1 & 1 & 7 & 3 & 0 & 0 & 0 & -3 & 0 & 0 & 0 & 0 \\
$c_5$          &   &   &   &   & -2 & 0 & 0 & 1 & 0 & 0 & 1 & 10 & -1 & 0 & 2 & -2 & 0 & -2 & -2 & 12 & 0 & 1 & 0 & 1 & 0 & -1 & 1 & 7 & -1 & 0 & 2 \\
$c_6$          &   &   &   &   &   & 0 & 1 & 0 & -6 & -1 & 1 & 0 & 1 & -3 & 2 & -3 & -6 & -2 & 0 & -3 & 23 & 0 & 1 & 0 & -6 & -1 & 2 & 0 & 0 & -3 & 3 \\
$c_7$          &   &   &   &   &   &   & -5 & 0 & 0 & 0 & 2 & 0 & 0 & -4 & 1 & -1 & -6 & 0 & 5 & 1 & 0 & 83 & -4 & 0 & 0 & 0 & 2 & 0 & 0 & -4 & 1 \\
\hline\hline
\end{tabular}
}
\label{tab:sta_optimal_corela}
\end{center}
\end{table*}

\begin{table*}
\caption{ The correlation coefficients for the systematic uncertainties (\%) between the $c_i$, $s_i$, $c^{\prime}_i$ and $s^{\prime}_i$ parameters for the optimal binning scheme.}
\begin{center}
\resizebox{!}{4.8cm}{
\begin{tabular}{l|cccccccccccccccccccccccccccccccc}
\hline\hline
   & $c_2$          & $c_3$          & $c_4$          & $c_5$          & $c_6$          & $c_7$          & $c_8$          & $s_1$          & $s_2$          & $s_3$          & $s_4$          & $s_5$          & $s_6$          & $s_7$          & $s_8$          & $c^{\prime}_1$ & $c^{\prime}_2$ & $c^{\prime}_3$ & $c^{\prime}_4$ & $c^{\prime}_5$ & $c^{\prime}_6$ & $c^{\prime}_7$ & $c^{\prime}_8$ & $s^{\prime}_1$ & $s^{\prime}_2$ & $s^{\prime}_3$ & $s^{\prime}_4$ & $s^{\prime}_5$ & $s^{\prime}_6$ & $s^{\prime}_7$ & $s^{\prime}_8$ \\
\hline
$c_1$          & 35 & 44 & 49 & 78 & 71 & 33 & 40 & 7 & 6 & 4 & -6 & 14 & 14 & 9 & 8 & 28 & 21 & 30 & 15 & 16 & 16 & 30 & 25 & 7 & 7 & 5 & 1 & 14 & 15 & 11 & 2 \\
$c_2$          &   & 14 & 28 & 30 & 27 & 13 & 21 & 12 & -54 & 3 & -12 & 14 & 2 & 11 & 11 & 13 & 94 & 8 & 17 & 12 & 7 & 13 & 15 & 12 & -48 & 5 & -6 & 13 & 3 & 10 & 9 \\
$c_3$          &   &   & 26 & 41 & 37 & 21 & 24 & 4 & 11 & -48 & -3 & 5 & 9 & 6 & 6 & 17 & 10 & 95 & 9 & 15 & 15 & 20 & 19 & 4 & 9 & -43 & 0 & 5 & 9 & 7 & 2 \\
$c_4$          &   &   &   & 42 & 36 & 26 & 22 & 18 & 4 & 9 & 2 & 23 & 20 & 21 & 13 & 26 & 23 & 19 & 45 & 16 & 14 & 27 & 16 & 18 & 8 & 12 & 9 & 25 & 22 & 21 & 8 \\
$c_5$          &   &   &   &   & 65 & 27 & 38 & 5 & 7 & 4 & -4 & 10 & 12 & 9 & 15 & 17 & 19 & 28 & 15 & 11 & 19 & 25 & 23 & 5 & 5 & 1 & 1 & 7 & 12 & 9 & 9 \\
$c_6$          &   &   &   &   &   & 25 & 29 & 3 & 3 & 9 & -5 & 10 & 37 & 3 & 2 & 14 & 16 & 25 & 9 & 9 & 51 & 22 & 16 & 3 & 3 & 8 & -1 & 9 & 34 & 5 & -4 \\
$c_7$          &   &   &   &   &   &   & 14 & 12 & 6 & 5 & 11 & 9 & 13 & -2 & 7 & 17 & 9 & 16 & 14 & 11 & 11 & 93 & 11 & 12 & 10 & 10 & 18 & 12 & 15 & -1 & 4 \\
\hline\hline
\end{tabular}
}
\label{tab:sys_optimal_corela}
\end{center}
\end{table*}

\begin{table*}
\caption{ The correlation coefficients (\%) of the statistical uncertainties between the strong-phase parameters for the modified optimal binning scheme.}
\begin{center}
\resizebox{!}{4.5cm}{
\begin{tabular}{l|cccccccccccccccccccccccccccccccc}
\hline\hline
      & $c_2$          & $c_3$          & $c_4$          & $c_5$          & $c_6$          & $c_7$          & $c_8$          & $s_1$          & $s_2$          & $s_3$          & $s_4$          & $s_5$          & $s_6$          & $s_7$          & $s_8$          & $c^{\prime}_1$ & $c^{\prime}_2$ & $c^{\prime}_3$ & $c^{\prime}_4$ & $c^{\prime}_5$ & $c^{\prime}_6$ & $c^{\prime}_7$ & $c^{\prime}_8$ & $s^{\prime}_1$ & $s^{\prime}_2$ & $s^{\prime}_3$ & $s^{\prime}_4$ & $s^{\prime}_5$ & $s^{\prime}_6$ & $s^{\prime}_7$ & $s^{\prime}_8$ \\
\hline
$c_1$          & -2 & -1 & -5 & -1 & 0 & -2 & 1 & -1 & 1 & 0 & 0 & 0 & 0 & 0 & 0 & 27 & -2 & 0 & -4 & 0 & 1 & -1 & 1 & -1 & 1 & 0 & 0 & 0 & 0 & 0 & 0 \\
$c_2$          &   & -7 & 12 & 1 & -2 & -3 & -2 & 1 & -6 & 0 & -2 & 0 & -1 & 0 & 1 & 0 & 93 & -11 & 8 & 1 & -2 & -3 & -2 & 1 & -6 & 0 & -2 & 0 & -1 & 0 & 1 \\
$c_3$          &   &   & 0 & -3 & -3 & -2 & 0 & -2 & 10 & 1 & 1 & 0 & 3 & 2 & -4 & 1 & -7 & 31 & -1 & -1 & -1 & -1 & 1 & -2 & 10 & 1 & 0 & 0 & 4 & 2 & -5 \\
$c_4$          &   &   &   & -3 & -1 & 5 & 3 & 1 & 3 & 0 & -6 & -1 & -1 & 2 & 0 & -3 & 12 & 0 & 63 & -1 & 0 & 8 & 3 & 1 & 3 & 0 & -5 & -1 & -1 & 2 & 0 \\
$c_5$          &   &   &   &   & -2 & 0 & 1 & -2 & 0 & 3 & 2 & 13 & 1 & -2 & 0 & -1 & 1 & -3 & -2 & 19 & -2 & 1 & 0 & -2 & 0 & 1 & 2 & 11 & 1 & -1 & 1 \\
$c_6$          &   &   &   &   &   & -3 & 1 & 2 & -6 & -2 & -1 & 0 & 1 & -3 & 2 & 1 & -2 & -2 & -1 & -3 & 37 & -2 & 1 & 2 & -6 & -4 & -1 & 0 & 0 & -3 & 3 \\
$c_7$          &   &   &   &   &   &   & -4 & 0 & 0 & 0 & 1 & 0 & 0 & -2 & 0 & -1 & -3 & -3 & 4 & 1 & -3 & 51 & -3 & 0 & 0 & 0 & 1 & 0 & 0 & -3 & 0 \\
\hline\hline
\end{tabular}
}
\label{tab:sta_mdoptimal_corela}
\end{center}
\end{table*}

\begin{table*}
\caption{ The correlation coefficients (\%) of the systematic uncertainties between the $c_i$, $s_i$, $c^{\prime}_i$ and $s^{\prime}_i$ parameters for the modified optimal binning scheme.}
\begin{center}
\resizebox{!}{4.8cm}{
\begin{tabular}{l|cccccccccccccccccccccccccccccccc}
\hline\hline
          & $c_2$          & $c_3$          & $c_4$          & $c_5$          & $c_6$          & $c_7$          & $c_8$          & $s_1$          & $s_2$          & $s_3$          & $s_4$          & $s_5$          & $s_6$          & $s_7$          & $s_8$          & $c^{\prime}_1$ & $c^{\prime}_2$ & $c^{\prime}_3$ & $c^{\prime}_4$ & $c^{\prime}_5$ & $c^{\prime}_6$ & $c^{\prime}_7$ & $c^{\prime}_8$ & $s^{\prime}_1$ & $s^{\prime}_2$ & $s^{\prime}_3$ & $s^{\prime}_4$ & $s^{\prime}_5$ & $s^{\prime}_6$ & $s^{\prime}_7$ & $s^{\prime}_8$ \\
\hline
$c_1$          & 36 & 70 & 40 & 53 & 63 & 52 & 40 & 14 & 16 & 17 & 4 & 17 & 24 & 18 & 5 & 42 & 34 & 33 & 31 & 30 & 31 & 38 & 23 & 17 & 15 & 16 & 4 & 17 & 27 & 17 & 2 \\
$c_2$          &   & 38 & 34 & 28 & 35 & 30 & 25 & 5 & -3 & 15 & 2 & 11 & 15 & 16 & 9 & 28 & 89 & 23 & 29 & 24 & 24 & 26 & 18 & 6 & -5 & 14 & 2 & 12 & 17 & 15 & 7 \\
$c_3$          &   &   & 43 & 65 & 77 & 53 & 53 & 1 & 19 & 2 & 3 & 19 & 18 & 12 & 7 & 31 & 35 & 44 & 31 & 26 & 34 & 33 & 31 & 1 & 18 & 0 & 3 & 19 & 19 & 10 & 4 \\
$c_4$          &   &   &   & 30 & 38 & 40 & 22 & 0 & 30 & 19 & 14 & 13 & 25 & 26 & 3 & 36 & 34 & 32 & 62 & 31 & 25 & 37 & 14 & 3 & 31 & 21 & 15 & 14 & 29 & 25 & -2 \\
$c_5$          &   &   &   &   & 59 & 35 & 44 & 12 & 18 & 6 & 15 & 24 & 21 & 7 & 11 & 17 & 26 & 20 & 27 & 16 & 29 & 18 & 25 & 11 & 16 & 2 & 15 & 24 & 18 & 7 & 11 \\
$c_6$          &   &   &   &   &   & 48 & 51 & -2 & 14 & 10 & -5 & 9 & 19 & 8 & 7 & 28 & 33 & 29 & 22 & 23 & 51 & 31 & 33 & -1 & 13 & 8 & -4 & 9 & 21 & 7 & 4 \\
$c_7$          &   &   &   &   &   &   & 26 & 3 & 15 & 23 & 6 & 13 & 26 & 14 & 1 & 41 & 29 & 33 & 33 & 32 & 29 & 64 & 16 & 6 & 15 & 22 & 7 & 14 & 30 & 9 & -2 \\
\hline\hline
\end{tabular}
}
\label{tab:sys_mdoptimal_corela}
\end{center}
\end{table*}

\begin{table*}
\caption{The measured strong-phase difference parameters $c_i$, $s_i$,  $c^{\prime}_i$ and $s^{\prime}_i$ where the results reported in Ref.~\cite{prd82_112006} are used as a constraint.  }
\begin{center}
\begin{tabular}{ccccccc} \hline
\hline
    & \multicolumn{4}{c}{Equal binning scheme } \\
    &      $c_i$       &       $s_i$        &      $c^{\prime}_i$       &       $s^{\prime}_i$             \\
\hline
 1 & $~~0.699\pm0.020$ & $~~0.091\pm0.063$ & $~~0.800\pm0.023$ & $~~0.101\pm0.065$ \\
 2 & $~~0.643\pm0.036$ & $~~0.300\pm0.110$ & $~~0.823\pm0.037$ & $~~0.266\pm0.116$ \\
 3 & $~~0.001\pm0.047$ & $~~1.000\pm0.075$ & $~~0.186\pm0.047$ & $~~0.946\pm0.083$ \\
 4 & $-0.608\pm0.052$ & $~~0.660\pm0.123$ & $-0.512\pm0.055$ & $0.730\pm0.129$ \\
 5 & $-0.955\pm0.023$ & $-0.032\pm0.069$ & $-0.961\pm0.027$ & $-0.060\pm0.079$ \\
 6 & $-0.578\pm0.058$ & $-0.545\pm0.122$ & $-0.371\pm0.069$ & $-0.610\pm0.131$ \\
\hline\hline
\end{tabular}
\label{tab:cisifit_cmbcleo}
\end{center}
\end{table*}

\begin{table*}
\caption{ The correlation coefficients (\%) between the strong-phase parameters in the equal binning scheme, where the results reported in Ref.~\cite{prd82_112006} are used as a constraint.  }
\begin{center}
\resizebox{!}{4.8cm}{
\begin{tabular}{l|cccccccccccccccccccccccccccccccc}
\hline\hline
                 & $c_2$          & $c_3$          & $c_4$          & $c_5$          & $c_6$          & $c_7$          & $c_8$          & $s_1$          & $s_2$          & $s_3$          & $s_4$          & $s_5$          & $s_6$          & $s_7$          & $s_8$          & $c^{\prime}_1$ & $c^{\prime}_2$ & $c^{\prime}_3$ & $c^{\prime}_4$ & $c^{\prime}_5$ & $c^{\prime}_6$ & $c^{\prime}_7$ & $c^{\prime}_8$ & $s^{\prime}_1$ & $s^{\prime}_2$ & $s^{\prime}_3$ & $s^{\prime}_4$ & $s^{\prime}_5$ & $s^{\prime}_6$ & $s^{\prime}_7$ & $s^{\prime}_8$ \\
\hline
$c_1$          & 1 & 12 & 9 & 12 & 12 & 15 & 5 & 2 & 1 & 0 & 1 & 1 & 1 & 1 & 2 & 44 & 1 & 5 & 6 & 10 & 3 & 4 & 0 & 2 & 1 & 1 & 1 & 2 & 1 & 1 & 3 \\
$c_2$          &   & 4 & 6 & 9 & 8 & 8 & 9 & 2 & 1 & -1 & 1 & 3 & 1 & 0 & 1 & -3 & 87 & -1 & 3 & 7 & 2 & 2 & 6 & 2 & 1 & 0 & 1 & 3 & 1 & 1 & 1 \\
$c_3$          &   &   & 4 & 10 & 8 & 14 & 13 & 1 & -4 & 18 & 3 & -2 & 4 & 6 & 4 & 9 & 3 & 34 & 3 & 8 & 3 & 3 & 8 & 0 & -3 & 15 & 3 & -2 & 4 & 5 & 2 \\
$c_4$          &   &   &   & 5 & 6 & 7 & 11 & 3 & -1 & 1 & 3 & 4 & 2 & 0 & 2 & 8 & 6 & 2 & 34 & 6 & 3 & 0 & 8 & 3 & -1 & 1 & 3 & 3 & 2 & -1 & 1 \\
$c_5$          &   &   &   &   & 4 & 11 & 12 & 5 & 1 & 3 & 1 & -1 & 4 & 1 & 7 & 18 & 10 & 8 & 8 & 60 & 4 & 4 & 11 & 4 & 1 & 3 & 0 & 0 & 5 & 0 & 6 \\
$c_6$          &   &   &   &   &   & 7 & 11 & 5 & 2 & 1 & 0 & 0 & 2 & 1 & 5 & 9 & 7 & 4 & 5 & 4 & 31 & 0 & 8 & 6 & 3 & 2 & 0 & 0 & 2 & 2 & 4 \\
$c_7$          &   &   &   &   &   &   & 9 & 7 & 3 & 2 & 0 & 1 & 0 & -2 & 6 & 9 & 8 & 7 & 5 & 9 & 2 & 33 & 5 & 7 & 3 & 3 & -1 & 0 & 0 & -2 & 6 \\
\hline\hline
\end{tabular}
}
\label{tab:systot_equal_cmbcleo}
\end{center}
\end{table*}

\begin{table*}
\caption{ The correlation coefficients (\%) between the strong-phase parameters in the optimal binning scheme, where the results reported in Ref.~\cite{prd82_112006} are used as a constraint.  }
\begin{center}
\resizebox{!}{4.8cm}{
\begin{tabular}{l|cccccccccccccccccccccccccccccccc}
\hline\hline
                 & $c_2$          & $c_3$          & $c_4$          & $c_5$          & $c_6$          & $c_7$          & $c_8$          & $s_1$          & $s_2$          & $s_3$          & $s_4$          & $s_5$          & $s_6$          & $s_7$          & $s_8$          & $c^{\prime}_1$ & $c^{\prime}_2$ & $c^{\prime}_3$ & $c^{\prime}_4$ & $c^{\prime}_5$ & $c^{\prime}_6$ & $c^{\prime}_7$ & $c^{\prime}_8$ & $s^{\prime}_1$ & $s^{\prime}_2$ & $s^{\prime}_3$ & $s^{\prime}_4$ & $s^{\prime}_5$ & $s^{\prime}_6$ & $s^{\prime}_7$ & $s^{\prime}_8$ \\
\hline
$c_1$          & 6 & 5 & 6 & 11 & 8 & 5 & 8 & 0 & 5 & 3 & -1 & 1 & 4 & 3 & -3 & 27 & 5 & 3 & 2 & 1 & 3 & 5 & 4 & 0 & 6 & 2 & 0 & 1 & 4 & 4 & -5 \\
$c_2$          &   & -4 & 11 & 6 & 2 & 0 & 5 & 3 & -7 & 1 & -3 & 2 & 0 & 2 & 1 & 0 & 79 & -6 & 7 & 3 & -2 & 0 & 3 & 2 & -6 & 1 & -2 & 2 & 0 & 2 & 0 \\
$c_3$          &   &   & 6 & 6 & 5 & 5 & 5 & 1 & 5 & -5 & 0 & 0 & 4 & 2 & -2 & -3 & -6 & 83 & 3 & 2 & 4 & 5 & 4 & 1 & 4 & -5 & 0 & 0 & 5 & 2 & -2 \\
$c_4$          &   &   &   & 5 & 6 & 11 & 7 & 3 & 1 & 1 & -3 & 3 & 2 & 3 & 1 & 3 & 11 & 6 & 32 & 2 & 5 & 12 & 6 & 2 & 0 & 1 & -2 & 3 & 2 & 2 & 0 \\
$c_5$          &   &   &   &   & 7 & 5 & 7 & 1 & 0 & 0 & -1 & 5 & 1 & 1 & 3 & 1 & 4 & 4 & 2 & 13 & 4 & 5 & 4 & 1 & 0 & -1 & 0 & 4 & 1 & 0 & 2 \\
$c_6$          &   &   &   &   &   & 4 & 6 & 1 & -6 & -2 & 2 & 1 & 7 & -2 & 3 & -1 & 1 & 3 & 2 & -1 & 29 & 4 & 4 & 1 & -6 & -3 & 2 & 0 & 5 & -2 & 3 \\
$c_7$          &   &   &   &   &   &   & 0 & 2 & 1 & 1 & 1 & 2 & 2 & -3 & 1 & 1 & -1 & 4 & 8 & 2 & 2 & 87 & 1 & 1 & 1 & 1 & 2 & 2 & 2 & -2 & 0 \\
\hline\hline
\end{tabular}
}
\label{tab:systot_optimal_cmbcleo}
\end{center}
\end{table*}

\begin{table*}
\caption{ The correlation coefficients (\%) between the strong-phase parameters in the modified binning scheme, where the results reported in Ref.~\cite{prd82_112006} are used as a constraint. }
\begin{center}
\resizebox{!}{4.8cm}{
\begin{tabular}{l|cccccccccccccccccccccccccccccccc}
\hline\hline
                 & $c_2$          & $c_3$          & $c_4$          & $c_5$          & $c_6$          & $c_7$          & $c_8$          & $s_1$          & $s_2$          & $s_3$          & $s_4$          & $s_5$          & $s_6$          & $s_7$          & $s_8$          & $c^{\prime}_1$ & $c^{\prime}_2$ & $c^{\prime}_3$ & $c^{\prime}_4$ & $c^{\prime}_5$ & $c^{\prime}_6$ & $c^{\prime}_7$ & $c^{\prime}_8$ & $s^{\prime}_1$ & $s^{\prime}_2$ & $s^{\prime}_3$ & $s^{\prime}_4$ & $s^{\prime}_5$ & $s^{\prime}_6$ & $s^{\prime}_7$ & $s^{\prime}_8$ \\
\hline
$c_1$          & 6 & 12 & 3 & 6 & 8 & 8 & 6 & 0 & 3 & 2 & 0 & 2 & 2 & 2 & 0 & 41 & 6 & 5 & 3 & 3 & 5 & 7 & 4 & 0 & 3 & 2 & 1 & 2 & 3 & 2 & -1 \\
$c_2$          &   & 8 & 18 & 7 & 7 & 7 & 4 & 1 & -4 & 2 & -2 & 2 & 2 & 2 & 1 & 7 & 91 & 0 & 15 & 5 & 5 & 7 & 2 & 1 & -5 & 2 & -1 & 2 & 2 & 2 & 1 \\
$c_3$          &   &   & 10 & 9 & 12 & 12 & 10 & -1 & 10 & 1 & 1 & 3 & 5 & 2 & -3 & 9 & 8 & 32 & 8 & 2 & 6 & 10 & 7 & -1 & 10 & -1 & 0 & 3 & 5 & 2 & -4 \\
$c_4$          &   &   &   & 3 & 6 & 13 & 6 & 0 & 6 & 2 & -3 & 2 & 3 & 3 & 0 & 6 & 18 & 8 & 60 & 3 & 5 & 15 & 5 & -1 & 6 & 2 & -3 & 2 & 3 & 3 & -1 \\
$c_5$          &   &   &   &   & 5 & 6 & 6 & 0 & 1 & 1 & 0 & 12 & 2 & -1 & 1 & 2 & 6 & 2 & 4 & 19 & 2 & 5 & 4 & 0 & 1 & 1 & 0 & 10 & 1 & 0 & 1 \\
$c_6$          &   &   &   &   &   & 6 & 8 & 1 & -4 & -1 & 0 & 1 & 2 & -1 & 3 & 5 & 7 & 3 & 4 & 0 & 39 & 5 & 6 & 1 & -5 & -2 & 0 & 1 & 2 & -2 & 2 \\
$c_7$          &   &   &   &   &   &   & 2 & 0 & 2 & 3 & 1 & 2 & 3 & -1 & 0 & 8 & 6 & 5 & 12 & 5 & 3 & 57 & 1 & 0 & 2 & 2 & 1 & 2 & 4 & -2 & -1 \\
\hline\hline
\end{tabular}
}
\label{tab:systot_mdoptimal_cmbcleo}
\end{center}
\end{table*}

\end{document}